\documentclass[iop,apj,tighten,numberedappendix]{emulateapj}
\usepackage[breaklinks,colorlinks,urlcolor=blue,citecolor=blue,linkcolor=blue]{hyperref}
\usepackage{amsmath,amssymb}
\usepackage{cleveref}
\usepackage{graphicx}
\usepackage{bm}
\usepackage[toc,title,page]{appendix}
\usepackage{tocbibind}
\usepackage{xcolor,ulem}
\usepackage{color}

\usepackage{multirow}

\newcommand{\Msun}{{\rm M}_\odot}

\defcitealias{Thompson05}{TQM}

\usepackage{lineno}

\def\lsim{~\rlap{$<$}{\lower 1.0ex\hbox{$\sim$}}}
\def\gsim{~\rlap{$>$}{\lower 1.0ex\hbox{$\sim$}}}

\shorttitle{BH Mergers in AGNs}
\shortauthors{et al.}

\begin{document}
\title{
Properties of black hole mergers in disks of active galactic nuclei 
}

\author{Hiromichi Tagawa\altaffilmark{1}, 
Zolt\'an Haiman\altaffilmark{2,3,4}, 
Bence Kocsis\altaffilmark{5,6}
}

\affil{
\altaffilmark{1}Shanghai Astronomical Observatory, Shanghai, 200030, People$^{\prime}$s Republic of China\\
\altaffilmark{2}Institute of Science and Technology Austria, Am Campus 1, Klosterneuburg 3400, Austria\\
\altaffilmark{3}Department of Astronomy, Columbia University, 550 W. 120th St., New York, NY, 10027, USA\\
\altaffilmark{4}Department of Physics, Columbia University, 550 W. 120th St., New York, NY, 10027, USA\\
\altaffilmark{5}Rudolf Peierls Centre for Theoretical Physics, University of Oxford, Clarendon Laboratory, Parks Road, Oxford, OX1 3PU, UK\\
\altaffilmark{6}St Hugh’s College, University of Oxford, St Margaret’s Rd, Oxford, OX2 6LE, UK
}
\email{E-mail: htagawa@shao.ac.jp}

\begin{abstract} 
Ground-based gravitational wave (GW) observatories have detected approximately 200 binary black hole (BH) mergers. The astrophysical origin of these events are debated, with evidence suggesting that at least a subset originated from dynamic environments characterized by frequent close encounters. Accretion disks in active galactic nuclei (AGNs) are of particular interest, as certain observed features could be more readily produced within such environments.
In this paper, we investigate the 
expected properties 
of mergers in these environments, and their dependence on various parameters, using one-dimensional $N$-body simulations combined with a comprehensive semi-analytical model. 
In our fiducial model, the distributions of masses ($m_1,~m_2$) and mass ratios ($q\equiv m_2/m_1\leq1$) are similar to those observed. However, they depend strongly on the lifetime and density of the AGN disk and on the number and accretion efficiency of BHs, with higher masses predicted as these quantities increase. 
The most massive mergers, such as GW231123, can be produced either by efficient gas accretion or by hierarchical mergers among $\geq 3$ generations of BHs. 
The observed negative correlation between $q$ and the average effective spin ($\chi_{\rm eff}$), along with the positive correlation between $\chi_{\rm eff}$ and the chirp mass ($M_{\rm chirp}$), can be explained by a combination of efficient gas accretion, which promotes spin alignment, and hierarchical mergers, which produce high-$|\chi_{\rm eff}|$ and low-$q$ binaries. 
Hierarchical mergers can also explain the negative correlation between $q$ and the dispersion of $\chi_{\rm eff}$, 
as well as the positive correlation between $|\chi_{\rm eff}|$ and $M_{\rm chirp}$. 
We present a comprehensive study on how the expected distribution of each of these quantities depends on model parameters and assumptions, which will aid the interpretation of observed GW population properties.
\end{abstract}

\section{Introduction}

Gravitational waves (GW) from approximately 200 mergers among black holes (BHs) and neutron stars (NS) \citep{LIGO2025_O4aCatalog,LIGO2025_O4aProp} have been detected by LIGO \citep{2015CQGra..32g4001L}, Virgo \citep{2015CQGra..32b4001A}, and KAGRA \citep{2021PTEP.2021eA101A}. 
An outstanding question is the astrophysical pathway(s) leading to these mergers, which remains highly debated. 
Promising channels for their formation include isolated binary evolution \citep[e.g.][]{Dominik12,Kinugawa14,Belczynski16,Spera19,Tanikawa2022,Kirouglu2025a,Kirouglu2025b}, the evolution of triple or quadruple systems
\citep[e.g.][]{Silsbee17,Antonini17,Fragione19}, dynamical evolution within star clusters
\citep[e.g.][]{Banerjee17,Kumamoto18,Rastello18,PortegiesZwart00,Samsing14,Samsing17,OLeary16,Rodriguez16,ArcaSedda20,Hoang18,Rasskazov19,Fragione19_NSBH_GN_I}, and in disks of active galactic nuclei (AGN)
\citep[e.g.][]{Bartos17,Stone17,McKernan17,Tagawa19}. 

Dynamical channels can explain many of the unusual features reported for several detections. 
One distinguishing property is the eccentricity of merging binaries. High eccentricity has been suggested for at least one of the proposed GW events \citep{2022:Romero-Shaw:GWTC-3-ecc,Gayathri+2022,2024:Gupte:GWTC-3-ecc,Morras2025,deLlucPlanas2025,RomeroShaw2025,XuYumeng2025,Gupte2026}. 
Such eccentricity is typically expected in mergers occurring within dynamical environments \citep{Antonini2018,Rodriguez2018,Kremer:CMC:2020,Zevin2021,Tagawa20_ecc,RomeroShaw2025}. 
Another significant discovery is that of unusually massive BHs. 
Events like GW190521 \citep{LIGO20_GW190521} and GW231123 \citep{LIGO2025_GW231123} are notably massive, posing challenges for formation through isolated binary evolution \citep{Tanikawa2025,Stegmann2025} due to the mass gap associated with pair-instability supernovae (\citealt{Chatzopoulos12}, see also \citealt{Croon2025}). 
Additionally, GW190521 suggests either strong precession or high eccentricity \citep{2020:Romero-Shaw:GW190521,2020:Gayathri:GW190521,2023:Gamba:GW190521,2023:Romero-Shaw:Ecc-or-precc}, phenomena difficult to produce through isolated channels, but more naturally arising in dynamical channels.

Globular clusters are known to be important sites for the dynamical BH merger channel due to their high density and low velocity dispersion which leads to high scattering cross sections for binary formation and evolution. 
However, the total mass of globular clusters within a galaxy is roughly comparable to that of nuclear star clusters \citep{ArcaSedda2014,Fahrion2022}, particularly in abundant galaxies with stellar masses of $\lesssim 10^{\rm 11}~\Msun$ \citep{Madau2014}. 
This similarity suggests that the number of BHs in both environments may be comparable. Therefore, the efficiency of mergers in each environment is crucial in determining their relative contributions to GW events. While the number densities in nuclear star clusters can greatly surpass those in globular clusters at their centers, the velocity dispersion due to the central supermassive BH (SMBH) is also much larger, and dynamical binary formation scales with $\rho^{3}v^{-9}$ \citep[e.g.,][]{Aarseth76,Binney08}. 
This indicates that binary formation is more efficient in globular clusters during gas-less phases. 
On the other hand, during active phases when dense disks are present, 
\citet{Tagawa19} showed that binary formation and hierarchical mergers are frequent in nuclear star clusters. 
If mergers predominantly occur in galactic centers, \citet{Ford2022} argued that mergers during active phases surpass those during quiescent phases, 
as interactions with gas enhance the merger rate enormously. 
Consequently, a significant number of mergers could be expected to occur within AGN disks. 

A potential signature of mergers in AGN disks is electromagnetic counterparts. 
Recent observations suggest that gamma-ray \citep[but see][]{Connaughton2016,Bagoly2016,DeLaunay2024GCN.38351....1D,Zhang2025_S241125n} and optical flares \citep{Graham20,Graham2023} may be associated with GW events. 
These associations can be attributed to emission from merging BHs in AGN disks \citep{Tagawa2023_EM,Tagawa2023_SC,Chen2023_EM,ChenKen2025,McKernan2019_EM,Rodriguez-Ramirez2024,Rodriguez-Ramirez2025,Ma2025,Tagawa2026_EM}. 
If these associations are genuine \citep{Connaughton2018,Ashton2020,Palmese2021,CalderonBustillo21,Morton2023,Veronesi2025}, they could serve as key signatures of mergers occurring in AGN disks. 
Another potential signature is the spatial correlation between the (3D) locations of GW events and AGNs \citep{Bartos17NatCom,Corley2019,Veronesi2022,Veronesi2023,Veronesi2025_AGN,Moncrieff2025}. 
A recent analysis using LIGO/Virgo/KAGRA O4 data 
estimated that the fraction of mergers originating from AGNs is 
$\sim 0.1$--$0.8$ with a $90\%$ confidence level, 
excluding high-luminosity ($\geq 10^{45}~{\rm erg~s^{-1}}$) AGNs 
\citep{Zhu2025}. 
The alignment of BH spins with the orbital angular momentum for massive mergers, especially in the upper-mass gap, may also serve as a potential signature of mergers in AGN disks \citep{LiYinJie2025_alignment}. 
Therefore, there are compelling suggestions that some mergers originate in AGN disks. 
Given this promising pathway, 
it would be beneficial to further elucidate how BHs merge in AGN disks and to identify the 
distribution of the expected properties of these mergers.

In an AGN disk, 
compact objects (COs), including stellar-mass BHs, 
are embedded through capture via dynamical interactions between the nuclear star cluster and the AGN disk \citep{Ostriker1983,Syer1991,Bartos17,McKernan17,Generozov2023,Wang2023_capture,Whitehead2025,Rowan2025}. 
Additionally, stars form in outer regions of the AGN disk where the self-gravity of gas becomes significant, contributing to the presence of COs in the disk 
\citep{Levin2003,Goodman2004,Milosavljevic2004,Thompson05,Nayakshin07,Stone17,Gilbaum2022,Derdzinski23,ChenY2023,EpsteinMartin2025}.

COs in the AGN disk migrate by interacting with the gas disk, 
and are predicted to accumulate in regions such as migration traps and gap-forming regions 
\citep{Bellovary16,Masset2017,Hankla2020,Yang19b_PRL,Yang19a,Secunda18,Secunda2020,Tagawa19,Derdzinski19,Grishin2023,Peng2021,Pan2021,Gilbaum2025,Xue2025,Vaccaro2024,EpsteinMartin2025b,Moncrieff2026}. 
Within these regions, they efficiently encounter one another and 
form binaries under the influence of gaseous torques \citep{Goldreich02,Tagawa19,Secunda2019,Secunda2020,LiJiaru2022,Rowan2022,DeLaurentiis2023,Rozner2023,Whitehead2023,Rowan2023,Qian2023,Whitehead_2023b_novae,Dodici2024} and GW emission \citep{Tagawa20_ecc,Samsing2022,Boekholt2023,LiJiaru2023}. 
After their formation, 
these binaries are hardened 
(or softened) 
by gaseous torques \citep{Grobner2020,LiYanPing2021,LiYaPing2022_hotdisk,LiRixin2022_1,LiRixin2023_eos2,LiRixin2023_viscosity3,Dempsey2022,Dittmann2023,ChenYiXian2022,LiYaPing2022_Spin_ecc,Calcino2024,Ishibashi2024,Rozner2024,Dittmann2025,Rowan2023,Rowan2025_Inclination,Whitehead2025} 
and by binary-single interactions \citep{Heggie1975,Leigh18,Tagawa19,Ginat2021,Ginat2023,Samsing2022,Su2025,Xue2025,Rowan2025_BS,WangMengye2025_BSI,WangMengye2025_BSII}, before GW emission \citep{Peters64} drives 
them to merge.

Several groups have studied the evolution of a population of BHs in AGN disks using Monte Carlo-like methods. 
\citet{McKernan19}, \citet{McKernan2025}, \citet{Cook2024}, and \citet{Delfavero2025} investigated the properties of mergers occurring in both migration traps and other locations, while simplifying many processes (see \S~\ref{sec:spin_mass}). 
\citet{Yang19a}, \citet{Yang19b_PRL}, and \citet{Vaccaro2024} focused specifically on the properties of mergers within migration traps. 
Our models focus on mergers outside migration traps and, 
for the first time, incorporate binary formation via gaseous torques and three-body interactions \citep{Tagawa19}, GW capture processes \citep{Tagawa20_ecc}, binary-single interactions \citep{Tagawa20_MassGap}, the effects of kicks due to mergers and interactions, gap formation (with an updated prescription), as well as the evolution of spin and binary angular momentum due to gas accretion \citep{Tagawa20b_spin}, to predict the properties of BH mergers in AGN disks.

We found that most BH mergers occur in the inner regions, 
where gaps often form 
and binary-single interactions significantly facilitate mergers \citep{Tagawa19}. 
Consequently, the mass ratio for the merging binaries is predicted to be close to unity \citep{2021ApJ...920L..42G}, and the effective spin parameter is distributed around zero \citep{Tagawa20b_spin}. Moreover, the merging masses are characterized by frequent hierarchical mergers, due to the high escape velocity from the gravitational potential of the SMBH \citep{Tagawa20_MassGap}. 
Comparisons between observed BH masses, spins, and merger rates suggest that a substantial fraction of the observed mergers may indeed originate in AGN disks \citep{2021ApJ...920L..42G,Tagawa20_MassGap,Ford2022,McKernan2025}. 
It is also suggested that various parameters and processes, such as migration \citep{Grishin2023}, interactions \citep{Trani2024}, 
eccentricity \citep{Samsing2022}, 
and accretion \citep{Begelman2017, Roupas2025,Roupas2026}, 
may affect the distributions of observables.

In this paper, we investigate the expected properties of mergers in these environments and their dependence on various parameters, using one-dimensional $N$-body simulations combined with a semi-analytical model. 
We have updated and varied the prescriptions for encounters, interactions, migration, accretion onto BHs, and AGN structures. 
Additionally, we explore the impact of key parameters, including AGN duration, BH number density, SMBH mass, and the accretion rate onto the SMBH. 
We compare the predicted properties of BH mergers with observations from LIGO/Virgo/KAGRA (LVK) and 
discuss the parameter space 
in which the AGN channel could significantly contribute to the observed events.

This paper is organized as follows: 
In \S~\ref{sec:method}, we detail the methods used in our models,  
\S~\ref{sec:result} presents our main findings, and 
\S~\ref{sec:conclusions} summarizes our key conclusions.


\section{Methods}

\label{sec:method}

\subsection{Overview of models}

\label{sec:overview}

We utilize the models employed in \citet{Tagawa20_MassGap}, 
which track $N$-body particles representing BHs. 
Here, we provide an overview of the processes involved in the evolution and mergers of BHs in an AGN disk in this model.

\paragraph{Disk model}
For the AGN accretion disk we assume a flat geometrically thin optically thick radiatively efficient \citet{Shakura73} model assuming local thermal equilibrium between viscous heating and radiative cooling. The model is extended into the outer gravitationally unstable region assuming $Q=1$ Toomre-parameter due to star formation following \citet{Thompson05} hereafter \citetalias{Thompson05} in our fiducial model, 
where the opacity model is given by \citet{Bell94}. 
We use a stationary passive disk model, i.e. the AGN disk properties are not affected by the presence of BHs or the nuclear star cluster during the simulation, neglecting possible feedback effects, and employ analytic prescriptions for accretion, migration, and spin changes independently for all embedded objects using the local physical properties of the disk. 
Gap formation and its influence are taken into account, reflecting the individual BH mass; for simplicity, the gap depth is assumed to be unaffected by other BHs orbiting in the same annulus\footnote{
According to our test calculations, in which the gap depth is determined by the most massive BH in each grid, this simplified prescription has a moderate effect on the merging mass distribution. 
The number of mergers at $t=3~{\rm Myr}$ and the top $1\%$ masses among merging binaries among 10 simulations are $2\times 10^3$ and $220~\Msun$, respectively. We find there are no noticeable differences in the observable properties resulting from this change. 
}.
We assume that the efficiency of angular momentum transport due to global gravitational torques in the outer regions is $m_{\rm AM}=0.15$ \citepalias{Thompson05}. 
The outer boundary from which gas inflows is set at $r_{\rm out}=5~{\rm pc}$, 
the pressure ratio parameter is assumed to be $\xi=1$, and the conversion efficiency of star formation to radiation is $\epsilon=1.5\times 10^{-4}$ \citepalias{Thompson05}. 
In the inner regions, the efficiency due to local viscous torque is $\alpha_{\rm SS}=0.1$ \citep{Shakura73}. 
In the transition between the inner and outer regions, the degree of angular momentum transport is adjusted to keep $Q=1$. 
This transition is based on the consideration that the mechanisms for angular momentum transfer likely differ depending on location \citep{Thompson05,Collin2008}. We also investigate cases where viscosity is prescribed uniformly 
(see Sec.~\ref{sec:updates} below).

\paragraph{BH processes}
Initially, BHs reside within the nuclear star cluster surrounding the SMBH \citep{Miralda-Escude2005,Alexander2009,OLeary09,Rasskazov19}. 
We assume that the initial density distribution of all BHs is slightly steeper than the single-mass Bahcall-Wolf distribution by 0.25 in the power-law index, 
which is suggested by numerical simulations \citep[e.g.,][]{Hopman06,Freitag06,OLeary09}. 
These BHs are gradually captured by the AGN disk through accretion torques and gas dynamical friction \citep{WangY2024_capture}. 
BHs can also form in the outer regions of the AGN disk, where the gas becomes gravitationally unstable \citep{Thompson05,EpsteinMartin2025}. 

Within the AGN disk, 
BHs migrate toward the SMBH due to type I/II torques exerted by the AGN disk \citep[e.g.][]{Gilbaum2025}, and binaries form through gas dynamical friction during two-body encounters \citep{Goldreich02} as well as through dynamical interactions during three-body encounters \citep[e.g.,][]{Aarseth76}. 
After their formation, 
the angular momentum of these binaries evolves 
due to gas dynamical friction and torques from circum-single and binary disks \citep{Bartos2017}. 
Additionally, binaries can interact with single stars, other BHs, and other BH binaries. 
Following binary-single (or binary-binary) interactions,
the semi-major axes and velocities of the binaries evolve, as described in \citet{Leigh18}, 
and we assume that the directions of the orbital angular momenta are randomized, and the eccentricity distribution is thermalized. 
The two most massive BHs are assumed to remain as binary components after binary-single and binary-binary interactions. 
The velocities of all BHs evolve due to scattering with other objects \citep{Binney08}. 
Highly eccentric BH binaries can also form via GW capture (GWC) in single-single encounters \citep[e.g.,][]{OLeary09} or through binary-single interactions \citep{Samsing14,Tagawa20_ecc}.

In later phases, the semi-major axes and eccentricities of the binaries decrease due to GW emission \citep{Peters64}. 
When the pericenter distance of a binary becomes smaller than the innermost stable orbit, we assume that the BHs merge, assigning a kick velocity \citep{Lousto12}, accounting for mass loss due to GW emission \citep{Barausse12}, and modeling the spin evolution at merger \citep{Rezzolla08}. 
BH masses and spins grow through mergers and Eddington-limited accretion \citep{Tagawa2020_spin}.

\subsection{Updates}

\label{sec:updates}

\paragraph{Radiative feedback}
\citet{Xue2025} performed semi-analytic simulations that adopt physical prescriptions similar to those in \citet{Tagawa19}, while treating the evolution of background objects in a probabilistic manner. By comparing the results of \citet{Xue2025}, we found that radiative feedback was not effectively considered in \citet{Tagawa19} due to computational bugs (mistakes in unit conversion). 
To address this oversight, we turn off gas dynamical friction when specific conditions are met (\S~3.3 in \citealt{Tagawa19}). In such cases, the gas capture rate by the BH is capped at the Eddington rate, and the accretion drag reduces due to the decreased capture rate. 

\paragraph{Dynamical encounter probabilities}
In \citet{Tagawa19}, the encounter probabilities for several mechanisms
--such as gas-capture binary formation, 
three-body binary formation, 
GW capture binary formation, 
and binary-single interactions--were overestimated. 
The local two-body encounter timescale is given by $t_{\rm enc,local}=1/(n \sigma v)$, 
where $n$ is the number density of BHs in the AGN disk, 
$\sigma =\Delta z \Delta r$ is the encounter cross section, 
$v$ is the relative velocity between BHs, 
and $\Delta r$ and $\Delta z$ are the radial and vertical scales over which encounters can occur, 
influenced by several factors including the tidal force from the SMBH. 
Their specific values are 
provided in \S~3.3.5, \S~3.3.7, \S~3.3.8 in \citealt{Tagawa19} and \S~2.2.1 in \citealt{Tagawa20_ecc} for binary-single interactions, three-body binary formation, gas-capture binary formation, and GW capture binary formation, respectively. 
Here, $v$ is taken as the maximum of the migration, shear, and random velocities from our previous studies \citep[e.g.,][]{Tagawa19}. 
However, this expression for the timescale is valid only when more than one BH resides within annuli within $\Delta r$ from one another, $N_{\Delta r}=2\pi r n \sigma >1$. 
In other cases, the timescale for migration into an annulus containing other BHs also needs to be considered. 
Thus, the encounter timescale is given as the maximum of (i) the above local encounter timescale ($t_{\rm enc,local}$) and (ii) the migration timescale to reach an annulus containing another BH, given by $t_{\rm mig-BH}=1/2\pi r  \Delta z n v_{\rm mig,rel}$, where $v_{\rm rel,mig}$ is the relative migration velocity of the BH with respect to another BH, considering gap formation \citep{Tagawa19}. 
Furthermore, if the crossing timescale within $\Delta r$, given by $t_{\rm cross}=\Delta r/ v_{\rm mig,rel}$, is shorter than the encounter timescale ($t_{\rm enc,local}$), then BHs needs to cross $N_{\rm cross}=t_{\rm enc,local}/t_{\rm cross}$ annuli containing BHs for an encounter to occur. Therefore, the total encounter timescale is expressed as: 
\begin{align}
t_{\rm enc} =
\left\{
\begin{array}{cl}
&t_{\rm enc,local}
 ~~~\mathrm{for}~~N_{\Delta r}\geq1\\
&{\rm max}(t_{\rm enc,local},t_{\rm mig-BH})\\
&~~~~~~\mathrm{for}~~N_{\Delta r}<1~\&~t_{\rm cross}>t_{\rm enc,local}\\
&{\rm max}\left(t_{\rm enc,local},N_{\rm cross}t_{\rm mig-BH}\right)\\
&~~~~~~\mathrm{for}~~N_{\Delta r}<1~\&~t_{\rm cross}<t_{\rm enc,local}.\\
\end{array}
\right.
\end{align}
In this way, this prescription for the encounter timescale is updated 
for gas-capture binary formation, three-body binary formation, GW capture binary formation, and binary-single interactions.

\paragraph{Exchange interactions}
We also discovered that the exchange of binary components, as implemented in \citet{Tagawa20_MassGap}, was not functioning correctly 
(as spin values were occasionally not properly conserved), 
and we have corrected this bug. 

\paragraph{Accretion rate}
In our fiducial model, the accretion rate onto the SMBH is reduced to $0.01$ times the Eddington rate, compared to $0.1$ times the Eddington rate used in \citet{Tagawa19}. 
This change reflects the findings that most mergers likely originate from lower-luminosity AGNs \citep{Tagawa19}, given their abundance \citep{Greene07} and the weak dependence of the merger rate on the accretion rate, as long as thin disks form around the SMBH \citep{Rowan2025_Inclination,Whitehead2025}. 
Consequently, models with low accretion rates onto SMBHs should be more common among BH mergers in AGN.

\paragraph{Black hole mass distribution}
While the upper mass limit for first-generation BHs in \citet{Tagawa19} is set to $15~\Msun$, 
motivated by the super-solar metallicity of stars in the Galactic center \citep{Do18}, 
we adopt a model in which the BH mass is multiplied by a factor of 2, as prescribed in \citet{Tagawa20_MassGap}. 
This adjustment accounts for the lower metallicity in nuclear star clusters around lower-mass SMBHs \citep{Fahrion2021}, which are likely formed through the accretion of globular clusters \citep{ArcaSedda2014,Fahrion2022}. 
Moreover, high initial BH masses are required to account for the change in spin magnitude at high masses through the AGN channel (see $\S~\ref{sec:spin_mass}$).

To reduce computational costs, 
we focus solely on the evolution of BHs and stars, excluding neutron stars, unlike in \citet{Tagawa20_MassGap}. 

\paragraph{Natal BH spin}
Based on the distribution of BH spin magnitudes inferred from the LVK observations \citep{LIGO2025_O4_Prop,Adamcewicz2025,Guttman2026}, 
we adopt a natal spin magnitude of $a=0.1$, instead of $a=0$ used in our previous works. 

\paragraph{Inner boundary}
Finally, we adjusted the inner radius for BH removal from $10^{-4}~{\rm pc}$ to $10^{-6}~{\rm pc}$ to ensure that we do not miss mergers occurring close to the SMBH. 

The parameters of our fiducial model (M1) are listed in Table~\ref{table:parameter_model}, which are consistent with those in 
\citet{Tagawa20_MassGap}, except for the aforementioned adjustments.

\begin{table*}
\begin{center}
\caption{Fiducial values of our model parameters. 
}
\label{table:parameter_model}
\hspace{-5mm}
\begin{tabular}{p{8cm}|p{4cm}|p{4cm}}
\hline 
Parameter & Fiducial value & References\\
\hline\hline
Spatial directions in which BS interactions occur&  isotropic in 3D&\citet{Tagawa20_ecc,Samsing2022}\\\hline
Number of temporary binary BHs formed during a BS interaction&  $N_\mathrm{int}=20$&\citet{Samsing14}\\\hline
Initial BH spin magnitude & $|{\bm a}|=0.1$&\citet{LIGO2025_O4_Prop}\\\hline
Angular momentum directions of circum-BH disks & $\hat{{\bm J}}_\mathrm{CBHD}=\hat{{\bm J}}_\mathrm{AGN}$ for single BHs, 
$\hat{{\bm J}}_\mathrm{CBHD}=\hat{{\bm J}}_\mathrm{bin}$ for BHs in binaries&e.g., \citet[][]{Lubow1999,Moody19}\\\hline
Ratio of viscosity responsible for warp propagation over that for transferring angular momentum 
& $\nu_2/\nu_1=10$& \citet{Ogilvie99,Lodato2013} \\\hline
Alignment efficiency of the binary orbital angular momentum
due to gas capture 
& $f_\mathrm{rot}=1$
(Eq.~14 in \citealt{Tagawa2020_spin})&\citet{Dittmann2023,Fabj2025}\\
\hline 
Mass of the central SMBH & $M_\mathrm{SMBH}=4\times 10^6\,\Msun$ &\citet{Ghez2008,Genzel2010}\\\hline
Gas accretion rate at the outer radius of the simulation ($5\,\mathrm{pc}$)
& ${\dot M}_\mathrm{out}=0.01\,{\dot M}_\mathrm{Edd}$ with $\eta=0.1$&\citet{Greene07,Tagawa19}\\\hline
Fraction of pre-existing binaries in 3D star cluster& $f_\mathrm{pre}=0.15$ &\citet{Pfuhl14,Stephan2016}\\\hline
Power-law exponent for the initial density profile for BHs & $\gamma_\mathrm{\rho}=0$ &\citet{Hopman06,OLeary09}\\\hline
Initial velocity anisotropy parameter such that $\beta_\mathrm{v}v_\mathrm{kep}(r)$ is the BH velocity dispersion  
& $\beta_\mathrm{v}=0.2$& \citet{Yelda14,Szolgyen18}\\\hline
Efficiency of angular momentum transport in the $\alpha$-disk & $\alpha_\mathrm{SS}=0.1$ &\citet{King07}\\\hline
Stellar mass within 3 pc &$M_\mathrm{star,3pc}=10^7\,\Msun$&\citet{Feldmeier14,Schodel14}\\\hline 
Stellar initial mass function slope & $\delta_\mathrm{IMF}=2.35$&\citet{Bartko10,Lu13}\\\hline
Angular momentum transfer parameter in the outer star forming regions 
&$m_\mathrm{AM}=0.15$ 
(Eq.~C8 in \citealt{Thompson05}) 
&\citet{Thompson05,Collin2008}\\
\hline
Accretion rate in Eddington units onto stellar-mass BHs with a radiative efficiency $\eta=0.1$
&$\Gamma_\mathrm{Edd,cir}=1$&e.g., \citet{Fragile2025}\\\hline
Numerical time-step parameter &$\eta_t=0.1$&\citet{Tagawa19}\\\hline
Number of radial cells storing physical quantities &$N_\mathrm{cell}=120$&\citet{Tagawa19}\\\hline
Minimum and maximum distances from an SMBH for the initial BH distribution&  $r_\mathrm{in,BH}=10^{-6}$ pc, $r_\mathrm{out,BH}=3$ pc & \citet{Feldmeier14}\\\hline
Initial number of BHs within 3 pc from an SMBH &$N_\mathrm{BH,ini}=2\times 10^4$ &\citet{Bartko10,Lu13}\\\hline
\end{tabular}
\end{center}
\end{table*}

\subsection{Model variation}

To understand the parameter dependence of observables, 
we have performed simulations using 20 models, labeled M1--M20. 
Model~M1 corresponds to the fiducial model (Table~\ref{table:parameter_model}), 
while the parameters or prescriptions are varied in models~M2--M20 (Table~\ref{table_results}). 

\paragraph{Velocity dispersion}
In model~M2, the velocity dispersion of BHs follows an isotropic distribution, and $\beta_\mathrm{v}$ is not used. 
Note that the isotropic distribution differs from the Gaussian distribution adopted in the fiducial model, and stars typically have higher inclination angles (see \citealt{Xue2025}). 
In model~M3, the velocity dispersion of BHs is increased to $\beta_{\rm v}=1$, where $\beta_\mathrm{v}v_\mathrm{kep}(r)$ represents the BH velocity dispersion and $v_\mathrm{kep}(r)$ is the Keplerian velocity at distance $r$ from the SMBH. 

\paragraph{BH numbers}
In model~M4, the initial number of BHs is reduced from $N_{\rm BH,ini}=2\times 10^4$ to $7\times 10^3$. 

\paragraph{Geometry -- 2D model}
In model~M5, binary-single interactions are assumed to occur in a two-dimensional geometry, 
co-planar with the AGN disk \citep{Samsing2022,Tagawa20_ecc}.

\paragraph{SMBH mass}
In model~M6, the SMBH mass is increased to $M_{\rm SMBH}=4\times 10^7~{\Msun}$. 

\paragraph{Accretion rate}
In model~M7, the gas inflow rate at the outer radius of the simulation is set to be ${\dot M}_{\rm out}=0.1~{\dot M}_{\rm Edd}(M_{\rm SMBH})$ 
to investigate the dependence on the gas inflow rate, 
where ${\dot M}_{\rm Edd}(M)$ is the Eddington accretion rate onto a BH of mass $M$. 

\paragraph{BH masses}
In models~M8, the initial BH mass is multiplied by $3/2$ relative to the fiducial model, resulting in maximum BH masses of $45~\Msun$, 
considering mass gaps caused by pair-instability supernovae \citep[e.g.,][]{Woosley2021}.

\paragraph{Migration}
In model~M9, BHs do not migrate within the AGN disk. In model~M10, thermal torque is included to calculate migration \citep{Grishin2023}. 

\paragraph{Radiation feedback}
In model~M11, the effects of radiation feedback on gaseous torques are turned off, 
to examine the case 
effectively prescribed in \citet{Tagawa19} due to a bug. 

\paragraph{Binary-single interactions}
In model~M12, the orbital angular momentum directions of binaries are assumed to be unaffected by binary-single interactions, 
while the semimajor axis and eccentricity are influenced in the same way as in the fiducial model. 
This represents an extreme case of inefficient randomization suggested by \citet{Trani2024} and \citet{Fabj2025}.

\begin{figure}
    \centering
    \includegraphics[width=1\linewidth]{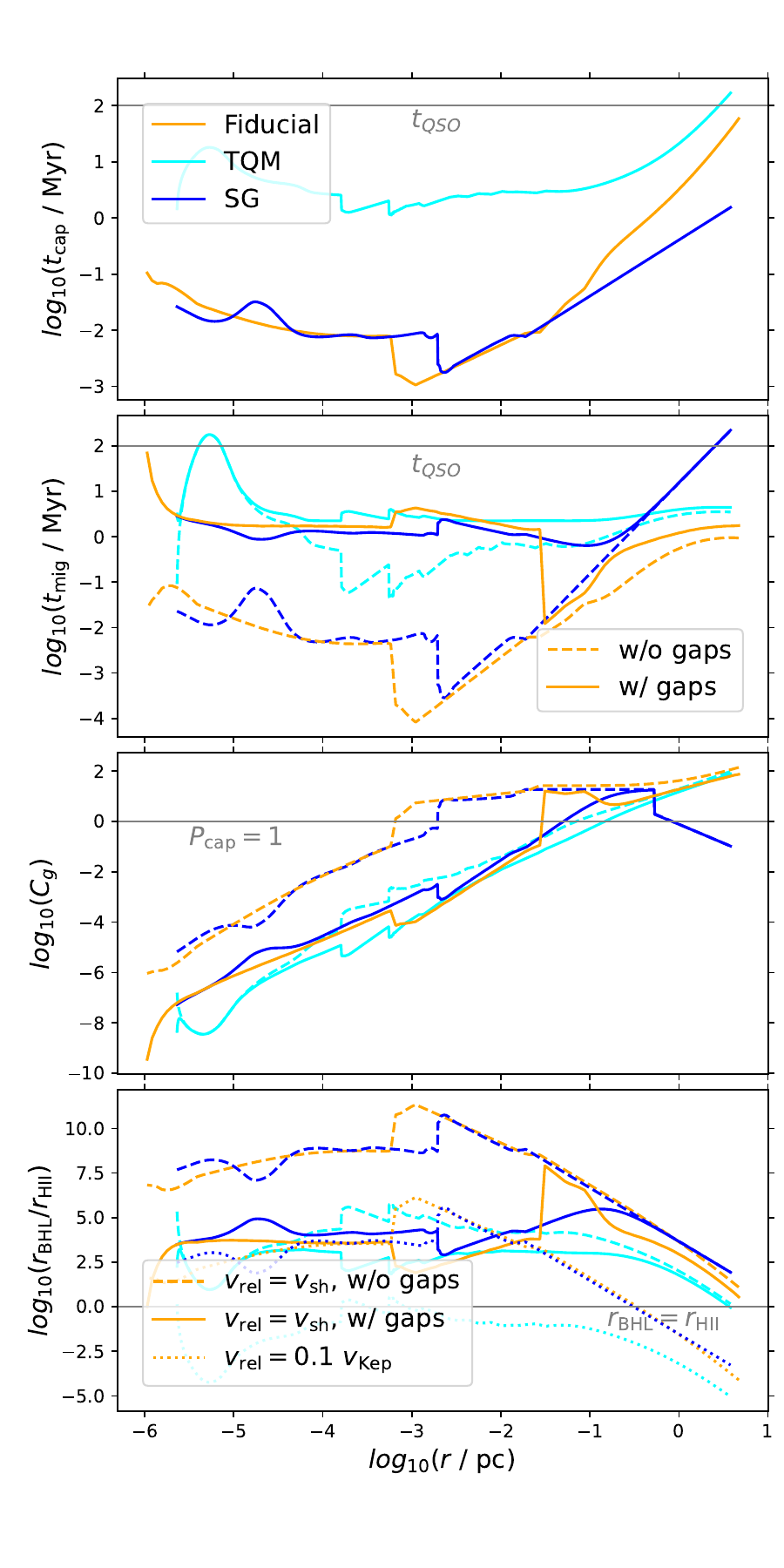}
    \caption{
    The timescales and conditions characterizing the evolution of $30~\Msun$ BHs in AGN disks. 
    The orange, cyan, and blue lines represent the fiducial, TQM, and SG models, respectively. In the second to fourth panels, dashed and solid lines indicate results without and with gaps, respectively. 
    The top panel shows the capture timescale for a BH with a relative velocity equal to $0.1 \times$ the local Keplerian velocity with respect to the local AGN gas. 
    The second panel shows radial migration timescales.     
    The gray lines in the first and second panels denote the typical upper limit for the lifetime of AGNs per galaxy, approximately $100~{\rm Myr}$ \citep{Martini04}.     
    The third panel shows the probability that two encountered bodies form a binary via gas-capture.     
    The gray line indicates $P_{\rm cap}=1$, above which binaries always form after an encounter. 
    The fourth shows the ratio of the Bondi-Hoyle-Lyttleton radius to the size of an H~II region. When this ratio exceeds unity, gas dynamical friction can operate for capture to the AGN disk, binary formation, and binary hardening, without being suppressed by radiative feedback. 
    For the solid and dashed lines, the relative velocity between a BH and the gas is given by the shear velocity considering gas-capture binary formation, 
    while for the dashed lines, it is given by $0.1$ times the local Keplerian velocity considering capture by the AGN disk. 
    The gray line represents $r_{\rm BHL}=r_{\rm HII}$, above which radiation feedback is inefficient.     
    }
    \label{fig:times}
\end{figure}

\paragraph{AGN model}
In models~M13 and M14, the AGN disk model proposed by \citet[][hereafter SG]{Sirko03} is employed, with the accretion rates onto the SMBH set to 
${\dot M}_{\rm SMBH}=0.1~{\dot M}_{\rm Edd}(M_{\rm SMBH})$ and $0.01~{\dot M}_{\rm Edd}(M_{\rm SMBH})$, respectively. 
In models~M15 and M16, the AGN disk model proposed by \citetalias[][]{Thompson05} is utilized with ${\dot M}_{\rm SMBH}=0.1~{\dot M}_{\rm Edd}(M_{\rm SMBH})$ and $0.01~{\dot M}_{\rm Edd}(M_{\rm SMBH})$, respectively. 
The parameters in models~M13--M16 are set to the fiducial values in \citet{Gangardt2024}. In the models~M15 and M16, the efficient angular momentum parameter is applied across all regions including the inner regions, which differs from the fiducial model (\S~\ref{sec:overview}, Fig.~\ref{fig:times}).

\paragraph{BH accretion}
In models~M17--M20, 
we modify the prescriptions for accretion onto BHs. 
In models~M17 and M18, 
the accretion rate onto a BH is given by
\begin{align}
\label{eq:m_acc}
{\dot M}_{\rm BH}=
{\dot M}_{\rm BHL}\left(\frac{r_{\rm in}}{r_{\rm wind}}\right)^s 
\end{align}
where $r_{\rm wind}$ and $r_{\rm in}$ are the radii inside and outside of which wind mass loss becomes significant, 
and ${\dot M}_{\rm BHL}$ is the gas capture rate by the BH (\citealt{Tagawa19} see also \citealt{Kaaz2023}). 
The inflow rate decreases with radius following a power-law due to winds mass loss, as predicted by adiabatic inflow-outflow solutions \citep[ADIOS,][]{Blandford1999}, observations \citep{Poutanen2007}, and numerical simulations \citep[e.g.][]{Hu2022,Fragile2025}. 
We set $r_{\rm in}= 5~r_{\rm g}$ \citep{Pan2021_accretion}, where $r_{\rm g}$ is the gravitational radius of the BH, and 
\begin{align}
r_{\rm wind}={\rm min}(r_{\rm circ},r_{\rm trap}),
\end{align}
where $r_{\rm circ}$ is the circularization radius, at which the gas becomes circularized, 
and $r_{\rm trap}\approx ({\dot M}_{\rm BHL}/{\dot M}_{\rm Edd}(m_{\rm BH})\eta_{\rm rad}) r_{\rm g}$ is the spherical photon trapping radius \citep[e.g.][]{Kato2008}, within which photons are advected inwards faster than they can diffuse outward, 
$m_{\rm BH}$ is the mass of the stellar mass BH, and 
$\eta_{\rm rad}$ is the radiative efficiency. 
Following the numerical simulations by \citet{Sagynbayeva2025}, we set 
\begin{eqnarray}
\label{eq:r_circ}
r_{\rm circ} &= &f_{\rm circ}r_{\rm Hill}
\end{eqnarray}
where $r_{\rm Hill}=R_{\rm BH}(m_{\rm BH}/3M_{\rm SMBH})^{1/3}$ is the Hill radius, 
$R_{\rm BH}$ is the distance from the SMBH to the BH, and 
$f_{\rm circ}$ is a factor estimated as 
\begin{align}
\label{eq:f_circ}
f_{\rm circ}\sim {\rm min}\{1,0.01(m_{\rm BH}/M_{\rm SMBH})^2(R_{\rm BH}/H_{\rm AGN})^6\}, 
\end{align}
where $H_{\rm AGN}$ is the scale height of the AGN disk. 
We set $s=0.5$ for model~M16 and $s=0.3$ for model~M17, 
considering the uncertainties suggested by numerical simulations \citep[e.g.,][]{Hu2022,Toyouchi2024,Fragile2025}.

In models~M19--M20, we adopt the prescription suggested by \citet{Begelman2017}, 
where the transition between accretion states occurs 
based on the relationship between $r_{\rm circ}$ and $r_{\rm trap}$.  
This prescription is motivated by observations of accreting systems, such as tidal disruption events \citep{Coughlin2014} and ultraluminous X-ray sources \citep{Poutanen2007}, and is consistent with 
3D general relativistic radiation magnetohydrodynamic simulations \citep{Sadowski2016,Fragile2025}. 
When $r_{\rm circ}$ exceeds $r_{\rm trap}$, outflows reduce the accretion rate near the Eddington limit (Eq.~\ref{eq:m_acc}), as predicted by the ADIOS model \citep{Blandford1999}. 
Conversely, if $r_{\rm circ}$ is smaller than $r_{\rm trap}$, 
shocked gas couples strongly to radiation 
due to efficient advection relative to diffusion, 
resulting in less efficient mass loss. 
This state is referred to as a zero-Bernoulli accretion (ZEBRA) flow \citep{Coughlin2014}. 
This process facilitates weak mass loss while allowing for super-Eddington accretion. 
Reflecting this suggestion, 
we assume that the accretion rate onto a BH is given by Eq.~\eqref{eq:m_acc} with $s=1$ when $r_{\rm circ}>r_{\rm trap}$, and by the gas capture rate ${\dot M}_{\rm BHL}$ when $r_{\rm circ}<r_{\rm trap}$. 
These slopes are supported by the numerical simulations \citep{Sadowski2016,Fragile2025}. 
Note that this prescription is also motivated by the possibility of producing strong electromagnetic counterparts to GW events, as suggested in \citet{Connaughton2016}, \citet{Bagoly2016}, \citet{Graham20}, \citet{Graham2023}, and \citet{Zhang2025_S241125n}, 
without causing the overgrowth problem \citep{Tagawa2026_EM}. 
This problem arises 
when BHs grow too rapidly by continuously accreting most of the captured gas, 
which likely conflicts with Soltan's argument and the dynamics of S-stars \citep{Yu2002,Tagawa2022_BHFeedback}. 
In model~M20, we reduce the accretion rate by a factor of 10 in both cases, considering a duty cycle of $\sim 0.1$, as roughly estimated in \citet{Tagawa2022_BHFeedback}. 
Note that we do not model the transition to the ZEBRA state associated with kicks, despite that it is predicted, because its influence on the final mass is estimated to be minor \citep{Tagawa2026_EM}.

To calculate the evolution of the BH spin direction, 
we estimate the warp radius (Eq.~6 in \citealt{Tagawa2020_spin}), which is the radius at which the circum-BH disk aligns with the BH spin. 
For this calculation, we use the inflow rate at the warp radius in models~M17--M20, 
unlike the fiducial model, which assumes the Eddington accretion rate regardless of radius. 
To account for the torque exerted by the circum-binary disk, 
we enhance its magnitude to reflect the actual accretion rate onto the BH.

\begin{table*}
\begin{center}
	\caption{
    The assumptions, motivation, and results in different models. 
    The input columns show the model number, and the differences with respect to the fiducial model, and the motivation for investigating each model. 
    The output columns list the number of mergers at $t=3~{\rm Myr}$ ($N_{\rm mer,10}$) 
    and the top $1\%$ of masses among merging binaries ($M_{\rm top1}$) among 10 simulations . 
        }
\label{table_results}
\begin{tabular}{p{0.8cm}|p{6.0cm}|p{6.0cm}||p{1.3cm}|p{1.6cm}}
\hline
\multicolumn{3}{c}{input} \vline& \multicolumn{2}{c}{output} \\\hline
Model&Parameter&Motivation of parameter search
&$N_\mathrm{mer,10}$&$M_{\rm top1}~[\Msun]$
\\\hline

M1&Fiducial&-&
$3\times 10^3$&180
\\\hline

M2&Isotropic velocity distribution&
Dependence on velocity structure of an NSC
&
$10^3$&43
\\\hline

M3&$\beta_\mathrm{v}=1$&
"
&
$10^3$&110
\\\hline

M4&$N_\mathrm{BH,ini}=6000$&
Dependence on the number of BHs in an NSC
&
$8.0\times 10^2$&120
\\\hline

M5&Binary-single interactions in 2D&
Influence of aligned interactions \citep{Samsing2022}
&
$6\times 10^3$&350
\\\hline

M6&
$M_{\rm SMBH}=4\times 10^7~\Msun$
&
Dependence on the SMBH mass
&
$3\times 10^3$&210
\\\hline

M7&
${\dot M}_{\rm out}=0.1~{\dot M}_{\rm Edd}$
&
Dependence on the AGN inflow rate
&
$4\times 10^3$&560
\\\hline

M8&
The initial BH masses are multiplied by $3/2$
&
Dependence on the initial BH mass
&
$3\times 10^3$&220
\\\hline

M9&
No migration
&
Influence of migration
&
$2\times 10^3$&120
\\\hline

M10&
With thermal torques
&
Influence of thermal torques
&
$2\times 10^3$&180
\\\hline

M11&
No radiation feedback
&
Influence of radiation feedback
&
$3\times 10^3$&170
\\\hline

M12&
No angular momentum change at binary-single interactions
&
Influence of inefficient changes suggested in \citet{Trani2024} and \citet{Fabj2025}
&
$2\times 10^3$&200
\\\hline

M13&
The SG model with ${\dot M}_{\rm out}=0.1~{\dot M}_{\rm Edd}$
&
Dependence on the disk model and the inflow rate
&
$7\times 10^3$&2000
\\\hline

M14&
The SG model with ${\dot M}_{\rm out}=0.01~{\dot M}_{\rm Edd}$
&
"
&
$4\times 10^3$&180
\\\hline

M15&
The TQM model with ${\dot M}_{\rm out}=0.1~{\dot M}_{\rm Edd}$
&
"
&
$3\times 10^3$&140
\\\hline

M16&
The TQM model with ${\dot M}_{\rm out}=0.01~{\dot M}_{\rm Edd}$
&
"
&
$7\times 10^2$&57
\\\hline

M17&
Super-Eddington accretion is allowed with $s=0.5$
&
Influence of super-Eddington accretion
&
$3\times 10^3$&290
\\\hline

M18&
Super-Eddington accretion is allowed with $s=0.3$
&
"
&
$3\times 10^3$&460
\\\hline

M19&
Consider transition to the ZEBRA state
&
Influence on the change in accretion prescription
&
$3\times 10^3$&840
\\\hline

M20&
Consider transition to the ZEBRA state and $f_{\rm active}=0.1$
&
"
&
$3\times 10^3$&200
\\\hline

\end{tabular}
\end{center}
\end{table*}

\subsection{Disk prescriptions in previous studies}

We briefly discuss previous studies that present suggestions different from our setup. 
Some studies claim that gap formation is inefficient \citep[e.g.,][]{Pan2021}, while updated prescriptions \citep{Duffell14,Kanagawa15,Kanagawa18} tend to favor gap formation \citep[e.g.,][]{Gilbaum2025}. 
In the presence of gaps, binary-single interactions have been found to be efficient \citep{Tagawa19}, influencing observable outcomes. 

\citet{Tomar2026} estimated that gaseous processes may become inefficient if magnetic fields 
are strong, as inferred by \citet{Hopkins2024}. However, in this model, the inflow rate onto the SMBH is quite high, at approximately $0.1$--$1$ times the star formation rate in the host galaxy, 
which may be inconsistent with the SMBH and galaxy (bulge) mass relation or typical Eddington rates of AGNs.

\citet{Hopkins2026} pointed out that the gas mass around SMBHs is constrained by the rotation speed of the gas, which disfavors the model that AGN disks with an $\alpha$ viscosity of $\alpha \lesssim 1$ extend to parsec scales, as proposed by \citet{Sirko03}. More efficient angular momentum transfer is required in the outer regions, as modeled by \citetalias{Thompson05} and \citet{EpsteinMartin2025}. This motivates our use of \citetalias{Thompson05} in the fiducial model M1.

\begin{figure*}
    \centering    \includegraphics[width=1.05\linewidth]{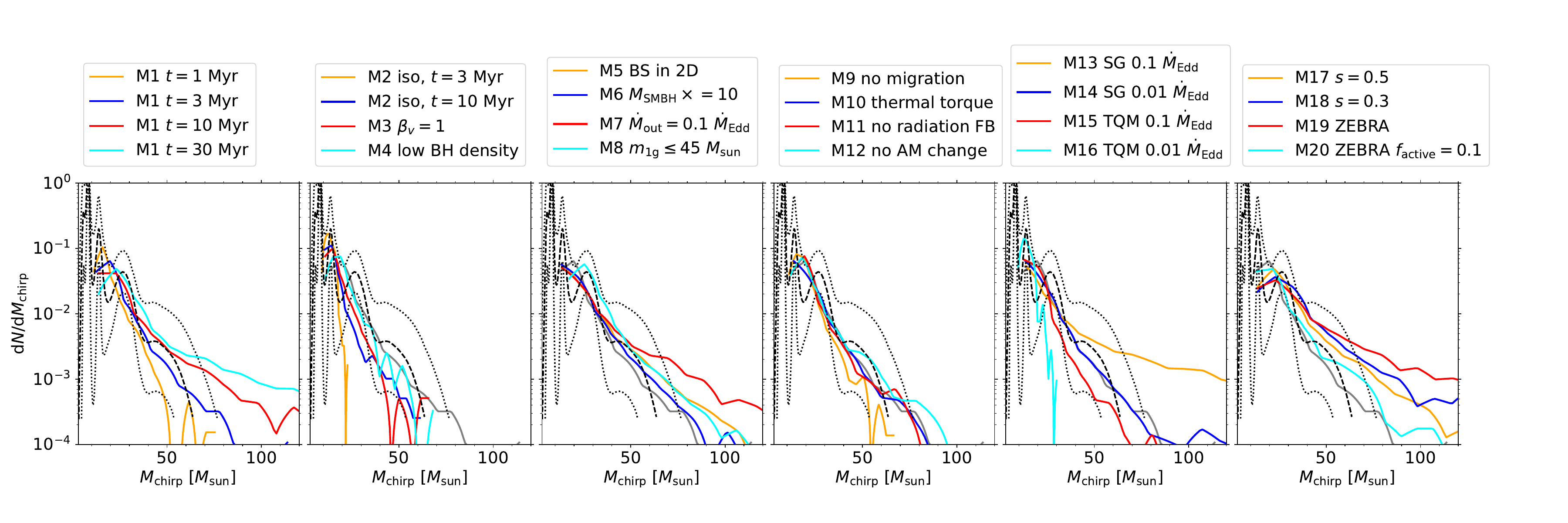}\\\vspace{-10pt}   \includegraphics[width=1.05\linewidth]{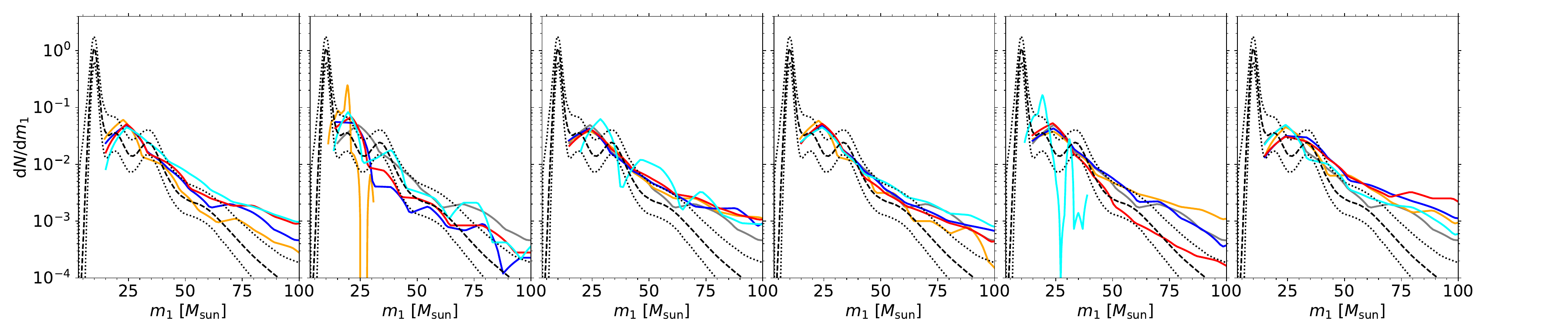}
    \caption{
    The linearly interpolated distribution of chirp mass (top row) 
    and the primary BH mass (bottom row)
    of the merging binaries for 
    models~M1--M20. 
    For model~M1 (leftmost panel), the results at $t=1~{\rm Myr}$--$30~{\rm Myr}$ are presented, while those at $t=3~{\rm Myr}$ are shown for the other models (other panels). 
    The dashed and dotted black lines represent the median and the $90\%$ credible intervals inferred from the LVK O1--O3 data \citep{Tiwari2024}.
    The gray lines show the results for model~M1 at $t=3~{\rm Myr}$ for reference. 
    The dashed and dotted black  lines in the bottom row of panels represent the median and the $90\%$ credible intervals for LVK O1--O4a data estimated using B-SPINE model in \citet{LIGO2025_O4_Prop}. 
    }    \label{fig:mc1_dist}
\end{figure*}

\begin{figure*}
    \centering    \includegraphics[width=1.05\linewidth]{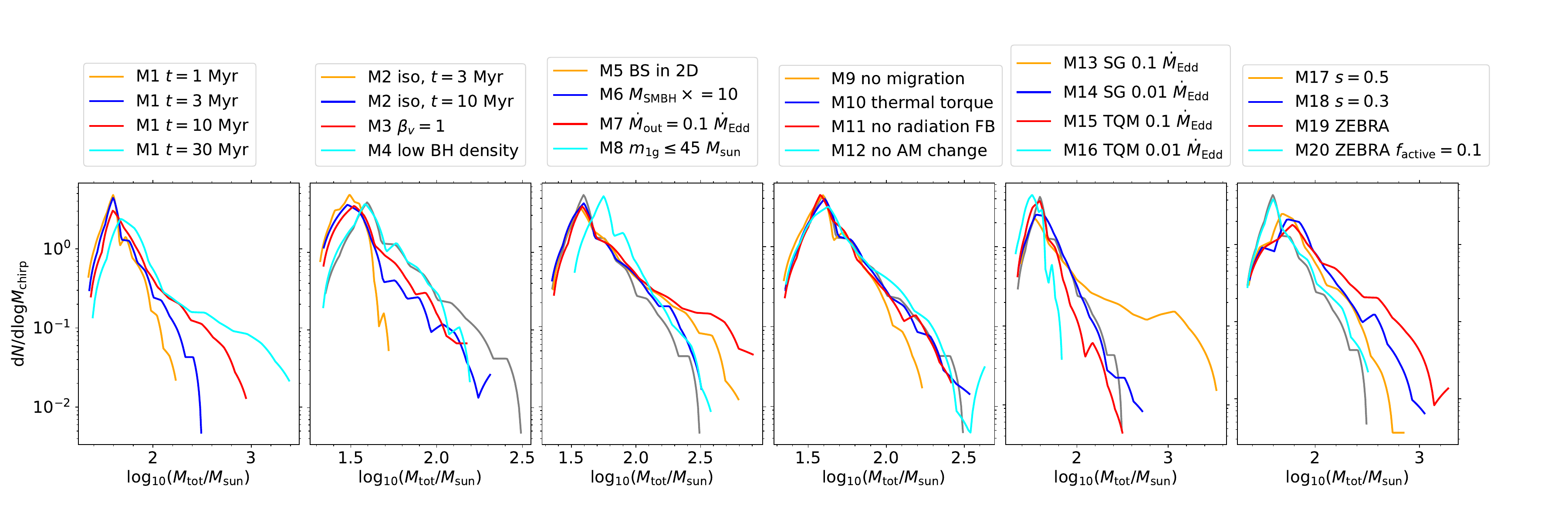}\\\vspace{-10pt}    \includegraphics[width=1.05\linewidth]{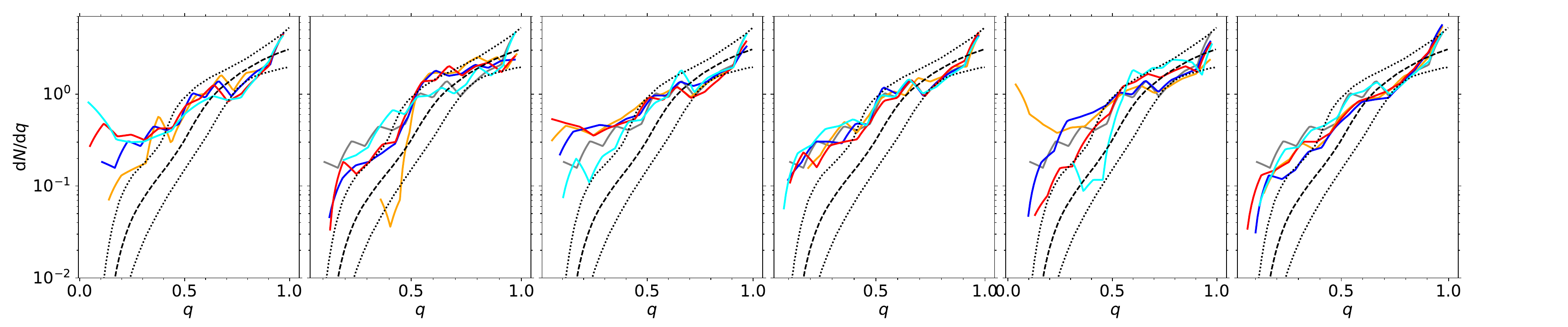}
    \caption{
    Similar to Fig.~\ref{fig:mc1_dist}, but showing the binary mass ($M_{\rm tot}$ top row) and mass ratio (bottom row) distribution on logarithmic and linear scales, respectively. 
    Dashed/dotted black lines represent the median and the $90\%$ credible intervals for LVK O1--O4a data estimated using the BROKEN POWER LAW $+$ PEAK model in \citet{LIGO2025_O4_Prop}. 
    }
    \label{fig:mq_dist}
\end{figure*}

\section{Results}

\label{sec:result}

\subsection{Bulk of mergers}

We present the distributions of observables in the fiducial model, 
and 
their dependence on various parameters. 
To demonstrate the distributions in this section, 
we performed ten simulations with different realizations of random numbers characterizing the initial BH distributions, unless otherwise specified.

The evolution of BHs can be broadly understood from Fig.~\ref{fig:times}, which summarizes 
the timescales and conditions relevant to gaseous processes. 
In the early phase, BHs with low velocities relative to the local AGN gas are gradually captured by the disk from the inner regions; the capture timescale is shown in the upper panel of Fig.~\ref{fig:times} 
(computed from Eq.~26 of \citealt{Tagawa19}\footnote{The capture timescale may be overestimated, particularly in the inner regions of $r\lesssim 10^{-2}~{\rm pc}$, where the Hill mass becomes smaller than $\sim 0.1 \times$ the BH mass \citep{Rowan2025_Inclination,Whitehead2025}.}). 
Once captured, BHs migrate within the AGN disk on the timescale shown in the second panel of Fig.~\ref{fig:times} (computed from Eqs.~17 and 18 of \citealt{Tagawa19}). 
During migration, BHs may form binaries upon encountering other BHs; the probability of binary formation is higher in the outer regions, as shown in the third panel of Fig.~\ref{fig:times}. 
This probability is computed as the ratio of the azimuthal passing timescale to the damping timescale of the relative velocity due to gas dynamical friction (Eq.~64 of \citealt{Tagawa19}), a criterion supported by recent numerical simulations \citep[e.g.,][]{Dodici2024}. 
Gaseous processes--capture by the AGN disk, binary formation and hardening--are effective only when the Bondi-Hoyle-Lyttleton radius ($r_{\rm BHL}$) exceeds the size of the H~II region 
(the possible influence of X-ray photons is ignored, \citealt{EpsteinMartin2025}); 
this ratio is shown in the fourth panel of Fig.~\ref{fig:times} (e.g.~\S~3.3.3 of \citealt{Tagawa19}). 
We calculate $r_{\rm BHL}=Gm_{\rm BH}/(v_{\rm rel}^2+c_s^2)$, where $G$ is the gravitational constant, $v_{\rm rel}$ is the relative velocity between a BH and the local gas, and $c_s$ is the sound speed of the local AGN gas. 
For gas-capture binary formation, $v_{\rm rel}$ is approximately equal to the shear velocity, while in binary hardening, it is determined by the rotation velocity of the binary. 
Gas-capture binary formation is typically unaffected by radiative feedback 
(solid lines in the bottom panel of Fig.~\ref{fig:times}), 
whereas gas dynamical friction on hard binaries with a rotation velocity significantly higher than the shear velocity can be less effective 
as $r_{\rm BHL}\propto v_{\rm rel}^{-2}$. 
In the outer regions, radiation feedback often prevents the capture of BHs into the AGN disk, particularly when the relative velocity is high. 
It is noteworthy that the reduction of gas density due to gap formation facilitates mergers driven by binary-single interactions, while mergers are primarily driven by gaseous torques when gap formation is turned off in the fiducial model, 
according to our test calculation. 
Examples of the time evolution of binary parameters can be found in previous studies \citep{Tagawa19,Tagawa2020_spin}, which are similar in the updated models adopted here. We present histograms of the BH population to quantify their evolution and merger properties next.

\subsubsection{Merging BH masses and mass ratio}

\begin{figure*}
    \centering
    \includegraphics[width=1\linewidth]{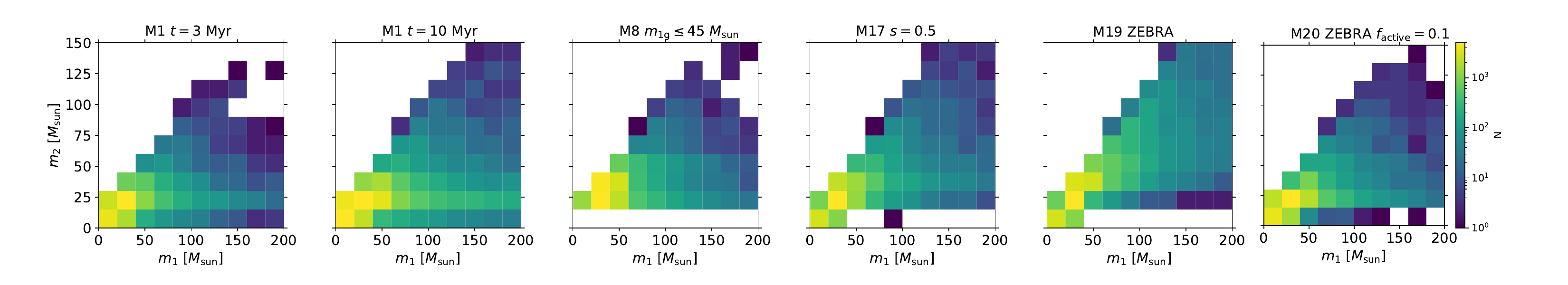}
    \caption{
    The distribution of primary and secondary BH masses in mergers. 
    To reduce Poisson noise, we performed 100 simulations. 
    }
    \label{fig:m_mpms}
\end{figure*}

Fig.~\ref{fig:mc1_dist} shows the chirp and primary BH mass distributions of the mergers in models~M1--M20 on a linear scale, respectively, 
Fig.~\ref{fig:mq_dist} shows the total mass and mass ratio distributions on logarithmic and linear scales, respectively, 
and Fig.~\ref{fig:m_mpms} shows the 2D distribution as a function of the primary and secondary BH mass in the merging binaries.

The results for the fiducial model (model~M1) at $t=1$, $3$, $10$, and $30~{\rm Myr}$ are displayed in the leftmost panels. 
For the other models, results at $t=3~{\rm Myr}$ are shown unless the duration is explicitly stated. 
These one-dimensional histograms are linearly interpolated to enhance visualization, when error regions are not included.

In the fiducial model, 
accretion is capped at the Eddington rate, and the initial BH mass is below $30~\Msun$. In the early phases, much shorter than the Salpeter timescale of $\sim 40~{\rm Myr}$, massive BHs of $\gtrsim 40~\Msun$ grow primarily through mergers. As time passes, the frequency of massive mergers increases (see the leftmost panels in Figs.~\ref{fig:mc1_dist}--\ref{fig:mq_dist}), mainly due to hierarchical mergers. 
For example, more than $90\%$ of the mass of merging BHs with $m_{\rm BH}>100~\Msun$ is attributed to hierarchical mergers 
for $t\leq 10~{\rm Myr}$, 
while the remaining mass ($\lesssim 10\%$ of the total merging masses) comes from accretion. 
For $t\gtrsim 10~{\rm Myr}$ (red and cyan lines), 
the model overproduces high-mass mergers ($M_{\rm chirp}\gtrsim 70~{\Msun}$) relative to the intrinsic merging-mass distribution as estimated by LVK observations after accounting for detection efficiency (dashed and dotted lines). 
Hence, in the fiducial model, the predicted mass distribution agrees with the observed distribution only for $t\lesssim 3~{\rm Myr}$.

The second panels in Figs.~\ref{fig:mc1_dist}--\ref{fig:mq_dist} show  
the merging mass distributions for models M2--M5. 
These models explore variations in the the number of BHs within an AGN disk by altering the initial velocity distribution of BHs, 
which affects the timescale for capture by the disk 
(orange, blue, and red lines),
or by changing the total number of BHs (cyan lines). 
As the number of BHs decreases, 
the rate of mergers among massive BHs also declines. 
This decrease can be offset by extending the duration $t$ (blue line). 
Therefore, although the effects of the number of BHs in an AGN disk and the duration of AGN phases on the merging mass distribution are degenerate, their combination can be constrained by comparing with the observed mass distributions.

The third panels of Figs.~\ref{fig:mc1_dist}--\ref{fig:mq_dist} depict variations in 
the geometry of binary-single interactions (orange lines), 
the SMBH mass (blue lines), 
the gas inflow rate toward the SMBH (red lines), 
and the initial BH masses (cyan lines). 
These parameters 
influence the merging mass distribution. 
In model~M5, the two-dimensional binary-single interactions efficiently drive mergers with fewer binary-single interactions. 
In model~M6 where the SMBH mass is high, 
the capture time to the AGN disk is extended due to the large Keplerian velocity, reducing the total number of mergers. However, a prolonged migration time in the inner region increases the fraction of hierarchical mergers. 
Model~M7 shows that a higher gas density results in efficient BH capture by the AGN disk, thus increasing the number of mergers. 
In model~8, higher initial masses for first-generation BHs generally lead to higher merging masses, enhanced by a certain factor. 
By comparing the predictions with the observed mass distribution, 
we can potentially constrain these parameters.

The fourth panels show 
the impact of varying prescriptions for the capture, migration, and binary-single interactions. 
In the absence of migration (orange lines), BHs are inhibited from accumulating, which somewhat suppresses mergers among higher-generation BHs. 
However, 
the effects of changes in migration (as in models~M9 and M10, orange and blue lines) on the mass distribution are 
not drastic, 
since migration typically takes $\gtrsim {\rm Myr}$ for many BHs in the fiducial setting, partly due to the low ${\dot M}_{\rm out}$. 
The influence of the radiation feedback (red lines) on merging masses is also not significant, 
as mergers can still occur through binary-single interactions even when gas hardening by dynamical friction becomes inefficient. 
In the model with no change in orbital angular momentum during binary-single interactions (cyan lines), mergers among high-mass BHs become moderately more frequent. This is because binaries tend to merge with their orbital angular momentum (anti-)aligned with BH spins, which reduces the recoil kick velocity during mergers among higher-generation BHs, thereby increasing the frequency of these mergers.

In the fifth panels, 
the dependence on disk models are examined. 
The frequency of highly massive mergers is higher in the SG model (orange and blue lines) and lower 
in the TQM models (red and cyan lines). 
In the SG model, 
mergers involving massive BHs are more frequent, 
because the higher gas surface density in the outer regions, 
resulting from less efficient angular momentum transfer, facilitates BH capture into the AGN disk and binary formation (the upper panel of Fig.~\ref{fig:times}). 
In contrast, the TQM model exhibits the opposite trend. 
Similarly, higher inflow rates lead to more frequent massive BH mergers (orange and red lines in the fifth panels of Figs.~\ref{fig:mc1_dist}--\ref{fig:mq_dist}), 
as also seen in model~M7. 
Note that in the fiducial model, the angular momentum transfer efficiency is comparable to that in the TQM and SG models in the outer and inner regions, respectively, reflecting the transition from global to local torques. 
Thus, the disk model significantly impacts the merging-mass distribution, primarily due to differences in gas density.

The sixth panels explore variations in accretion prescriptions. 
For the models with $s\leq 0.5$ (orange and blue lines), 
and those considering the transition to the ZEBRA state (red lines), 
the rate of massive BH mergers significantly exceeds observations. 
Therefore, if this channel contributes to a portion of the observed events, the efficiency of BH growth via accretion should be moderate. 
On the other hand, when the duty cycle of accretion is reduced to $0.1$ (cyan lines), as roughly estimated by \citet{Tagawa2022_BHFeedback}, 
the merging mass distribution aligns more closely with observations, even when accounting for the transition to the ZEBRA state.

The bottom panels in Fig.~\ref{fig:mq_dist} show the mass ratio distributions, 
along with their observed distribution, which is also model-dependent (see \citealt{LIGO2025_O4_Prop}; \citealt{Colloms2025}). 
Models with chirp mass distributions that are inconsistent with observations (specifically, model~M1 at $t\gtrsim 10~{\rm Myr}$ and models M5, M6, M7, and M13 at $t=3~{\rm Myr}$) also show similar discrepancies in the mass ratio distribution. 
This occurs because frequent hierarchical mergers tend to produce massive remnants that pair with lower-mass objects, resulting in low-mass ratio mergers. 
Such very low-$q$ mergers are less frequent in cluster models due to the lower escape velocity of the system \citep[e.g.][]{Rodriguez16}, preventing the formation of massive BHs through hierarchical mergers. 
An exception to this trend is observed in the highly accreting models (M17--M20), where the BH masses grow substantially due to gas accretion. 
This increases both the component masses and the mass ratios, producing frequent high-mass-ratio mergers that are broadly consistent with the observed distribution.

Overall, 
the merging mass and mass ratio distributions are influenced by several parameters, particularly the duration of the AGN phase, the AGN gas density, the BH number density within the AGN disk, and the accretion prescriptions. 
If a substantial fraction of mergers originates from AGN disks, observational data can be used to constrain the combination of these parameters.

Although we do not alter several parameters, 
the key parameters affecting the merger rate and the maximum BH masses found in \citet{Xue2025} 
are related to the BH number and AGN disk densities, as explored in this study. 
Table~\ref{table_results} lists the number of mergers and the top $1\%$ of masses among merging binaries. 
The dependence on these parameters is mostly consistent with the results from \citet{Xue2025}, 
while the maximum mass is higher by a factor of a few in this study. 
This discrepancy is presumably due to the implementation of binary component exchanges and the evolution of background BHs through mergers and accretion, 
which tends to produce more massive binaries and merger remnants. 
Specifically, the absence of exchanges during binary-single interactions and the growth of background objects through mergers can, respectively, reduce the mass enhancement at mergers by up to a factor of $\sim 3$ (reflecting the initial BH mass range) and by a factor of the expected number of generations of merging background BHs.

In the fiducial model, the AGN lifetime needs to be less than $\sim 3~{\rm Myr}$, 
limited primarily by BH number density. Currently, the fiducial model aligns with the observed constraints. 
If the accretion rate onto BHs is characterized by a super-Eddington rate, 
the duty cycle of the accretion phase should be low as investigated in model~M20, 
or substantial wind losses may regulate accretion \citep[e.g.,][]{Sadowski2016}. 
These variations among models could be further constrained through comparisons with other observables.

\begin{figure*}
    \centering \includegraphics[width=1\linewidth]{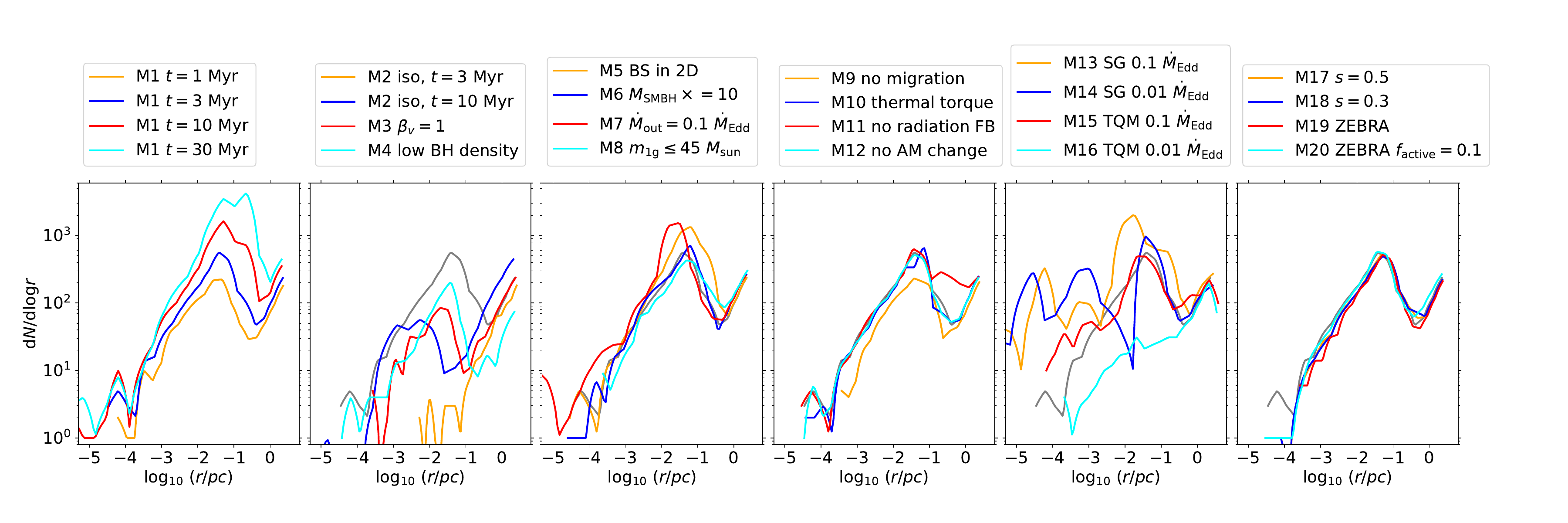}
    \caption{
    The linearly interpolated distribution of the distance from the SMBH of the mergers for models~M1--M20. 
    }
    \label{fig:r_dist}
\end{figure*}

\begin{figure*}
    \centering
    \includegraphics[width=1\linewidth]{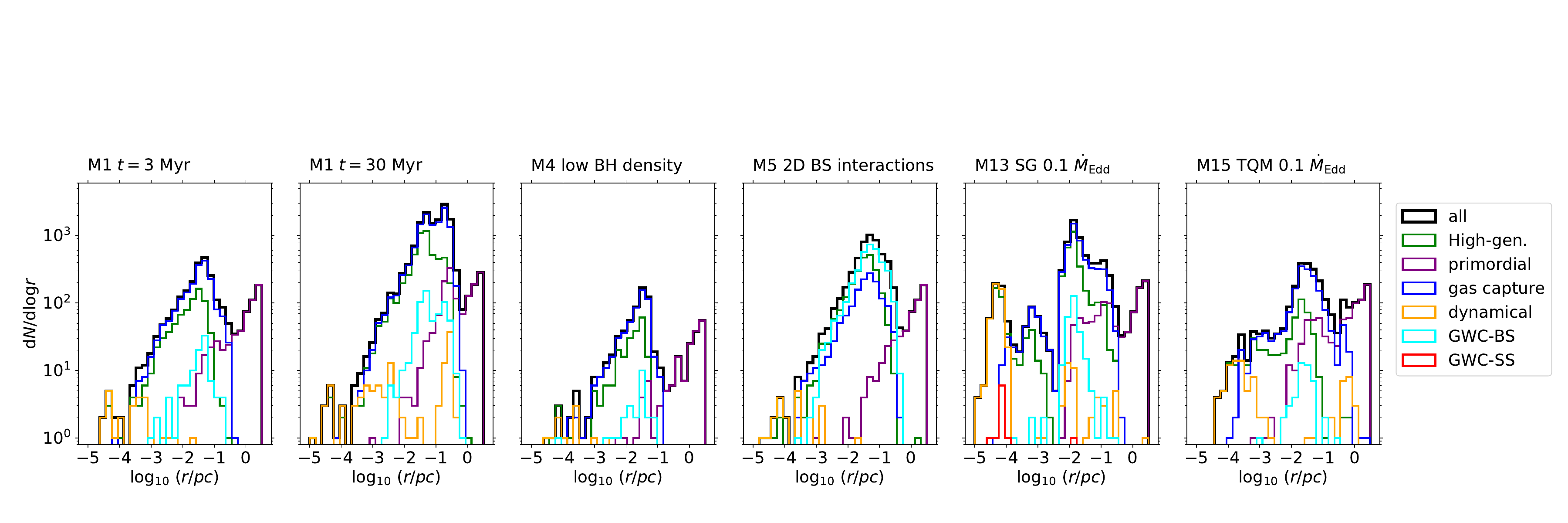}
    \caption{
    The distribution of the distance of mergers from the SMBH for different binary formation mechanisms in models~M1, M4, M5, M13, and M15. 
    The purple, blue, orange, red, and cyan lines represent 
    mergers from binaries formed as primordial binaries, through gas capture processes, dynamical interactions, 
    GW capture during single-single interactions, and GW capture mechanisms 
    during binary-single interactions, respectively. 
    We also present the merger locations for binaries containing higher-generation BHs by the green line. 
    }
    \label{fig:r_type}
\end{figure*}

\begin{figure*}
    \centering
    \includegraphics[width=1.05\linewidth]{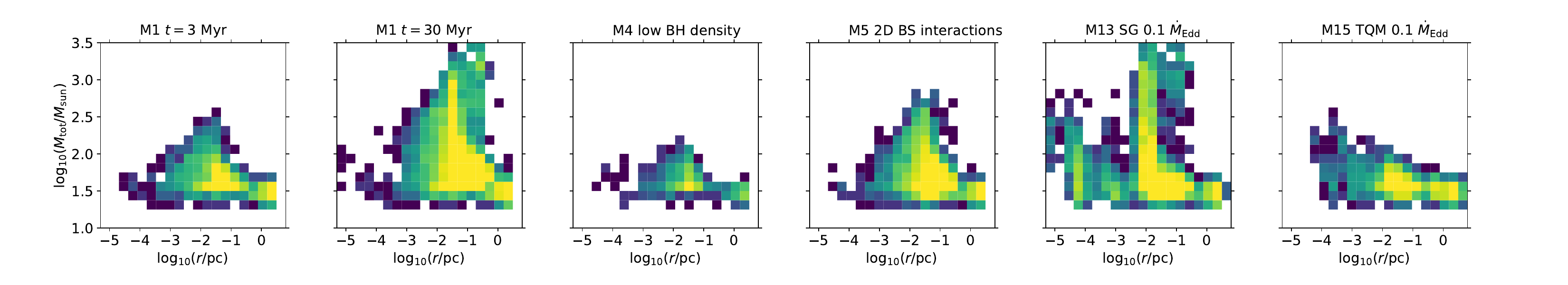}
    \caption{
    The distribution of the binary mass and their distance from the SMBH for all merging BHs in models~M1, M4, M5, M13, and M15. 
        }
    \label{fig:rm_dist}
\end{figure*}

\subsubsection{Merger location}

\label{sec:location}

Figs.~\ref{fig:r_dist} and \ref{fig:r_type} shows the distribution of radial distances of the mergers ($r$) from the SMBH, 
and Fig.~\ref{fig:rm_dist} presents the distributions for both the $r$ and the binary mass at merger. 
Specifically, Fig.~\ref{fig:r_dist} shows the overall distribution, while Fig.~\ref{fig:r_type} presents the distribution for different binary formation mechanisms. 
Figs.~\ref{fig:r_type} and \ref{fig:rm_dist} depict results 
across models~M1, M4, M5, M13, and M15, selected because they exhibit variations in merger locations. 
In both the fiducial and many other models, 
two peaks appear near $\sim 0.01$--$0.05~{\rm pc}$ and $\sim 3~{\rm pc}$.

The outer peak is dominated by primordial binaries (purple lines in Fig.~\ref{fig:r_type}). 
These mergers occur slowly due to the low gas density, and 
are typically between a pair of first-generation BHs. 
Remnant BHs merged at $\sim 1~{\rm pc}$ often escape by GW kicks, resulting from the shallower gravitational potential of an SMBH. They are inefficiently recaptured by AGN disks due to lower gas density, leading to infrequent hierarchical mergers (Fig.~\ref{fig:rm_dist}).

Generally the mergers near the inner peak $\sim 0.01$--$0.05~{\rm pc}$ are dominated by gas-assisted binary formation. In this region BHs are captured by the disk earlier due to the higher gas surface densities there and more frequent disk crossings (the top panel of Fig.~\ref{fig:times}). 
After their capture, BHs migrate, form binaries due to gas, and merge rapidly, with most hierarchical mergers occurring in this region (Fig.~\ref{fig:rm_dist}). 
Over time, the radial location of mergers involving higher-generation BHs gradually extends to the $\sim {\rm pc}$ regions (green line in the second panel of Fig.~\ref{fig:r_type}) as the BHs get captured in the disk in these regions as well.

In the inner regions, binary-single interactions take place, and merging 
binaries form through a GW capture process during these interactions (cyan lines in Fig.~\ref{fig:r_type}). 
Since the fraction of mergers resulting from this process ($\sim 5$--$10\%$, \citealt{Tagawa20_ecc}) is comparable to that in cluster models \citep[e.g.,][]{Rodriguez18bR,Zevin2021}, the frequency of binary-single interactions before mergers is likely to be similar. 
Note that this rate might be significantly enhanced by dense gas accumulated around a binary, as the ejection of this gas during interactions can significantly harden the three-body system \citep{Rowan2025_BS,Tagawa2025}. 
Mergers via this process become highly frequent if binary-single interactions occur in a 2D plane, in that case dominating over the gas-assisted binary formation pathway (cyan lines in the fourth panel of Fig.~\ref{fig:r_type}). 
Unlike in our previous papers \citep[e.g][]{Tagawa20_ecc}, 
binary formation and mergers through single-single GW capture encounters are rare in all of our models, due to the updated encounter probabilities described in $\S~\ref{sec:updates}$.

In cases of low BH number density (models~M2--M4), 
mergers are less frequent in the inner regions, as the chances of encounters with other BHs and the binary formation are reduced. 
Hence, mergers among massive BHs are infrequent in these models (e.g. Fig.~\ref{fig:rm_dist}).

The location of mergers is significantly influenced by the choice of AGN disk model. 
In the SG model, 
mergers via the gas-capture process become inefficient 
around several $\times 10^{-3}~{\rm pc}$ (dip features in orange and blue lines in the fifth panel of Fig.~\ref{fig:r_dist}). 
This is due to enhanced opacity from dust sublimation, which increases the scale height and reduces the gap depth, thereby shortening the migration timescale (solid blue line in the second panel of Fig.~\ref{fig:times}). 
Note that this change in opacity appears in the inner regions of $\lesssim 10^{-3}~{\rm pc}$ in the fiducial model, though its effects are not particularly prominent. 
Furthermore, binaries often merge through three-body encounters and GW capture during single-single interactions at distances of several $\times 10^{-5}~{\rm pc}$. This is attributed to the iron opacity bump \citep{Jiang2016}, which is implemented only in the SG model developed by \citet{Gangardt2024} among the disk models we used. 
In this region, the migration timescale increases especially 
for smaller BHs, leading to elevated merger rates. 
In the TQM model with ${\dot M}_{\rm SMBH}=0.1~{\dot M}_{\rm Edd}$, mergers are relatively frequent within $10^{-3}~{\rm pc}$ because the inward migration timescale increases there. This effect is less prominent in the TQM model with ${\dot M}_{\rm SMBH}=0.01~{\dot M}_{\rm Edd}$.

Merger locations may be directly constrained by the detection of Doppler acceleration for inspiraling sources using future GW facilities \citep{Meiron17,Inayoshi17b,Wong2019,Han_Yang2024,Hendriks2024,Zwick2025,Tagawa2025,Hendriks2026} and/or using the detection or nondetection of a lensed secondary GW signal by the central SMBH \citep{Gondan_Kocsis2022}. Combining with the expected properties of merging BH population discussed in this paper, this could offer novel indirect constraints on the properties of AGN disks.

\begin{figure*}
    \centering
    \includegraphics[width=1\linewidth]{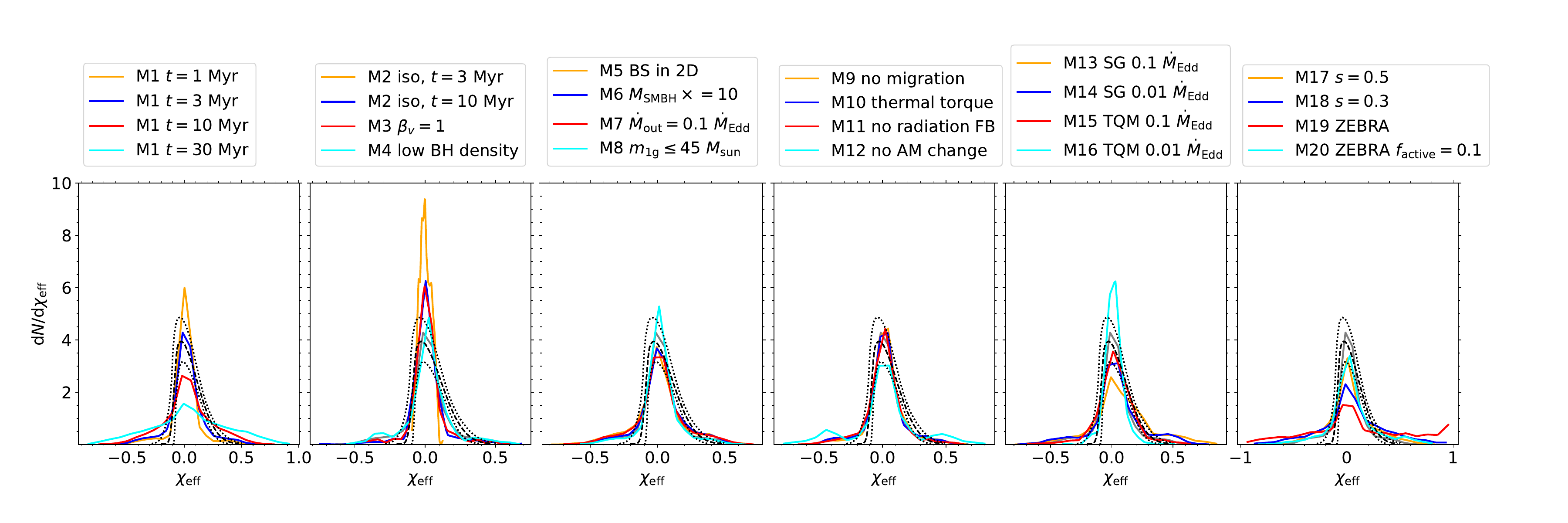}
    \caption{
    Similar to Fig.~\ref{fig:mc1_dist}, but 
    the distribution of $\chi_{\rm eff}$ is shown.  
    The dashed and dotted black lines represent the median and the $90\%$ credible intervals for LVK O1--O4a data estimated using the SKEW-NORMAL EFFECTIVE SPIN model in \citet{LIGO2025_O4_Prop}. 
    }
    \label{fig:xeff_dist}
\end{figure*}

\begin{figure*}
    \centering
    \includegraphics[width=1\linewidth]{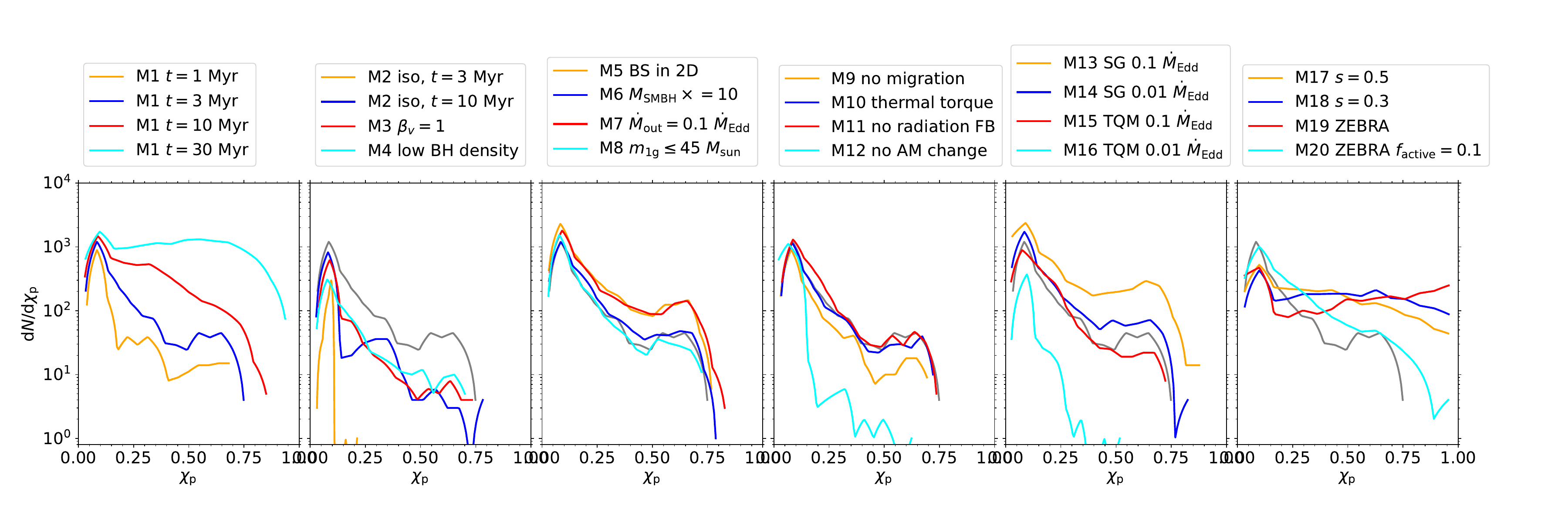}
    \caption{
    Similar to Fig.~\ref{fig:mc1_dist}, but 
    the distribution of $\chi_{\rm p}$ is shown.  
    }
    \label{fig:xp_dist}
\end{figure*}

\begin{figure*}
    \centering
    \includegraphics[width=1\linewidth]{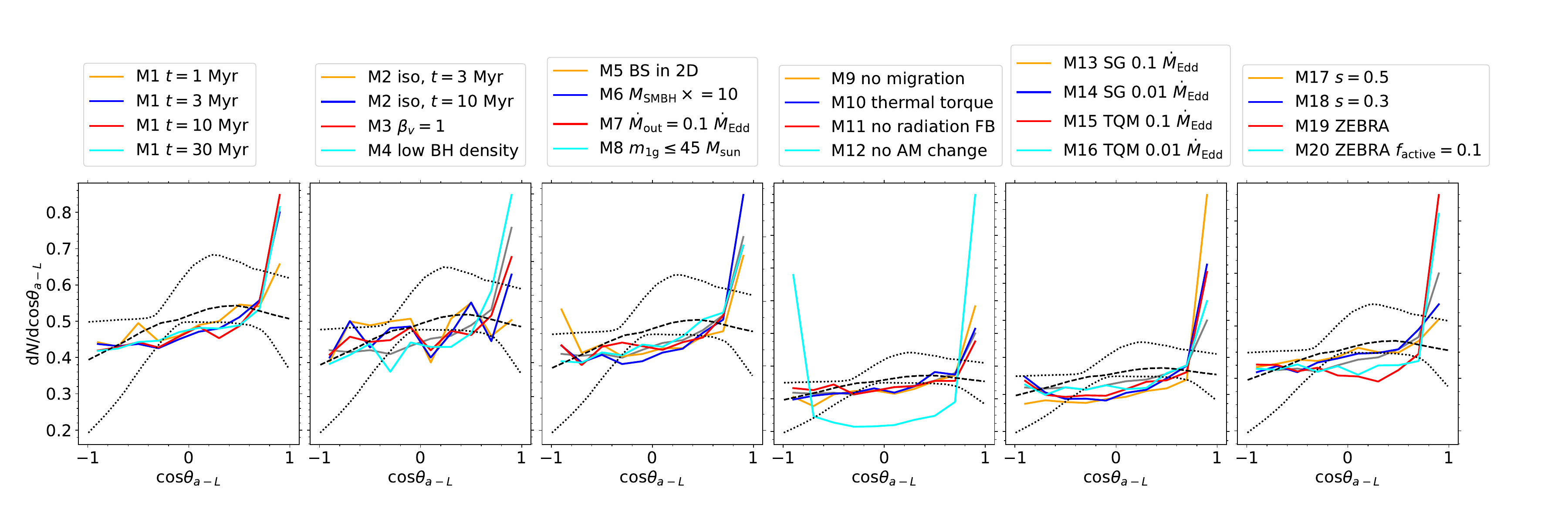}
    \caption{
    Similar to Fig.~\ref{fig:mc1_dist}, but 
    the distribution of $\rm{cos}(\theta_{a-L})$ is shown.  
    The dashed and dotted black lines represent the median and the $90\%$ credible intervals for LVK O1--O4a data estimated using the GAUSSIAN COMPONENT SPINS model. 
    }
    \label{fig:costh_al}
\end{figure*}

\subsubsection{Spin tilts}

The distributions of $\chi_{\rm eff}$ and $\chi_{\rm p}$ are constrained by observations, providing valuable information on the astrophysical models involving the mergers. 
Figs.~\ref{fig:xeff_dist} and \ref{fig:xp_dist} show the $\chi_{\rm eff}$ and $\chi_{\rm p}$ distributions for models~M1--M20. 
Fig.~\ref{fig:costh_al} shows the distributions of the angle between the BH spin and the orbital angular momentum direction of merging binaries ($\theta_{a-l}$), which is useful for understanding the $\chi_{\rm eff}$ and $\chi_{\rm p}$ distributions.

In the fiducial model, 
$\chi_{\rm eff}$ and $\chi_{\rm p}$ peak at 0, as most mergers involve first-generation BHs with spin magnitudes around $\sim 0.1$. 
A subset of these mergers is contributed by higher-generation BHs, which typically have spin magnitudes around $\sim 0.7$. 
The distribution of $\theta_{a-l}$ results from the interplay between alignment due to the Bardeen-Peterson effect and randomization by binary-single interactions, leading to a broad distribution around ${\rm cos}(\theta_{a-l})=1$. 
Consequently, 
the $\chi_{\rm eff}$ and $\chi_{\rm p}$ values for higher-generation mergers are spread over the ranges $\sim -0.7$ to $\sim 0.7$ and $0$ to $\sim 0.7$, respectively, with a weak asymmetry in the $\chi_{\rm eff}$ distribution.

Over time, the fraction of mergers involving higher-generation BHs increases, which broadens the $\chi_{\rm eff}$ and $\chi_{\rm p}$ distributions. 
In contrast, if the proportion of hierarchical mergers is low (e.g. models~M2--M4, M16), $\chi_{\rm eff}$ and $\chi_{\rm p}$ both become concentrated in a narrow range around 0.

In cases where the binary angular momentum direction is not randomized during binary-single interactions (model~M12), the angular momentum directions of binaries are typically aligned or anti-aligned with respect to the angular momentum direction of the AGN disk. 
For first-generation BHs, BH spins evolve through gas accretion, leading to a reduced $\theta_{a-l}$. 
For higher-generation BHs, the spin magnitude is significantly increased through mergers, causing their spin directions to align with the angular momentum direction of the merged binary. 
The angular momentum direction of binaries formed through gas capture (as well as GW capture during single-single interactions) is set to be either prograde or retrograde with respect to the AGN disk rotation, with equal probabilities. 
Consequently, the BH spin directions for high-generation BHs are typically (anti-)aligned with the binary angular momentum direction (two peaks in the fourth panel of Fig.~\ref{fig:costh_al}). 
This alignment results in a broader distribution of $\chi_{\rm eff}$ and a narrower distribution of $\chi_{\rm p}$, compared to other models.

When accretion is highly efficient, BH spin directions quickly align with the binary angular momentum directions via the Bardeen-Petterson effect (red and cyan lines in the sixth panel of Fig.~\ref{fig:costh_al}), while their spin magnitudes significantly increase in the model with efficient gas accretion (model~M19, e.g., the sixth panel of Fig.~\ref{fig:m_a_dist}). 
In model~M19, these effects produce a broad, asymmetric distribution of $\chi_{\rm eff}$ relative to 0, with higher values of $\chi_{\rm p}$. 
Furthermore, in the SG model with high accretion rates (model~M13), 
the alignment process operates efficiently due to the high gas density, particularly in the outer regions, 
leading to a reduction in $\theta_{a-l}$.

Observations constrain the distribution of $\theta_{a-l}$, 
suggesting it is broad. 
\citet{LIGO2025_O4aProp} estimated a possible peak in $\rm{cos}(\theta_{a-l})$ near zero (see also \citealt{LiYinJie2025_alignment,Farah2026}), although a peak near one cannot be ruled out; \citet{Vitale2025} noted that a peak near one tends to be underestimated. 
Consequently, the fiducial and many other models are 
not ruled out by the observed distribution, 
while model~M12, which prefer $\rm{cos}(\theta_{a-l})\sim -1$, is likely disfavored.

Although the observed $\chi_{\rm p}$ distribution is reported by \citet{LIGO2025_O4aProp}, 
it may contain a significant bias. 
This speculation arises because $\chi_{\rm p}$ should be lower than the spin magnitude of merging BHs, while the peak of the $\chi_{\rm p}$ distribution exceeds the peak of the spin magnitude distribution. 
Therefore, we refrain from directly comparing our results to the observed distribution.

For the distribution of $\chi_{\rm eff}$, the peak values appear to be consistent with the observed data; however, negative $\chi_{\rm eff}$ values are overproduced in some of the models. 
The fraction of mergers with $-0.3 \lesssim \chi_{\rm eff}\lesssim -0.2$ is model-dependent, while those with $\chi_{\rm eff}\lesssim -0.4$ are likely rare. Only a few events may exhibit such values, such as GW241110, which has $\chi_{\rm eff}= -0.28^{+0.23}_{-0.20}$ \citep{LVK26_GW241110}. 
We find that the fraction of mergers with $\chi_{\rm eff}<-0.4$ is $\sim 1\%$ in the fiducial model, which is roughly consistent with the observed $\chi_{\rm eff}$ distribution. 
However, it is higher, around $\sim 4\%$, in models~M12, M13, and M14. 
In models~M18, M1 at $t=30~{\rm Myr}$, and M19, 
the fractions are $8\%$, $10\%$, and $15\%$, respectively. 
As a result, the contribution from these models to the observed events can be constrained to be minor. 
Nevertheless, caution may be needed in comparison, as the observed $\chi_{\rm eff}$ distribution could also be somewhat biased \citep[e.g.,][]{Williamson2017}.

\begin{figure*}
    \centering
    \includegraphics[width=1\linewidth]{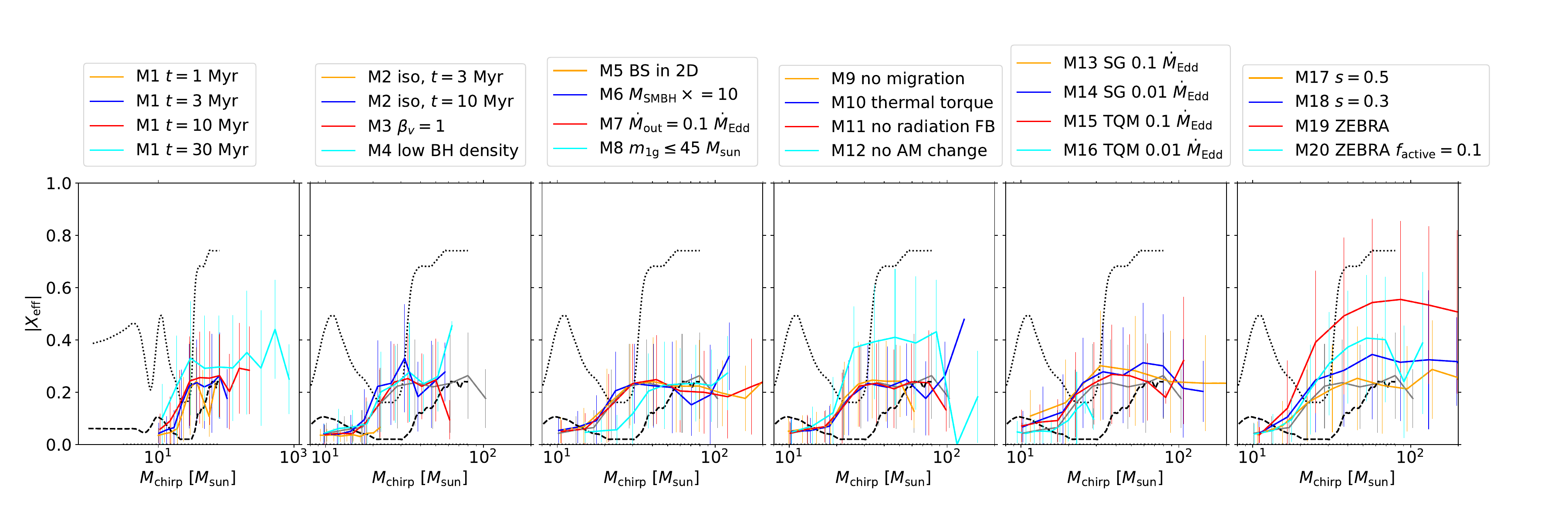}
    \caption{
    The average and dispersion of the absolute value of the effective spin parameter as a function of the chirp mass for models~M1--M20. 
    The dashed and dotted black lines represent the median and the $90\%$ credible intervals for the aligned spin magnitude $\chi_z$, as inferred from the LVK O1--O4a data estimated in  \citet{LIGO2025_O4_Prop}.    
    When the number of mergers in a bin is less than $2$, we do not present the dispersion. 
    }
    \label{fig:xm_dist}
\end{figure*}

\begin{figure*}
    \centering
    \includegraphics[width=1\linewidth]{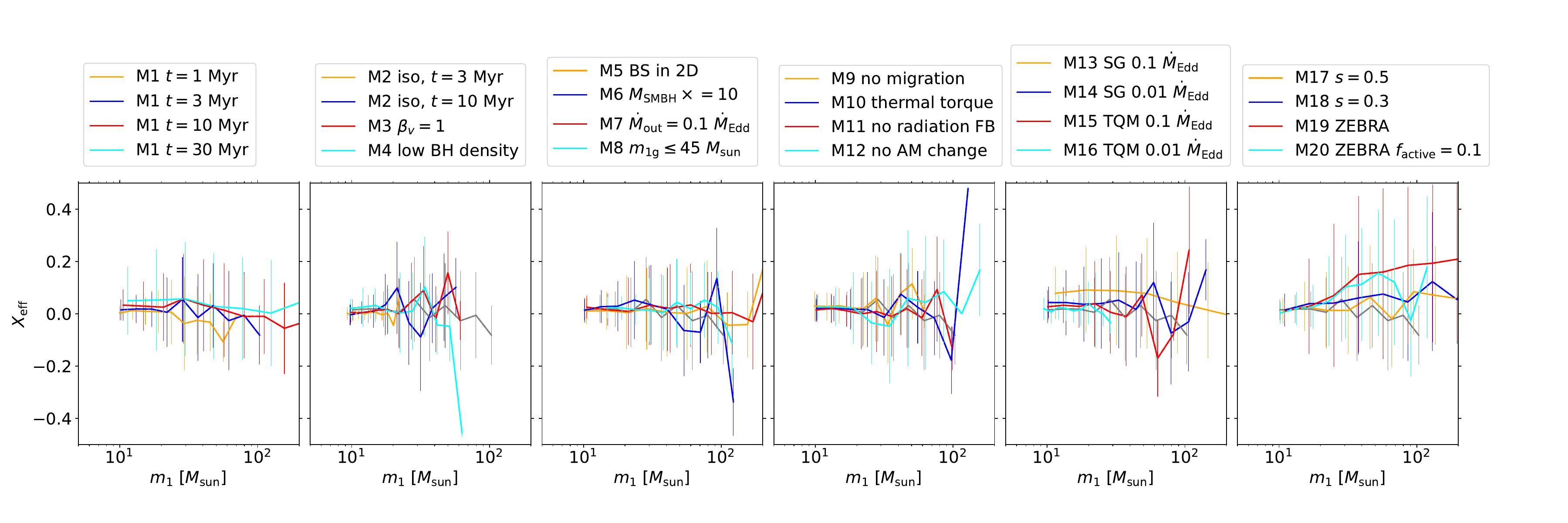}
    \caption{
    The average and dispersion of $\chi_{\rm eff}$ 
    as a function of $m_{\rm 1}$ 
    for models~M1--M20. 
    }
    \label{fig:xeffm_dist}
\end{figure*}

\begin{figure*}
    \centering
    \includegraphics[width=1\linewidth]{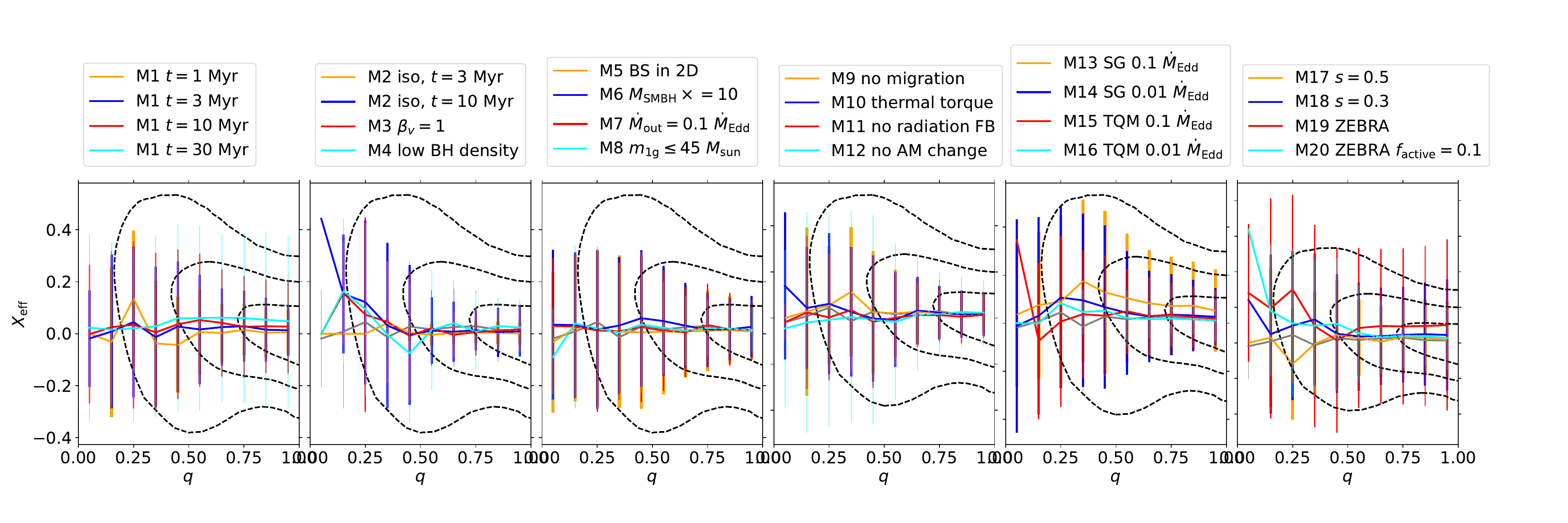}
    \caption{
    The average and dispersion of the effective spin parameter 
    as a function of the mass ratio for models~M1--M20. 
    The three black lines represent $50\%$, $90\%$, and $99\%$ credible regions inferred from the LVK O1--O4a data estimated using the LINEAR model in \citet{LIGO2025_O4_Prop}. 
    }
    \label{fig:xq_dist}
\end{figure*}

\begin{figure}
    \centering
    \includegraphics[width=1\linewidth]{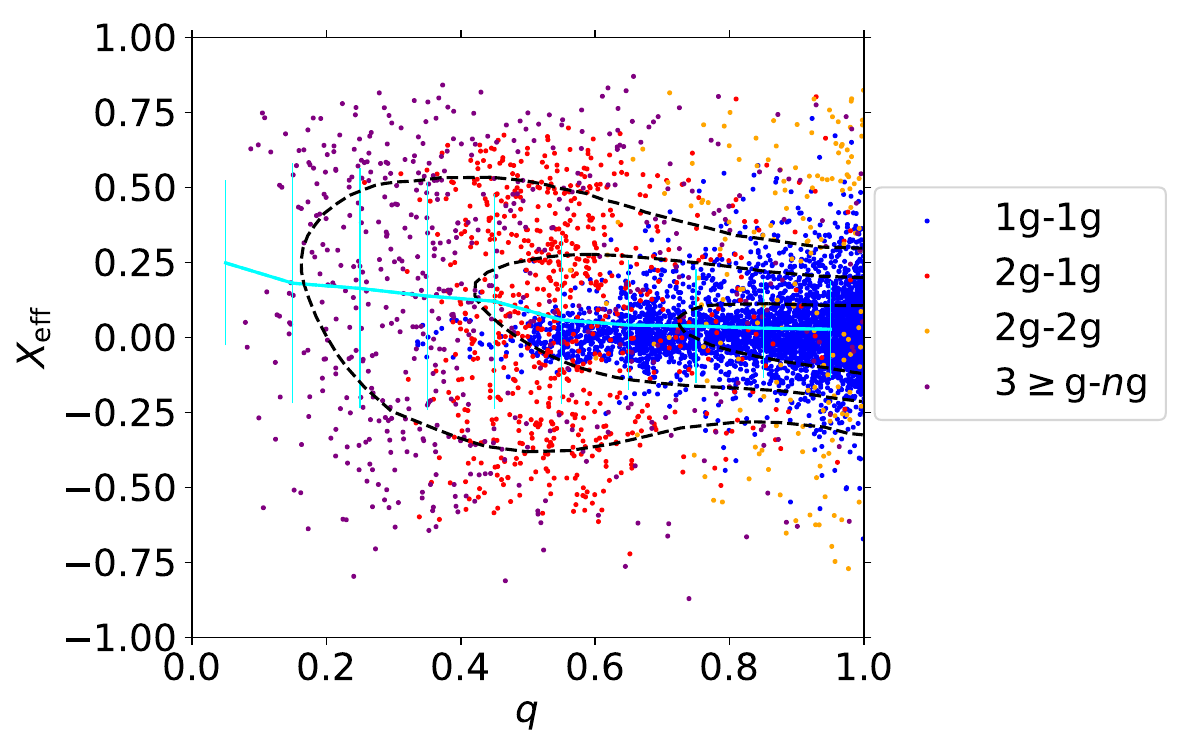}
    \caption{
    Same as Fig.~\ref{fig:xq_dist}, but for model~M20, with the distribution of mergers shown. 
    The median and dispersion of $\chi_{\rm eff}$ are calculated by performing 100 simulations, to reduce a bias caused by Poisson noise. 
    Mergers among first-generation (1g)-1g, 2g-1g, 2g-2g, and $3\geq$g-$n$g BHs are represented by blue, red, orange, and purple points, respectively, for 20 simulations (for visualization purposes). 
    }
    \label{fig:xq_dist2}
\end{figure}

\subsubsection{Spin and mass}

\label{sec:spin_mass}

Here, we present the correlation between mass and spin, 
as well as between mass ratio and spin, as suggested by observed GW data \citep{Callister2021,Tagawa2021_hierarchical,Adamcewicz2022,Adamcewicz2023,LiYinJie2024,Pierra2024,Antonini2025,LIGO2025_O4aProp,LiYinJie2025_alignment,LIGO2025_O4_Prop,WangYuanZhu2025,Tong2025,Plunkett2026,Vijaykumar2026,Farah2026}.

Fig.~\ref{fig:xm_dist} shows the average and dispersion of the absolute value of $\chi_{\rm eff}$ as a function of $M_{\rm chirp}$. 
In the higher $M_{\rm chirp}$ bins, the number of mergers is often small, making the estimates of the average and dispersion less robust. 
We observe an increase in $|\chi_{\rm eff}|$ at low masses, 
followed by a plateau at higher masses in most models. 
This plateau arises from hierarchical mergers, as the spin magnitudes of remnant BHs tend to cluster around $\sim 0.7$. 
The minimum mass of the plateau roughly corresponds to the maximum initial mass of first-generation BHs \citep{Tagawa2021_hierarchical}; 
this boundary shifts in model~M8, where the maximum initial BH mass is changed.

Above this mass, several models show larger mean $|\chi_{\rm eff}|$ ($|\chi_{\rm eff}|\gtrsim 0.3$) than the fiducial model ($|\chi_{\rm eff}|\sim 0.2$) at $t=3~{\rm Myr}$. 
This behavior appears in highly accreting models (models~M17--M19), SG models (models~M13 and M14), 
the model with no change in orbital angular momentum during binary-single interactions (model~M12), 
and in the fiducial model at $t=30~{\rm Myr}$. 
These differences reflect BH spin evolution due to accretion and smaller angles between BH spins and the binary orbital angular momentum (e.g., in model~M12). 
Thus, the plateau contains information useful for constraining accretion processes and interactions among BHs. 

An increase and plateau in the $M_{\rm chirp}$--spin plane has been implied from GW data \citep{Tagawa2021_hierarchical,LiYinJie2024,Antonini2025,Pierra2024,WangYuanZhu2025}, which can help constrain model parameters. 
Dashed and dotted lines represent the median and $90\%$ credible interval for aligned spin magnitude from \citet{LIGO2025_O4aProp}. 
Although current constraints are weak, they indicate a possible increase in $|\chi_{\rm eff}|$ at high masses, along with a slight increase around $\sim 10~\Msun$, 
likely contributed by 
hierarchical mergers within 
a subpopulation \citep{Tong2025,Plunkett2026,Farah2026}. 
Potential locations for these events include dense star clusters with high metallicity \citep{Ye2026}. 
Additionally, this subpopulation may consist of in-situ formed stars with higher metallicity, 
while more massive components could originate from the migration of globular clusters with lower metallicity in AGN environments \citep{Fahrion2021,Fahrion2022}. 
Comparing models to these intervals, except for the possible increase in low-mass regions, 
most models slightly exceed the $90\%$ credible region around $\sim 30~\Msun$, while the model with more massive first-generation BHs (model~M8) aligns better with the median. The observed median also suggests higher maximum first-generation BH masses above $\sim 45~\Msun$. With tighter observational constraints, 
the formation and evolution of first-generation BHs will be more tightly constrained. 
Hence, the correlation between mass and spins provides valuable insight into the properties of first-generation BHs, accretion mechanisms, and interactions between BHs.

\citet{LiYinJie2025_alignment} demonstrate a correlation between $\theta_{a-l}$ (or $\chi_{\rm eff}$) and the mass of merging BHs. Notably, an increase in $\theta_{a-l}$ (or $\chi_{\rm eff}$) is observed in mergers within the upper-mass gap regions, posing challenges to various models. 
They found that mergers with high spin magnitudes 
tend to populate around $\chi_{\rm eff}\sim 0.2$--$0.5$ (or $\rm{cos}(\theta_{a-l})\gtrsim 0.5$) and $m_1\gtrsim 40~{\Msun}$ 
(we do not show this in the figure).

To explore this correlation, 
Fig.~\ref{fig:xeffm_dist} shows the relationship between $\chi_{\rm eff}$ and $m_{\rm 1}$. 
For most models, no significant correlation is observed. 
However, models with high accretion rates 
show a correlation. 
This is because accretion torques the binary angular momentum toward the prograde direction of the AGN disk, resulting in merged remnants having a similar spin direction. When these remnants merge again, they tend to exhibit small $\theta_{a-l}$ or high $\chi_{\rm eff}$, 
reproducing the observed correlation. 
The mass range and typical magnitude of $\chi_{\rm eff}$ for the high $\chi_{\rm eff}$ events are roughly consistent with the findings of \citet{LiYinJie2025_alignment}, offering a plausible model that explains these observations.

Fig.~\ref{fig:xq_dist} shows the correlation between the mass ratio ($q$) and $\chi_{\rm eff}$. Contrary to the observational implications presented by \citet{Callister2021}, \citet{Adamcewicz2022}, \citet{Adamcewicz2023}, and \citet{LiYinJie2025_Xeffq}, 
most of our models do not show a clear correlation between $q$ and the average of $\chi_{\rm eff}$. 
This is because binary orbital angular momentum directions are randomized by binary-single interactions, and mergers usually occur before BH spins align with the binary orbital angular momentum via gas accretion. 
Note that in the model that does not randomize angular momentum directions during binary-single interactions (model~M12), the angular momentum direction of binaries typically remains unchanged from the initial direction. 
In this case, it is aligned (or anti-aligned) with the AGN disk angular momentum direction with a probability of $0.5$, which limits the efficient production of positive $\chi_{\rm eff}$.

However, models characterized by rapid accretion (models~M19--M20) and by high gas density (model~M13) show a correlation between $q$ and the average of $\chi_{\rm eff}$. 
This correlation arises from the enhancement of $\chi_{\rm eff}$ driven by spin alignment through accretion due to the Bardeen-Petterson effect, the asymmetric mass ratios resulting from mergers across different generation BHs (such as mergers between first- and second-generation BHs), and spin enhancements due to mergers. 
Note that while mass and spin growth through accretion contribute to a positive correlation, mass growth from hierarchical mergers leads to a negative correlation by producing asymmetric mass ratio mergers. 
Although alignment due to accretion is required for the negative correlation, 
the growth of mass and spin magnitude should be predominantly driven by mergers, with accretion playing a minor role.

Fig.~\ref{fig:xq_dist2} also shows the $q$--$\chi_{\rm eff}$ correlation and the distribution of individual mergers for model~M20. 
Mergers between first- and second-generation BHs are typically clustered around $q\sim 0.4$--$0.7$ (red points), while mergers involving third or higher-generation BHs contribute to $q\gtrsim 0.1$ (purple points). Here, we define an $n$th-generation BH as one composed of $n$ first-generation BHs. 
These results imply that hierarchical mergers with efficient alignment via gas accretion can produce a negative correlation between the average of $\chi_{\rm eff}$ and $q$. 
Moreover, 
high-generation mergers generally have large $\chi_{\rm eff}$, 
consistent with \citet{LiYinJie2025}. 
Thus, in the AGN channel, it is possible to generate a negative correlation between $q$ and the mean $\chi_{\rm eff}$, 
providing a promising observational signature of this model.

Similarly, a negative correlation between $q$ and the spin magnitude is suggested by \citet{Vijaykumar2026}. This correlation can be produced by hierarchical mergers in most models, unless gas accretion is efficient and significantly enhances $q$ values, as seen in model~M19.

On the other hand, \citet{LIGO2025_O4_Prop} suggested that by analyzing a larger number of GW events and using a more flexible model, 
the confidence level for the negative correlation between $q$ and the average of $\chi_{\rm eff}$ is reduced to $82\%$ credibility, while the negative correlation between $q$ and the dispersion of $\chi_{\rm eff}$ is found to have $95\%$ credibility. 
This correlation between $q$ and the dispersion of $\chi_{\rm eff}$ is clearly observed in most models, except for model~M1 at $t=30~{\rm Myr}$ and model~M19. 
This correlation is driven by hierarchical mergers, which result in high BH spin magnitude and low $q$ (if the generations of merging BHs are different), leading to mergers characterized by low $q$ and high dispersion of $\chi_{\rm eff}$. 
Since the efficient accretion likely diminishes this correlation, 
this comparison is useful for constraining the accretion process.

If the negative correlation between $q$ and the average of $\chi_{\rm eff}$ is real, 
a model characterized by efficient accretion and hierarchical mergers, 
such as model~M20, is favored (Figs.~\ref{fig:xq_dist} and \ref{fig:xq_dist2}). 
This model also produces a positive correlation between $m_{\rm 1}$ and the average of $\chi_{\rm eff}$ (Fig.~\ref{fig:xeffm_dist}), as suggested by \citet{LiYinJie2025_alignment}. Since this positive correlation is well reproduced only by highly accreting models (models~M19 and M20) while the model with the most efficient accretion (model~M19) fails to reproduce the mass distribution, 
we tentatively regard model~M20 as the most promising model. 
A potential problem is that the average of $|\chi_{\rm eff}|$ at high $M_{\rm chirp}$ appears larger than the observational median; if this constraint tightens to around the current median values, a model with less efficient accretion (e.g., model~M1) would be preferred. 
Further observations and analyses will help discriminate between these models.

Several studies have reproduced the $q$-$\chi_{\rm eff}$ correlation via the AGN channel \citep{McKernan2022,McKernan2024,McKernan2025,Cook2024}, although notable differences remain relative to our model. 
\citet{McKernan2022} and \citet{McKernan2024} explained that this correlation 
arises because massive BHs tend to be rapidly captured by the AGN disk and spun up more efficiently, enabling mergers with lower-mass massive BHs that retain random spins. 
Meanwhile, \citet{McKernan2025} and \citet{Cook2024} interpret the correlation in terms of spin alignment and hierarchical mergers; however, their resulting distributions differ significantly from those in our model, particularly the presence of bimodal peaks in $\chi_{\rm eff}$. 
These differences likely stem from several modelling choices in those works: 
no treatment of binary component exchanges or evolution of angular-momentum directions during binary-single interactions; 
the assumption that the circum-binary disk is always aligned with the AGN disk regardless of the binary's angular momentum; 
a fixed capture rate from the nuclear star cluster; 
a fixed hardening time of binaries due to gaseous torques in the unit of orbital timescale around the SMBH; 
no inclination excitation of kicked objects; 
the assumption that binaries form whenever two BHs come within the Hill radius 
(\citealt{McKernan2022}, \citealt{McKernan2025}, and \citealt{Cook2024}). 
Additionally, other channels involving isolated binary evolution have been proposed to produce similar correlations \citep{Zevin2022,Banerjee2024,Olejak2024,WangZiYuan2025}. 
However, to generate 
high $\chi_{\rm eff}$ values preferentially for massive BHs, as suggested by \citet{LiYinJie2025}, 
the AGN scenario is likely more favorable.

\begin{figure*}
    \centering
    \includegraphics[width=1\linewidth]{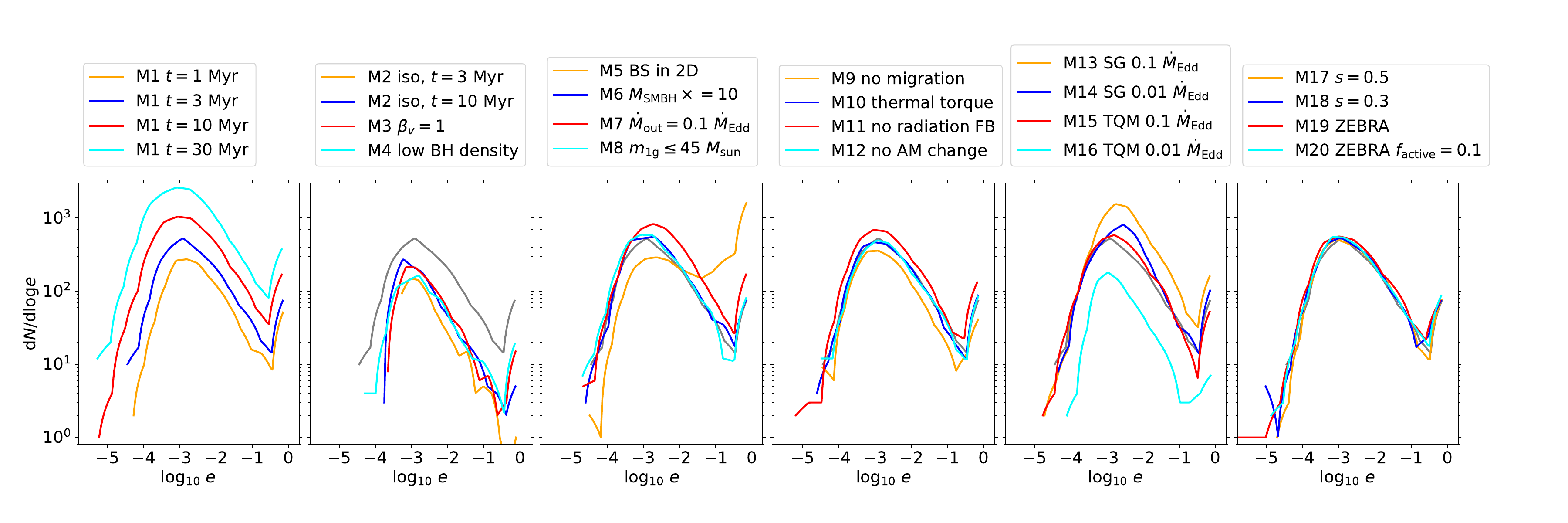}
    \caption{
    The linearly interpolated distribution of the orbital eccentricity of merging binaries for models~M1--M20. 
    }
    \label{fig:e_dist}
\end{figure*}

\subsubsection{Eccentricity}

Fig.~\ref{fig:e_dist} shows the distribution of eccentricities at $10~{\rm Hz}$. 
In the fiducial model, the eccentricity at this frequency is distributed at around $e_{\rm 10Hz}\sim 10^{-4}$--$0.1$ for mergers of binaries formed through gas-capture processes, three-body interactions, and primordial binaries. 
Non-zero eccentricities around $e\sim 10^{-3}$ are characterized by the competition between hardening due to gas dynamical friction and GW emission (with high initial values established during binary-single interactions or binary formation). 
Conversely, the highest eccentricities of $e_{\rm 10Hz}\gtrsim 0.1$
are primarily contributed by highly eccentric binaries formed through GW capture processes \citep{Tagawa20_ecc,Samsing2022,Gondan_Kocsis2022}. 
This process typically occurs during binary-single interactions. 
The minimum value of the eccentricity resulting from the GW capture mechanism ($\sim 0.1$) is determined by the relative velocity of the binary-single interactions \citep[e.g.,][]{Quinlan1989}, which depends on the semi-major axis of the binaries before the interactions. 
As demonstrated in previous studies \citep{Tagawa20_ecc,RomeroShaw2025}, 
the eccentricity distribution is sensitive to the treatment of binary-single interactions (model~M5, indicated by the orange line in the second panel), which significantly influences the merger rate through GW capture processes. 

Notably, highly eccentric mergers are prevalent in these models (model~M5). Approximately half of the mergers exhibit $e_{\rm 10Hz}\gtrsim 0.4$, compared to those with $e_{\rm 10Hz}\gtrsim 0.1$, as discussed in \citet{RomeroShaw2025}. Such highly eccentric mergers may correspond to several events with $e_{\rm 10Hz}\gtrsim 0.4$, as suggested by \citet{XuYumeng2025}. Given the difficulty of producing these highly eccentric mergers from typical interactions in gas-poor environments, they may serve as a promising signature of mergers occurring within AGN disks \citep[but see][]{Rasskazov19,Gondan_Kocsis2022}.

\begin{figure*}
    \centering
    \includegraphics[width=1\linewidth]{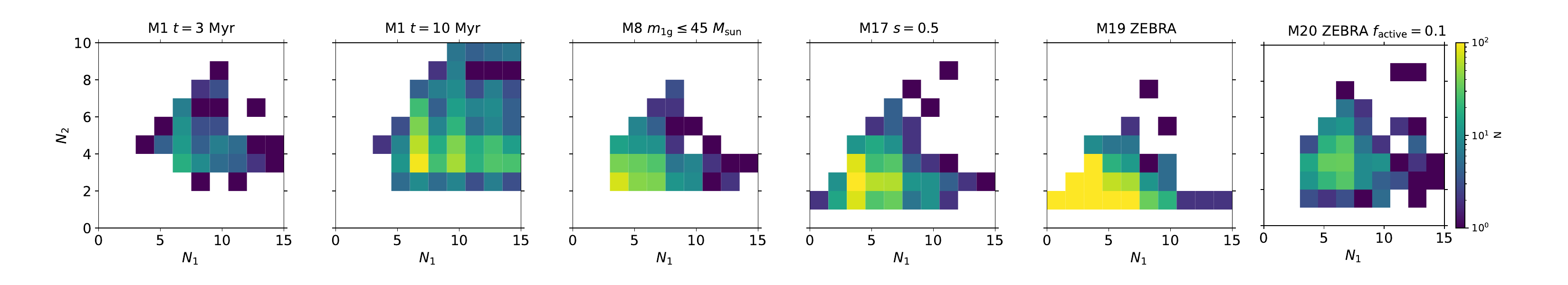}
    \caption{
    The distribution of the number of generations composing the primary and secondary BHs in mergers among massive BHs ($m_1\geq 100~\Msun$ and $m_2\geq 50~\Msun$).     
    To reduce Poisson noise, we performed 100 simulations. 
    }
    \label{fig:m_npns}
\end{figure*}

\begin{figure*}
    \centering
    \includegraphics[width=1\linewidth]{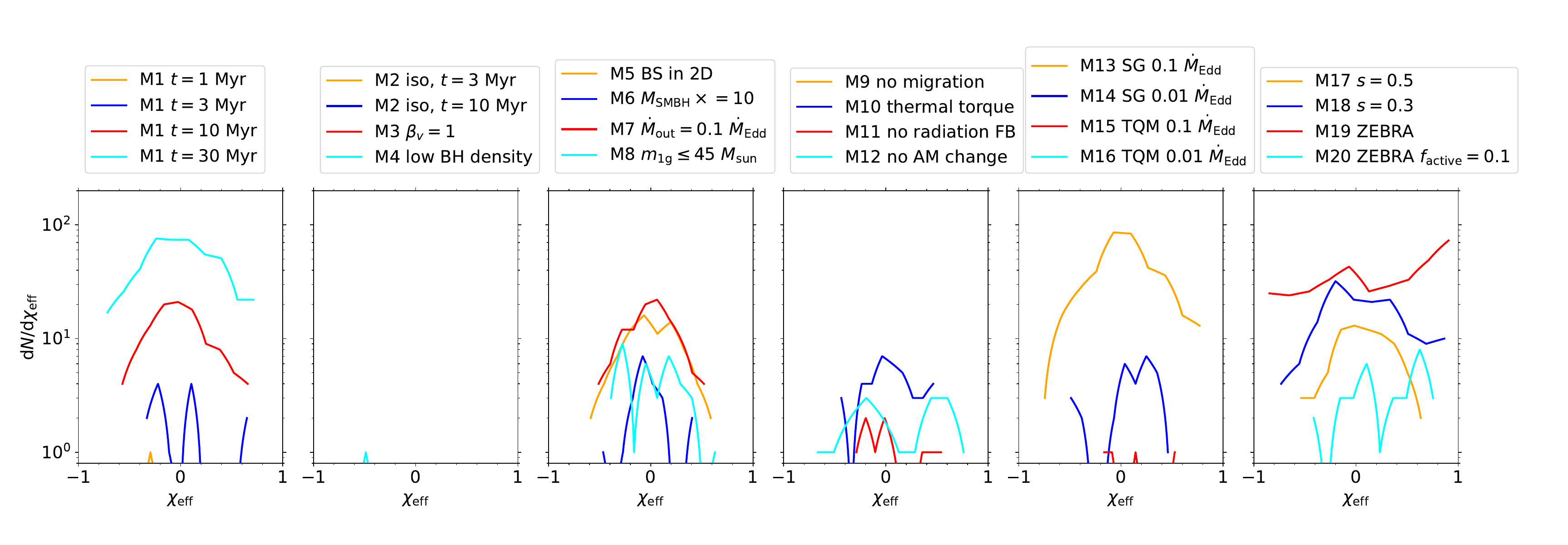}
    \caption{
    The linearly interpolated distribution of $\chi_{\rm eff}$ for mergers among massive BHs. 
    }
    \label{fig:m_xeff}
\end{figure*}

\begin{figure*}
    \centering
    \includegraphics[width=1\linewidth]{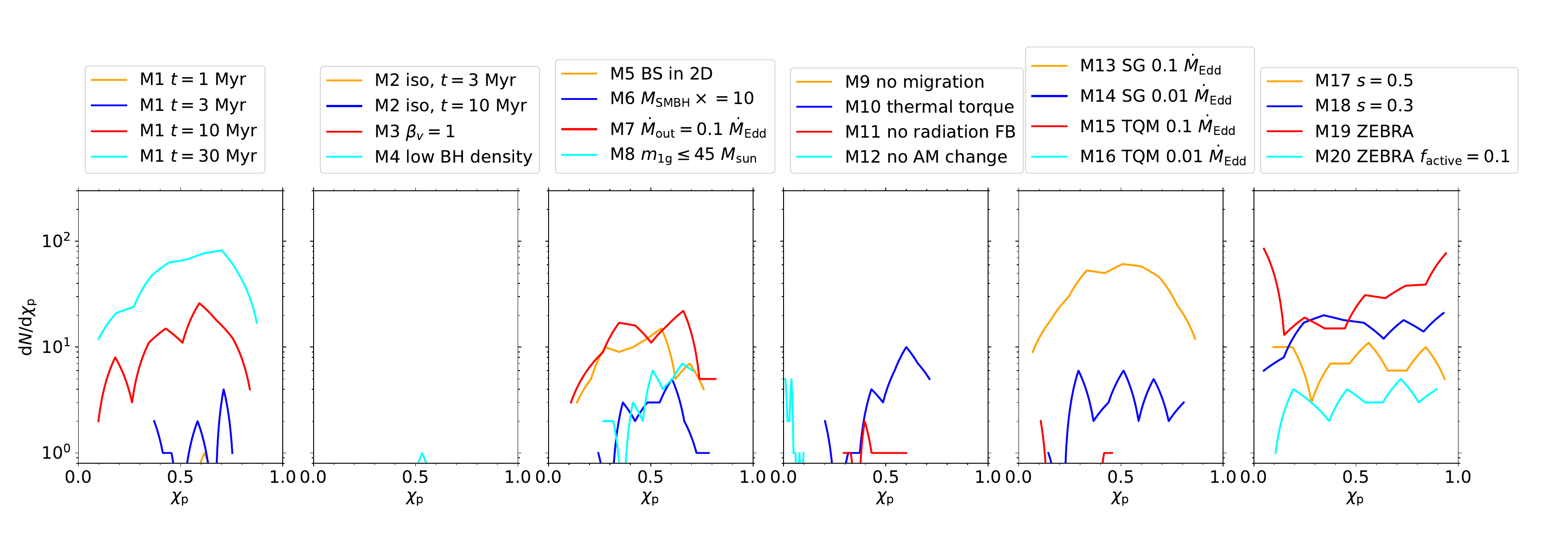}
    \caption{
    The linearly interpolated distribution of $\chi_{\rm p}$ for mergers among massive BHs. 
    }
    \label{fig:m_xp}
\end{figure*}

\begin{figure*}
    \centering
    \includegraphics[width=1\linewidth]{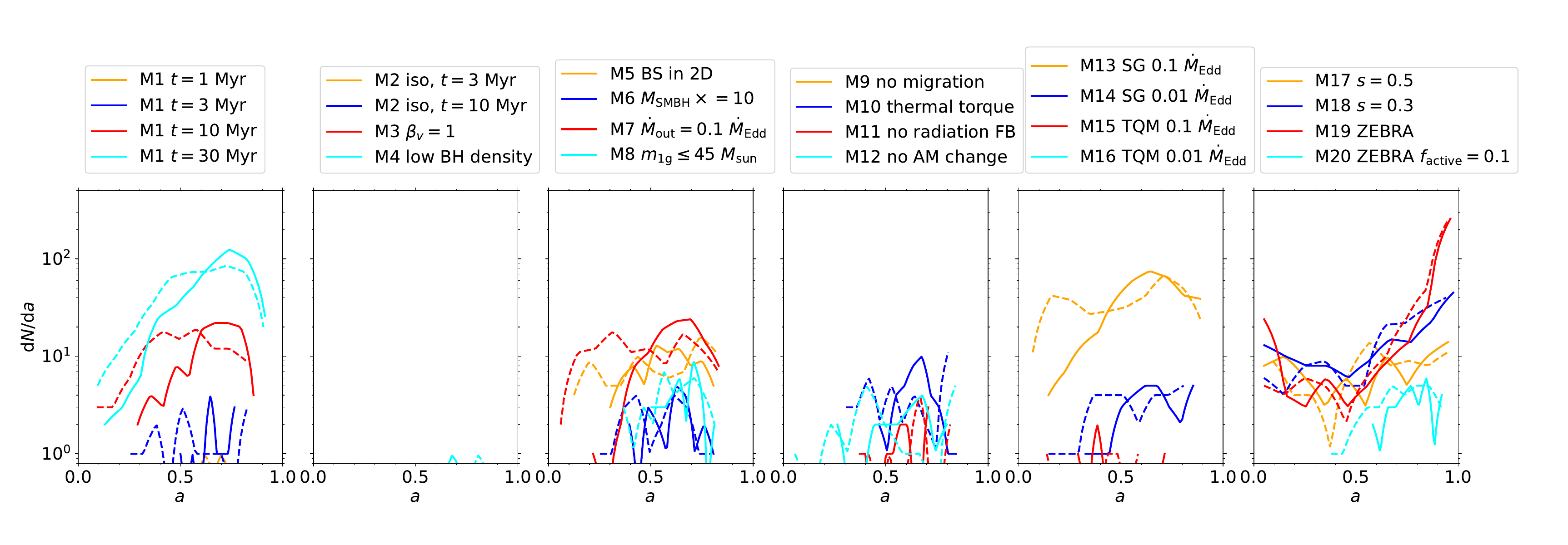}
    \caption{
    The linearly interpolated distribution of the spin magnitude of merging binaries composed of massive BHs.
    Solid and dashed lines represent distributions of the spin magnitudes of the primary and secondary BHs, respectively. 
    }
    \label{fig:m_a_dist}
\end{figure*}

\begin{figure*}
    \centering
    \includegraphics[width=1\linewidth]{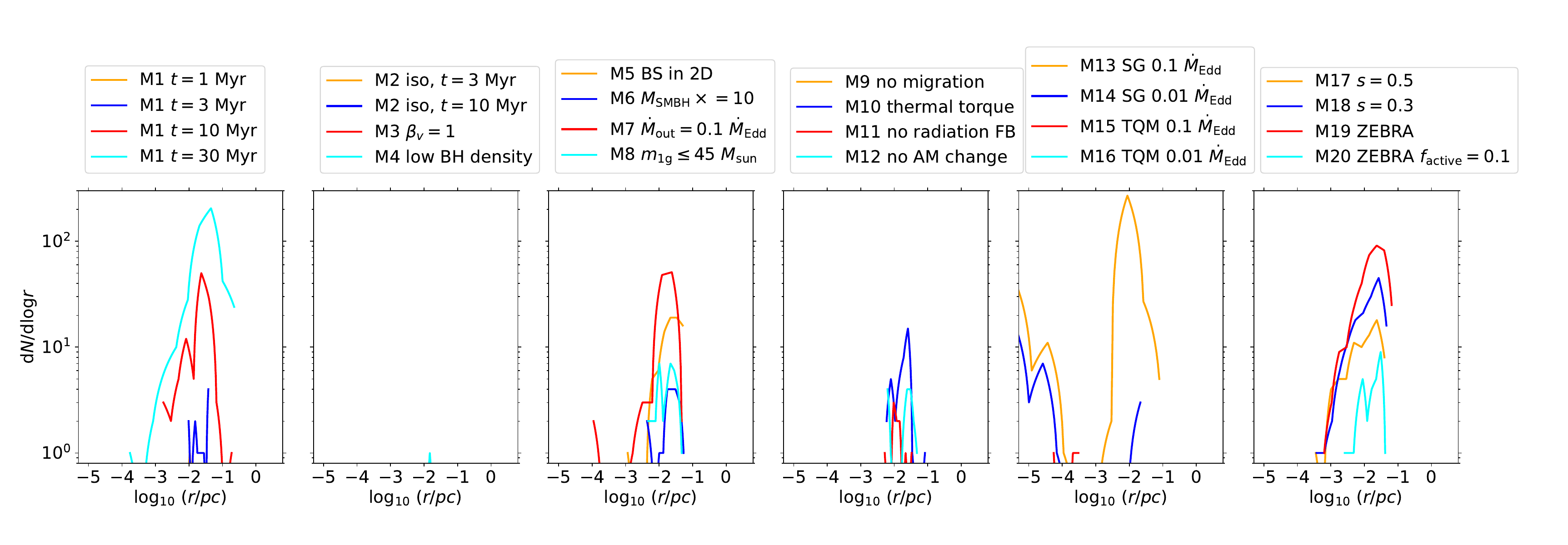}
    \caption{
    The linearly interpolated distribution of the distance from the SMBH of the mergers among massive BHs. 
    }
    \label{fig:m_r_dist}
\end{figure*}

\subsection{Massive mergers}

Reflecting the findings of massive BH mergers, 
such as 
GW190521 and GW231123, 
we investigate whether mergers with similar properties can be produced via the AGN channel. 
For GW190521 \citep{LIGO20_GW190521}, the paramerters are: 
$m_1=85^{+21}_{-14}~\Msun$, $m_2=66^{+17}_{-18}~\Msun$, 
$\chi_{\rm eff}=0.08^{+0.27}_{-0.36}$, 
$\chi_{\rm p}=0.68^{+0.25}_{-0.37}$, 
$a_{\rm 1}=0.69^{+0.27}_{-0.62}$, and 
$a_{\rm 2}=0.73^{+0.24}_{-0.64}$. 
For GW231123 \citep{LIGO2025_GW231123}, the parameters are: 
$m_1=137^{+22}_{-17}~\Msun$, $m_2=103^{+20}_{-52}~\Msun$, 
$\chi_{\rm eff}=0.31^{+0.24}_{-0.39}$, 
$\chi_{\rm p}=0.77^{+0.17}_{-0.19}$, 
$a_{\rm 1}=0.90^{+0.10}_{-0.19}$, and 
$a_{\rm 2}=0.80^{+0.20}_{-0.51}$. 
These massive mergers are characterized by high spin magnitudes and strong spin precession. 
Several studies discussed whether these masses \citep{Paiella2025,Liu2025,Tanikawa2025,LiuBin2025,LiYinJie2025,Goyal2025,Yang2025_lense,Shan2025,Chakraborty2025} and/or spins \citep{Passenger2025,Bartos2025,Croon2025,Gottlieb2025,DeLuca2025,Delfavero2025,Popa2025,Stegmann2025} can be explained by various models.

Focusing on massive BH mergers, 
Figs.~\ref{fig:m_npns}--\ref{fig:m_r_dist} 
show the distributions of properties for mergers involving a primary BH with $m_{\rm 1}\geq 100~{\Msun}$ 
and a secondary BH with $m_{\rm 2}\geq 50~{\Msun}$. 
In Figs.~\ref{fig:m_mpms} and \ref{fig:m_npns}, respectively, 
the merging masses and the generations of high-mass mergers are shown for models~M1 (at $t=3$ and $10~{\rm Myr}$), M8, M17, M19, and M20. 
These models were selected because their merging mass distributions typically differ (e.g. Figs.~\ref{fig:mc1_dist}--\ref{fig:mq_dist}). 
For models~M1 and M20, 
massive mergers occur among $\sim 5-10$ generation BHs, consistent with the model of \citet{LiYinJie2025}. 
In model~M8, where the initial BH mass is higher than in the fiducial model, 
massive mergers primarily involve $\sim 5$ generation BHs. 
Additionally, 
in models~M17 and M19, where accretion is more efficient, 
mergers among low-generation ($\lesssim 5$) BHs contribute significantly to the high-mass mergers, in line with the suggestions of \citet{Bartos2025}. 
On the other hand, in models with short duration(model~M1 at $t=1~{\rm Myr}$),
a lower number of BHs in AGN disks (models~M2--M4), 
no migration (model~M9), 
or low gas density (models~M15 and M16), 
massive mergers like those found in GW231123 are difficult to produce.

Figs.~\ref{fig:m_xeff}, \ref{fig:m_xp}, and \ref{fig:m_a_dist} 
show the distributions of $\chi_{\rm eff}$, $\chi_{\rm p}$, and BH spin magnitudes for massive mergers across models~M1--M20. 
The distribution of $\chi_{\rm eff}$ around $\sim 0$ in many models (Fig.~\ref{fig:m_xeff}) results from the frequent randomization of angular momentum directions due to binary-single interactions.

In models~M17--M19, where accretion is significant, the spin magnitudes of massive BHs are distributed over a wide range of values instead of being concentrated around $a\sim 0.7$ Fig.~\ref{fig:m_a_dist}). 
Moreover, 
$\theta_{\rm a-l}$ tends to decrease in these models, leading to lower $\chi_{\rm p}$ values (Fig.~\ref{fig:m_xp}), although $\theta_{\rm a-l}$ is still distributed over a wide range. 
Consequently, these models can readily produce the high spin magnitude and $\chi_{\rm p}$ suggested for GW231123. 
By contrast, when accretion is inefficient, as prescribed in the fiducial and many other models, 
the probability of producing spins $\gtrsim 0.7$ is lower. 
\citet{Passenger2025} estimate that, 
in globular clusters, where up to 2g-2g mergers can occasionally occur, 
there is only a $\sim 1\%$ chance that merger-remnant spins are consistent with those inferred for GW231123. 
Nevertheless, since higher-generation BHs tend to have larger spins \citep{Borchers2025}, 
BHs with spins consistent with GW231123 could still arise from mergers among higher-generation objects in the fiducial model.

Fig.~\ref{fig:m_r_dist} shows the distribution of the distance from the SMBH at which massive BH mergers occur. 
Most mergers primarily peak around $0.01$--$0.05~{\rm pc}$ from the SMBH, 
with no peak at $\sim 3~{\rm pc}$ unlike for low-mass mergers, consistent with the locations of hierarchical mergers as opposed to mergers of primordial binaries (c.f. Fig.~\ref{fig:r_type}).

Overall, massive BH mergers, such as GW231123, can be produced by 
hierarchical mergers among $\sim 3$--$10$ generation BHs 
or through efficient accretion. 
In the former case, the BH spins are typically distributed around $\sim 0.7$, while 
in the latter case, the BH spins tend to be closer to one.

\section{Conclusions}
\label{sec:conclusions}

We have thoroughly investigated the properties of BH mergers in AGN disks for a broad range of model parameters and prescriptions. To calculate the distribution of observables, we use one-dimensional $N$-body simulations combined with a semi-analytical model. 
Our findings are summarized as follows:

\begin{enumerate}

\item 
In our fiducial model, the distributions of masses and mass ratios are similar to those observed by LVK. However, they depend strongly on the lifetime and density of the AGN disk and on the number and accretion efficiency of BHs, with higher masses predicted as these quantities increase.

\item 
The negative correlation between $q$ and the average of $\chi_{\rm eff}$, as suggested by observed GW events, can arise in a model with efficient gas accretion 
and hierarchical mergers. 
In such a model, higher generation BHs produce a high spin magnitude and low $q$ (if the generations of merging BHs are different). Additionally, accretion reduces the angle between BH spins and the angular momentum direction of the binaries, resulting in mergers characterized by high $\chi_{\rm eff}$. 
On the other hand, the negative correlation between $q$ and the dispersion of $\chi_{\rm eff}$, which is estimated to be more significant, can arise in most models due to hierarchical mergers, provided that gas accretion is not overly efficient. 

\item 
The positive correlation between $\chi_{\rm eff}$ and $M_{\rm chirp}$ reported in recent GW analyses can be reproduced if accretion onto BHs is efficient. Accordingly, the model with moderately efficient accretion (model~M20) appears to be overall the most promising for explaining the observations, whereas the model with highly efficient accretion (model~M19) is unlikely, 
as it overproduces the high $M_{\rm chirp}$ tail. 

\item 
The positive correlation between the effective spin magnitude $|\chi_{\rm eff}|$ 
(instead of $\chi_{\rm eff}$ above) 
and $M_{\rm chirp}$ can be reproduced by hierarchical mergers, which can help constrain the properties of first-generation BHs, accretion processes, and the interactions between BHs. 

\item 
In the fiducial model, mergers frequently occur in the outer regions ($\gtrsim {\rm pc}$) due to the abundance of pre-existing binaries in these regions, as well as in the inner regions around $\sim 0.01$--$0.05~{\rm pc}$ from the SMBH where capture into an AGN disk and gas-capture binary formation are efficient. 
The merger location and binary formation processes are further influenced by the AGN disk models utilized. 

\item 
The eccentricity distribution at merger is predominantly influenced by the geometry of binary-single interactions 
(either isotropic or two-dimensional co-planar with the AGN disk), 
as the probability of the GW capture process is sensitive to this geometry.

\item 
Massive mergers as observed in GW231123 can be produced either through 
hierarchical mergers among $\geq 3$ generation BHs 
or via low-generation mergers with efficient gas accretion. 
In the former case, the BH spin is typically distributed around $\sim 0.7$, while in the latter case, it is closer to $\sim 1$.

\end{enumerate}

Based on all of the above, we tentatively regard model~M20 to be the most promising model overall. 
This model effectively reproduces 
the negative correlation between $q$ and the average of $\chi_{\rm eff}$, as well as 
a positive correlation between $m_{\rm 1}$ and the average of $\chi_{\rm eff}$. 
Both correlations highlight the necessity for efficient accretion and hierarchical mergers. 
However, a potential concern may arise regarding the average of $|\chi_{\rm eff}|$ at high $M_{\rm chirp}$, which seems to exceed the observational median. If this constraint tightens to align more closely with the current median values, a model with less efficient accretion, such as model~M1, would be favored to prevent excessive spin growth and alignment due to accretion. 
Given the parameter dependencies demonstrated in this work, 
further observations and more dedicated analyses (e.g., \citealt{Xue2025} and \citealt{Gayathri2025} for $N$ parameter cases) will be crucial for deriving a formal best-fit model in future studies.

Further modeling that includes various components and their interactions, 
the evolution of AGN disks, and a realistic treatment of processes using $N$-body and hydrodynamical simulations would be valuable for improving predictions. 
Additionally, the identification of possible electromagnetic counterparts would enhance our understanding of their evolution.

\section*{Acknowledgments}

H.T. is supported by the National Science and Technology Major Project of China (No. 2024ZD1100601) and 
the National Key R$\&$D Program of China (grant No.2024YFC2207700). 
Z.H. was supported by NASA grants 80NSSC22K0822 and 80NSSC24K0440. 
B.K. is supported by the Science and Technology Facilities Council Grant Number ST/W000903/1. 
Simulations were carried out on Cray XD2000 at the Center for Computational Astrophysics, National Astronomical Observatory of Japan.

\appendix

\bibliographystyle{aasjournal}
\bibliography{agn_bhm}

@ARTICLE{Ghez2008,
       author = {{Ghez}, A.~M. and {Salim}, S. and {Weinberg}, N.~N. and {Lu}, J.~R. and {Do}, T. and {Dunn}, J.~K. and {Matthews}, K. and {Morris}, M.~R. and {Yelda}, S. and {Becklin}, E.~E. and {Kremenek}, T. and {Milosavljevic}, M. and {Naiman}, J.},
        title = "{Measuring Distance and Properties of the Milky Way's Central Supermassive Black Hole with Stellar Orbits}",
      journal = {\apj},
         year = 2008,
        month = dec,
       volume = {689},
       number = {2},
        pages = {1044-1062},
          doi = {10.1086/592738},
archivePrefix = {arXiv},
       eprint = {0808.2870},
 primaryClass = {astro-ph},
       adsurl = {https://ui.adsabs.harvard.edu/abs/2008ApJ...689.1044G}
}

@ARTICLE{Genzel2010,
       author = {{Genzel}, Reinhard and {Eisenhauer}, Frank and {Gillessen}, Stefan},
        title = "{The Galactic Center massive black hole and nuclear star cluster}",
      journal = {Reviews of Modern Physics},
         year = 2010,
        month = oct,
       volume = {82},
       number = {4},
        pages = {3121-3195},
          doi = {10.1103/RevModPhys.82.3121},
archivePrefix = {arXiv},
       eprint = {1006.0064},
 primaryClass = {astro-ph.GA},
       adsurl = {https://ui.adsabs.harvard.edu/abs/2010RvMP...82.3121G}
}

@ARTICLE{Stephan2016,
       author = {{Stephan}, Alexander P. and {Naoz}, Smadar and {Ghez}, Andrea M. and {Witzel}, Gunther and {Sitarski}, Breann N. and {Do}, Tuan and {Kocsis}, Bence},
        title = "{Merging binaries in the Galactic Center: the eccentric Kozai-Lidov mechanism with stellar evolution}",
      journal = {\mnras},
         year = 2016,
        month = aug,
       volume = {460},
       number = {4},
        pages = {3494-3504},
          doi = {10.1093/mnras/stw1220},
archivePrefix = {arXiv},
       eprint = {1603.02709},
 primaryClass = {astro-ph.SR},
       adsurl = {https://ui.adsabs.harvard.edu/abs/2016MNRAS.460.3494S}
}

@ARTICLE{Lodato2013,
       author = {{Lodato}, G. and {Gerosa}, D.},
        title = "{Black hole mergers: do gas discs lead to spin alignment?}",
      journal = {\mnras},
         year = 2013,
        month = feb,
       volume = {429},
        pages = {L30-L34},
          doi = {10.1093/mnrasl/sls018},
archivePrefix = {arXiv},
       eprint = {1211.0284},
 primaryClass = {astro-ph.CO},
       adsurl = {https://ui.adsabs.harvard.edu/abs/2013MNRAS.429L..30L}
}

@ARTICLE{Kirouglu2025b,
       author = {{K{\i}ro{\u{g}}lu}, Fulya and {Kremer}, Kyle and {Biscoveanu}, Sylvia and {Gonz{\'a}lez Prieto}, Elena and {Rasio}, Frederic A.},
        title = "{Black Hole Accretion and Spin-up through Stellar Collisions in Dense Star Clusters}",
      journal = {\apj},
         year = 2025,
        month = feb,
       volume = {979},
       number = {2},
          eid = {237},
        pages = {237},
          doi = {10.3847/1538-4357/ada26b},
archivePrefix = {arXiv},
       eprint = {2410.01879},
 primaryClass = {astro-ph.HE},
       adsurl = {https://ui.adsabs.harvard.edu/abs/2025ApJ...979..237K}
}

@ARTICLE{Kirouglu2025a,
       author = {{K{\i}ro{\u{g}}lu}, Fulya and {Lombardi}, James C. and {Kremer}, Kyle and {Vanderzyden}, Hans D. and {Rasio}, Frederic A.},
        title = "{Spin─Orbit Alignment in Merging Binary Black Holes Following Collisions with Massive Stars}",
      journal = {\apjl},
         year = 2025,
        month = apr,
       volume = {983},
       number = {1},
          eid = {L9},
        pages = {L9},
          doi = {10.3847/2041-8213/adc263},
archivePrefix = {arXiv},
       eprint = {2501.09068},
 primaryClass = {astro-ph.HE},
       adsurl = {https://ui.adsabs.harvard.edu/abs/2025ApJ...983L...9K}
}

@ARTICLE{Ginat2023,
       author = {{Ginat}, Yonadav Barry and {Perets}, Hagai B.},
        title = "{Analytic modelling of binary-single encounters: non-thermal eccentricity distribution and gravitational-wave source formation}",
      journal = {\mnras},
         year = 2023,
        month = feb,
       volume = {519},
       number = {1},
        pages = {L15-L20},
          doi = {10.1093/mnrasl/slac145},
archivePrefix = {arXiv},
       eprint = {2205.15957},
 primaryClass = {astro-ph.HE},
       adsurl = {https://ui.adsabs.harvard.edu/abs/2023MNRAS.519L..15G}
}

@ARTICLE{Ginat2021,
       author = {{Ginat}, Yonadav Barry and {Perets}, Hagai B.},
        title = "{Analytical, Statistical Approximate Solution of Dissipative and Nondissipative Binary-Single Stellar Encounters}",
      journal = {Physical Review X},
         year = 2021,
        month = jul,
       volume = {11},
       number = {3},
          eid = {031020},
        pages = {031020},
          doi = {10.1103/PhysRevX.11.031020},
archivePrefix = {arXiv},
       eprint = {2011.00010},
 primaryClass = {astro-ph.GA},
       adsurl = {https://ui.adsabs.harvard.edu/abs/2021PhRvX..11c1020G}
}

@ARTICLE{EpsteinMartin2025b,
       author = {{Epstein-Martin}, Marguerite and {Stone}, Nicholas and {Becker}, Juliette},
        title = "{Mean Motion Resonances in AGN Disks}",
      journal = {arXiv e-prints},
         year = 2025,
        month = oct,
          eid = {arXiv:2510.12895},
        pages = {arXiv:2510.12895},
          doi = {10.48550/arXiv.2510.12895},
archivePrefix = {arXiv},
       eprint = {2510.12895},
 primaryClass = {astro-ph.HE},
       adsurl = {https://ui.adsabs.harvard.edu/abs/2025arXiv251012895E}
}

@ARTICLE{Moncrieff2026,
       author = {{Moncrieff}, Jordan W.~N. and {Grishin}, Evgeni and {Trani}, Alessandro A. and {Panther}, Fiona H. and {Pietrosanti}, Olga},
        title = "{Not all roads lead to merger: AGN disc properties influence the interactions of highly unequal mass black holes}",
      journal = {\mnras},
         year = 2026,
        month = jan,
       volume = {545},
       number = {3},
          eid = {staf2217},
        pages = {staf2217},
          doi = {10.1093/mnras/staf2217},
archivePrefix = {arXiv},
       eprint = {2511.09129},
 primaryClass = {astro-ph.HE},
       adsurl = {https://ui.adsabs.harvard.edu/abs/2026MNRAS.545f2217M}
}

@ARTICLE{Kaaz2023,
       author = {{Kaaz}, Nicholas and {Schr{\o}der}, Sophie Lund and {Andrews}, Jeff J. and {Antoni}, Andrea and {Ramirez-Ruiz}, Enrico},
        title = "{The Hydrodynamic Evolution of Binary Black Holes Embedded within the Vertically Stratified Disks of Active Galactic Nuclei}",
      journal = {\apj},
         year = 2023,
        month = feb,
       volume = {944},
       number = {1},
          eid = {44},
        pages = {44},
          doi = {10.3847/1538-4357/aca967},
archivePrefix = {arXiv},
       eprint = {2103.12088},
 primaryClass = {astro-ph.HE},
       adsurl = {https://ui.adsabs.harvard.edu/abs/2023ApJ...944...44K}
}

@ARTICLE{Woosley2021,
       author = {{Woosley}, S.~E. and {Heger}, Alexander},
        title = "{The Pair-instability Mass Gap for Black Holes}",
      journal = {\apjl},
         year = 2021,
        month = may,
       volume = {912},
       number = {2},
          eid = {L31},
        pages = {L31},
          doi = {10.3847/2041-8213/abf2c4},
archivePrefix = {arXiv},
       eprint = {2103.07933},
 primaryClass = {astro-ph.SR},
       adsurl = {https://ui.adsabs.harvard.edu/abs/2021ApJ...912L..31W}
}

@ARTICLE{Heggie1975,
       author = {{Heggie}, D.~C.},
        title = "{Binary evolution in stellar dynamics.}",
      journal = {\mnras},
         year = 1975,
        month = dec,
       volume = {173},
        pages = {729-787},
          doi = {10.1093/mnras/173.3.729},
       adsurl = {https://ui.adsabs.harvard.edu/abs/1975MNRAS.173..729H}
}

@ARTICLE{Su2025,
       author = {{Su}, Yubo and {Rowan}, Connar and {Rozner}, Mor},
        title = "{Gas meets Kozai: the influence of a gas-rich accretion disc on hierarchical triples undergoing von Zeipel─Lidov─Kozai oscillations}",
      journal = {\mnras},
         year = 2025,
        month = oct,
       volume = {543},
       number = {2},
        pages = {1864-1877},
          doi = {10.1093/mnras/staf1592},
archivePrefix = {arXiv},
       eprint = {2505.23889},
 primaryClass = {astro-ph.GA},
       adsurl = {https://ui.adsabs.harvard.edu/abs/2025MNRAS.543.1864S}
}

@ARTICLE{Rozner2024,
       author = {{Rozner}, Mor and {Perets}, Hagai B.},
        title = "{Soft No More: Gas Shielding Protects Soft Binaries from Disruption in Gas-rich Environments}",
      journal = {\apj},
         year = 2024,
        month = jun,
       volume = {968},
       number = {2},
          eid = {80},
        pages = {80},
          doi = {10.3847/1538-4357/ad4bdd},
archivePrefix = {arXiv},
       eprint = {2404.01384},
 primaryClass = {astro-ph.HE},
       adsurl = {https://ui.adsabs.harvard.edu/abs/2024ApJ...968...80R}
}

@ARTICLE{Ishibashi2024,
       author = {{Ishibashi}, W. and {Gr{\"o}bner}, M.},
        title = "{Gravitational wave mergers of accreting binary black holes in AGN discs}",
      journal = {\mnras},
         year = 2024,
        month = apr,
       volume = {529},
       number = {2},
        pages = {883-892},
          doi = {10.1093/mnras/stae569},
archivePrefix = {arXiv},
       eprint = {2412.01925},
 primaryClass = {astro-ph.HE},
       adsurl = {https://ui.adsabs.harvard.edu/abs/2024MNRAS.529..883I}
}

@ARTICLE{Grobner2020,
       author = {{Gr{\"o}bner}, M. and {Ishibashi}, W. and {Tiwari}, S. and {Haney}, M. and {Jetzer}, P.},
        title = "{Binary black hole mergers in AGN accretion discs: gravitational wave rate density estimates}",
      journal = {\aap},
         year = 2020,
        month = jun,
       volume = {638},
          eid = {A119},
        pages = {A119},
          doi = {10.1051/0004-6361/202037681},
archivePrefix = {arXiv},
       eprint = {2005.03571},
 primaryClass = {astro-ph.GA},
       adsurl = {https://ui.adsabs.harvard.edu/abs/2020A&A...638A.119G}
}

@ARTICLE{Roupas2025,
       author = {{Roupas}, Zacharias},
        title = "{Black hole mass function shift in proto-stellar clusters driven by gas accretion}",
      journal = {\aap},
         year = 2025,
        month = oct,
       volume = {702},
          eid = {A208},
        pages = {A208},
          doi = {10.1051/0004-6361/202556434},
archivePrefix = {arXiv},
       eprint = {2509.08448},
 primaryClass = {astro-ph.GA},
       adsurl = {https://ui.adsabs.harvard.edu/abs/2025A&A...702A.208R}
}

@ARTICLE{Roupas2026,
       author = {{Roupas}, Zacharias},
        title = "{Spin-up and spin distribution of stellar black holes grown by gas accretion in proto-stellar clusters}",
      journal = {\aap},
         year = 2026,
        month = apr,
       volume = {709},
          eid = {A5},
        pages = {A5},
          doi = {10.1051/0004-6361/202558435},
archivePrefix = {arXiv},
       eprint = {2603.18857},
 primaryClass = {astro-ph.GA},
       adsurl = {https://ui.adsabs.harvard.edu/abs/2026A&A...709A...5R}
}

@ARTICLE{Tagawa2026_EM,
       author = {{Tagawa}, Hiromichi and {Haiman}, Zolt{\'a}n and {Kimura}, Shigeo S. and {Yesuf}, Hassen M. and {Guo}, Hengxiao},
        title = "{Electromagnetic Flares from Compact-Object Mergers in AGN Disks: Signatures and Predictions}",
      journal = {arXiv e-prints},
         year = 2026,
        month = apr,
          eid = {arXiv:2604.05020},
        pages = {arXiv:2604.05020},
          doi = {10.48550/arXiv.2604.05020},
archivePrefix = {arXiv},
       eprint = {2604.05020},
 primaryClass = {astro-ph.HE},
       adsurl = {https://ui.adsabs.harvard.edu/abs/2026arXiv260405020T}
}

@ARTICLE{Gupte2026,
       author = {{Gupte}, Nihar and {Miller}, M. Coleman and {Udall}, Rhiannon and {Bini}, Sophie and {Buonanno}, Alessandra and {Gair}, Jonathan and {Gamboa}, Aldo and {Pompili}, Lorenzo and {Ramos-Buades}, Antoni and {Dax}, Maximilian and {Green}, Stephen R. and {Kofler}, Annalena and {Macke}, Jakob and {Sch{\"o}lkopf}, Bernhard},
        title = "{Eccentricity constraints disfavor single-single capture in nuclear star clusters as the origin of all LIGO-Virgo-KAGRA binary black holes}",
      journal = {arXiv e-prints},
         year = 2026,
        month = mar,
          eid = {arXiv:2603.29019},
        pages = {arXiv:2603.29019},
archivePrefix = {arXiv},
       eprint = {2603.29019},
 primaryClass = {astro-ph.HE},
       adsurl = {https://ui.adsabs.harvard.edu/abs/2026arXiv260329019G}
}

@ARTICLE{Gayathri2025,
       author = {{Gayathri}, V. and {Iorio}, Giuliano and {Tagawa}, Hiromichi and {Wysocki}, Daniel and {Anglin}, Jeremiah and {Bartos}, Imre and {Bhaumik}, Shubhagata and {Haiman}, Zolt'an and {Mapelli}, Michela and {O'Shaughnessy}, R. and {Xue}, LingQin},
        title = "{Reconstructing the origin of black hole mergers using sparse astrophysical models}",
      journal = {arXiv e-prints},
         year = 2025,
        month = sep,
          eid = {arXiv:2509.09647},
        pages = {arXiv:2509.09647},
          doi = {10.48550/arXiv.2509.09647},
archivePrefix = {arXiv},
       eprint = {2509.09647},
 primaryClass = {astro-ph.HE},
       adsurl = {https://ui.adsabs.harvard.edu/abs/2025arXiv250909647G}
}

@ARTICLE{Gayathri+2022,
       author = {{Gayathri}, V. and {Healy}, J. and {Lange}, J. and {O'Brien}, B. and {Szczepa{\'n}czyk}, M. and {Bartos}, Imre and {Campanelli}, M. and {Klimenko}, S. and {Lousto}, C.~O. and {O'Shaughnessy}, R.},
        title = "{Eccentricity estimate for black hole mergers with numerical relativity simulations}",
      journal = {Nature Astronomy},
         year = 2022,
        month = jan,
       volume = {6},
        pages = {344-349},
          doi = {10.1038/s41550-021-01568-w},
       adsurl = {https://ui.adsabs.harvard.edu/abs/2022NatAs...6..344G}
}

@ARTICLE{Williamson2017,
       author = {{Williamson}, A.~R. and {Lange}, J. and {O'Shaughnessy}, R. and {Clark}, J.~A. and {Kumar}, Prayush and {Calder{\'o}n Bustillo}, J. and {Veitch}, J.},
        title = "{Systematic challenges for future gravitational wave measurements of precessing binary black holes}",
      journal = {\prd},
         year = 2017,
        month = dec,
       volume = {96},
       number = {12},
          eid = {124041},
        pages = {124041},
          doi = {10.1103/PhysRevD.96.124041},
archivePrefix = {arXiv},
       eprint = {1709.03095},
 primaryClass = {gr-qc},
       adsurl = {https://ui.adsabs.harvard.edu/abs/2017PhRvD..96l4041W}
}

@ARTICLE{Collin2008,
       author = {{Collin}, S. and {Zahn}, J.-P.},
        title = "{Star formation in accretion discs: from the Galactic center to active galactic nuclei}",
      journal = {\aap},
         year = 2008,
        month = jan,
       volume = {477},
       number = {2},
        pages = {419-435},
          doi = {10.1051/0004-6361:20078191},
archivePrefix = {arXiv},
       eprint = {0709.3772},
 primaryClass = {astro-ph},
       adsurl = {https://ui.adsabs.harvard.edu/abs/2008A&A...477..419C}
}

@ARTICLE{Hopkins2024,
       author = {{Hopkins}, Philip F. and {Grudic}, Michael Y. and {Su}, Kung-Yi and {Wellons}, Sarah and {Angles-Alcazar}, Daniel and {Steinwandel}, Ulrich P. and {Guszejnov}, David and {Murray}, Norman and {Faucher-Giguere}, Claude-Andre and {Quataert}, Eliot and {Keres}, Dusan},
        title = "{FORGE'd in FIRE: Resolving the End of Star Formation and Structure of AGN Accretion Disks from Cosmological Initial Conditions}",
      journal = {The Open Journal of Astrophysics},
         year = 2024,
        month = mar,
       volume = {7},
          eid = {18},
        pages = {18},
          doi = {10.21105/astro.2309.13115},
archivePrefix = {arXiv},
       eprint = {2309.13115},
 primaryClass = {astro-ph.GA},
       adsurl = {https://ui.adsabs.harvard.edu/abs/2024OJAp....7E..18H}
}

@ARTICLE{Tomar2026,
       author = {{Tomar}, Yashvardhan and {Hopkins}, Philip F. and {Kremer}, Kyle},
        title = "{Thick Disks, Thin Hopes: Suppressed Capture and Merger Rates in AGN}",
      journal = {arXiv e-prints},
         year = 2026,
        month = jan,
          eid = {arXiv:2601.02487},
        pages = {arXiv:2601.02487},
          doi = {10.48550/arXiv.2601.02487},
archivePrefix = {arXiv},
       eprint = {2601.02487},
 primaryClass = {astro-ph.HE},
       adsurl = {https://ui.adsabs.harvard.edu/abs/2026arXiv260102487T}
}

@ARTICLE{Hopkins2026,
       author = {{Hopkins}, Philip F. and {Baron}, Dalya and {Piotrowska}, Joanna M.},
        title = "{Masers and Broad-Line Mapping Favor Magnetically-Dominated AGN Accretion Disks}",
      journal = {arXiv e-prints},
         year = 2026,
        month = jan,
          eid = {arXiv:2601.06253},
        pages = {arXiv:2601.06253},
          doi = {10.48550/arXiv.2601.06253},
archivePrefix = {arXiv},
       eprint = {2601.06253},
 primaryClass = {astro-ph.HE},
       adsurl = {https://ui.adsabs.harvard.edu/abs/2026arXiv260106253H}
}

@ARTICLE{Rowan2025_Inclination,
       author = {{Rowan}, Connar and {Whitehead}, Henry and {Fabj}, Gaia and {Kirkeberg}, Philip and {Pessah}, Martin E. and {Kocsis}, Bence},
        title = "{Hydrodynamic simulations of black hole evolution in AGN discs ─ I. Orbital alignment of highly inclined satellites}",
      journal = {\mnras},
         year = 2025,
        month = oct,
       volume = {543},
       number = {1},
        pages = {132-145},
          doi = {10.1093/mnras/staf1449},
archivePrefix = {arXiv},
       eprint = {2505.23739},
 primaryClass = {astro-ph.HE},
       adsurl = {https://ui.adsabs.harvard.edu/abs/2025MNRAS.543..132R}
}

@ARTICLE{Borchers2025,
       author = {{Borchers}, Angela and {Ye}, Claire S. and {Fishbach}, Maya},
        title = "{Gravitational-wave Kicks Impact the Spins of Black Holes from Hierarchical Mergers}",
      journal = {\apj},
         year = 2025,
        month = jul,
       volume = {987},
       number = {2},
          eid = {146},
        pages = {146},
          doi = {10.3847/1538-4357/addec6},
archivePrefix = {arXiv},
       eprint = {2503.21278},
 primaryClass = {astro-ph.HE},
       adsurl = {https://ui.adsabs.harvard.edu/abs/2025ApJ...987..146B}
}

@ARTICLE{Ye2026,
       author = {{Ye}, Claire S. and {Fishbach}, Maya and {Kremer}, Kyle and {Reina-Campos}, Marta},
        title = "{Mass Distribution of Binary Black Hole Mergers from Young and Old Dense Star Clusters}",
      journal = {\apj},
         year = 2026,
        month = feb,
       volume = {997},
       number = {2},
          eid = {267},
        pages = {267},
          doi = {10.3847/1538-4357/ae317f},
archivePrefix = {arXiv},
       eprint = {2507.07183},
 primaryClass = {astro-ph.HE},
       adsurl = {https://ui.adsabs.harvard.edu/abs/2026ApJ...997..267Y}
}

@ARTICLE{LVK26_GW241110,
       author = {{Abac}, A.~G. and {Abouelfettouh}, I. and {Acernese}, F. and {Ackley}, K. and {Adamcewicz}, C. and {Adhicary}, S. and {Adhikari}, D. and {Adhikari}, N. and {Adhikari}, R.~X. and {Adkins}, V.~K. and {Afroz}, S. and {Agapito}, A. and {Agarwal}, D. and {Agathos}, M. and {Aggarwal}, N. and {Aggarwal}, S. and {Aguiar}, O.~D. and {Ahrend}, I.-L. and {Aiello}, L. and {Ain}, A. and {Ajith}, P. and {Akutsu}, T. and {Albanesi}, S. and {Ali}, W. and {Al-Kershi}, S. and {All{\'e}n{\'e}}, C. and {Allocca}, A. and {Al-Shammari}, S. and {Altin}, P.~A. and {Alvarez-Lopez}, S. and {Amar}, W. and {Amarasinghe}, O. and {Amato}, A. and {Amicucci}, F. and {Amra}, C. and {Ananyeva}, A. and {Anderson}, S.~B. and {Anderson}, W.~G. and {Andia}, M. and {Ando}, M. and {Andr{\'e}s-Carcasona}, M. and {Andri{\'c}}, T. and {Anglin}, J. and {Ansoldi}, S. and {Antelis}, J.~M. and {Antier}, S. and {Antonini}, F. and {Aoumi}, M. and {Appavuravther}, E.~Z. and {Appert}, S. and {Apple}, S.~K. and {Arai}, K. and {Ara{\'u}jo-{\'A}lvarez}, C. and {Araya}, A. and {Araya}, M.~C. and {Arca Sedda}, M. and {Areeda}, J.~S. and {Aritomi}, N. and {Armato}, F. and {Armstrong}, S. and {Arnaud}, N. and {Arogeti}, M. and {Aronson}, S.~M. and {Arun}, K.~G. and {Ashton}, G. and {Aso}, Y. and {Asprea}, L. and {Assiduo}, M. and {Assis de Souza Melo}, S. and {Aston}, S.~M. and {Astone}, P. and {Aswathi}, P.~S. and {Attadio}, F. and {Aubin}, F. and {Aultoneal}, K. and {Avallone}, G. and {Avila}, E.~A. and {Babak}, S. and {Badger}, C. and {Bae}, S. and {Bagnasco}, S. and {Baiotti}, L. and {Bajpai}, R. and {Baka}, T. and {Baker}, A.~M. and {Baker}, K.~A. and {Baker}, T. and {Baldi}, G. and {Baldicchi}, N. and {Ball}, M. and {Ballardin}, G. and {Ballmer}, S.~W. and {Banagiri}, S. and {Banerjee}, B. and {Bankar}, D. and {Baptiste}, T.~M. and {Baral}, P. and {Baratti}, M. and {Barayoga}, J.~C. and {Barish}, B.~C. and {Barker}, D. and {Barman}, N. and {Barneo}, P. and {Barone}, F. and {Barr}, B. and {Barsotti}, L. and {Barsuglia}, M. and {Barta}, D. and {Bartoletti}, A.~M. and {Barton}, M.~A. and {Bartos}, I. and {Basalaev}, A. and {Bassiri}, R. and {Basti}, A. and {Bawaj}, M. and {Baxi}, P. and {Bayley}, J.~C. and {Baylor}, A.~C. and {Baynard}, II, P.~A. and {Bazzan}, M. and {Bedakihale}, V.~M. and {Beirnaert}, F. and {Bejger}, M. and {Belardinelli}, D. and {Bell}, A.~S. and {Bellie}, D.~S. and {Bellizzi}, L. and {Benoit}, W. and {Bentara}, I. and {Bentley}, J.~D. and {Ben Yaala}, M. and {Bera}, S. and {Bergamin}, F. and {Berger}, B.~K. and {Bernuzzi}, S. and {Beroiz}, M. and {Berry}, C.~P.~L. and {Bersanetti}, D. and {Bertheas}, T. and {Bertolini}, A. and {Betzwieser}, J. and {Beveridge}, D. and {Bevilacqua}, G. and {Bevins}, N. and {Bhandare}, R. and {Bhatt}, R. and {Bhattacharjee}, D. and {Bhattacharyya}, S. and {Bhaumik}, S. and {Biancalana}, V. and {Bianchi}, A. and {Bilenko}, I.~A. and {Billingsley}, G. and {Binetti}, A. and {Bini}, S. and {Binu}, C. and {Biot}, S. and {Birnholtz}, O. and {Biscoveanu}, S. and {Bisht}, A. and {Bitossi}, M. and {Bizouard}, M.-A. and {Blaber}, S. and {Blackburn}, J.~K. and {Blagg}, L.~A. and {Blair}, C.~D. and {Blair}, D.~G. and {Bode}, N. and {Boettner}, N. and {Boileau}, G. and {Boldrini}, M. and {Bolingbroke}, G.~N. and {Bolliand}, A. and {Bonavena}, L.~D. and {Bondarescu}, R. and {Bondu}, F. and {Bonilla}, E. and {Bonilla}, M.~S. and {Bonino}, A. and {Bonnand}, R. and {Borchers}, A. and {Borhanian}, S. and {Boschi}, V. and {Bose}, S. and {Bossilkov}, V. and {Bothra}, Y. and {Boudon}, A. and {Bourg}, L. and {Boyle}, M. and {Bozzi}, A. and {Bradaschia}, C. and {Brady}, P.~R. and {Branch}, A. and {Branchesi}, M. and {Braun}, I. and {Briant}, T. and {Brillet}, A. and {Brinkmann}, M. and {Brockill}, P. and {Brockmueller}, E.},
        title = "{GW241011 and GW241110: Exploring Binary Formation and Fundamental Physics with Asymmetric, High-spin Black Hole Coalescences}",
      journal = {\apjl},
         year = 2025,
        month = nov,
       volume = {993},
       number = {1},
          eid = {L21},
        pages = {L21},
          doi = {10.3847/2041-8213/ae0d54},
archivePrefix = {arXiv},
       eprint = {2510.26931},
 primaryClass = {astro-ph.HE},
       adsurl = {https://ui.adsabs.harvard.edu/abs/2025ApJ...993L..21A}
}

@ARTICLE{XuYumeng2025,
       author = {{Xu}, Yumeng and {Valencia}, Jorge and {Estell{\'e}s Estrella}, H{\'e}ctor and {Ramos Buades}, Antoni and {Husa}, Sascha and {Rossell{\'o}-Sastre}, Maria and {Llobera Querol}, Joan and {Ramis Vidal}, Felip and {de Lluc Planas Llompart}, Maria and {Colleoni}, Marta and {Hamilton}, Eleanor and {Montava Agudo}, Arnau and {Y{\'e}bana Carrilero}, Jes{\'u}s and {Heffernan}, Anna},
        title = "{Parameter estimation for the GWTC-4.0 catalog with phenomenological waveform models that include orbital eccentricity and an updated description of spin precession}",
      journal = {arXiv e-prints},
         year = 2025,
        month = dec,
          eid = {arXiv:2512.19513},
        pages = {arXiv:2512.19513},
          doi = {10.48550/arXiv.2512.19513},
archivePrefix = {arXiv},
       eprint = {2512.19513},
 primaryClass = {gr-qc},
       adsurl = {https://ui.adsabs.harvard.edu/abs/2025arXiv251219513X}
}

@ARTICLE{Hendriks2026,
       author = {{Hendriks}, Kai and {Zwick}, Lorenz and {Saini}, Pankaj and {Tak{\'a}tsy}, J{\'a}nos and {Samsing}, Johan},
        title = "{Towards gravitational wave parameter inference for binaries with an eccentric companion}",
      journal = {arXiv e-prints},
         year = 2026,
        month = jan,
          eid = {arXiv:2601.14918},
        pages = {arXiv:2601.14918},
          doi = {10.48550/arXiv.2601.14918},
archivePrefix = {arXiv},
       eprint = {2601.14918},
 primaryClass = {astro-ph.HE},
       adsurl = {https://ui.adsabs.harvard.edu/abs/2026arXiv260114918H}
}

@ARTICLE{WangYuanZhu2025,
       author = {{Wang}, Yuan-Zhu and {Li}, Yin-Jie and {Gao}, Shi-Jie and {Tang}, Shao-Peng and {Fan}, Yi-Zhong},
        title = "{A new group of low-spin $50-70M_\odot$ Black Holes and the high pair-instability mass cutoff}",
      journal = {arXiv e-prints},
         year = 2025,
        month = oct,
          eid = {arXiv:2510.22698},
        pages = {arXiv:2510.22698},
          doi = {10.48550/arXiv.2510.22698},
archivePrefix = {arXiv},
       eprint = {2510.22698},
 primaryClass = {astro-ph.HE},
       adsurl = {https://ui.adsabs.harvard.edu/abs/2025arXiv251022698W}
}

@ARTICLE{Farah2026,
       author = {{Farah}, Amanda M. and {Vijaykumar}, Aditya and {Fishbach}, Maya},
        title = "{The steep redshift evolution of the hierarchical binary black hole merger rate may cause the $z$-$χ_{\rm eff}$ correlation}",
      journal = {arXiv e-prints},
         year = 2026,
        month = jan,
          eid = {arXiv:2601.03456},
        pages = {arXiv:2601.03456},
          doi = {10.48550/arXiv.2601.03456},
archivePrefix = {arXiv},
       eprint = {2601.03456},
 primaryClass = {astro-ph.HE},
       adsurl = {https://ui.adsabs.harvard.edu/abs/2026arXiv260103456F}
}

@ARTICLE{Colloms2025,
       author = {{Colloms}, Storm and {Doctor}, Zoheyr and {Berry}, Christopher P.~L.},
        title = "{Can Big Black Holes Merge with the Smallest Black Holes?}",
      journal = {\apj},
         year = 2025,
        month = dec,
       volume = {995},
       number = {1},
          eid = {123},
        pages = {123},
          doi = {10.3847/1538-4357/ae1f09},
archivePrefix = {arXiv},
       eprint = {2508.14159},
 primaryClass = {astro-ph.HE},
       adsurl = {https://ui.adsabs.harvard.edu/abs/2025ApJ...995..123C}
}

@ARTICLE{Plunkett2026,
       author = {{Plunkett}, Cailin and {Callister}, Thomas and {Zevin}, Michael and {Vitale}, Salvatore},
        title = "{Signatures of a subpopulation of hierarchical mergers in the GWTC-4 gravitational-wave dataset}",
      journal = {arXiv e-prints},
         year = 2026,
        month = jan,
          eid = {arXiv:2601.07908},
        pages = {arXiv:2601.07908},
          doi = {10.48550/arXiv.2601.07908},
archivePrefix = {arXiv},
       eprint = {2601.07908},
 primaryClass = {gr-qc},
       adsurl = {https://ui.adsabs.harvard.edu/abs/2026arXiv260107908P}
}

@ARTICLE{Tong2025,
       author = {{Tong}, Hui and {Callister}, Thomas A. and {Fishbach}, Maya and {Thrane}, Eric and {Antonini}, Fabio and {Stevenson}, Simon and {Romero-Shaw}, Isobel M. and {Dosopoulou}, Fani},
        title = "{A subpopulation of low-mass, spinning black holes: signatures of dynamical assembly}",
      journal = {arXiv e-prints},
         year = 2025,
        month = nov,
          eid = {arXiv:2511.05316},
        pages = {arXiv:2511.05316},
          doi = {10.48550/arXiv.2511.05316},
archivePrefix = {arXiv},
       eprint = {2511.05316},
 primaryClass = {astro-ph.HE},
       adsurl = {https://ui.adsabs.harvard.edu/abs/2025arXiv251105316T}
}

@ARTICLE{Chakraborty2025,
       author = {{Chakraborty}, Aniruddha and {Mukherjee}, Suvodip},
        title = "{The First Model-Independent Upper Bound on Micro-lensing Signature of the Highest Mass Binary Black Hole Event GW231123}",
      journal = {arXiv e-prints},
         year = 2025,
        month = dec,
          eid = {arXiv:2512.19077},
        pages = {arXiv:2512.19077},
          doi = {10.48550/arXiv.2512.19077},
archivePrefix = {arXiv},
       eprint = {2512.19077},
 primaryClass = {gr-qc},
       adsurl = {https://ui.adsabs.harvard.edu/abs/2025arXiv251219077C}
}

@ARTICLE{Shan2025,
       author = {{Shan}, Xikai and {Yang}, Huan and {Mao}, Shude},
        title = "{GW231123: A Case for Binary Microlensing in a Strong Lensing Field}",
      journal = {arXiv e-prints},
         year = 2025,
        month = dec,
          eid = {arXiv:2512.19118},
        pages = {arXiv:2512.19118},
          doi = {10.48550/arXiv.2512.19118},
archivePrefix = {arXiv},
       eprint = {2512.19118},
 primaryClass = {astro-ph.GA},
       adsurl = {https://ui.adsabs.harvard.edu/abs/2025arXiv251219118S}
}

@ARTICLE{Yang2025_lense,
       author = {{Yang}, Qiyuan and {You}, Zhi-Qiang and {Fan}, Xilong},
        title = "{Considering lensing effect on gravitational wave signals from black holes in mass gap}",
      journal = {arXiv e-prints},
         year = 2025,
        month = dec,
          eid = {arXiv:2512.20890},
        pages = {arXiv:2512.20890},
          doi = {10.48550/arXiv.2512.20890},
archivePrefix = {arXiv},
       eprint = {2512.20890},
 primaryClass = {astro-ph.HE},
       adsurl = {https://ui.adsabs.harvard.edu/abs/2025arXiv251220890Y}
}

@ARTICLE{Goyal2025,
       author = {{Goyal}, Srashti and {Villarrubia-Rojo}, Hector and {Zumalacarregui}, Miguel},
        title = "{Across the Universe: GW231123 as a magnified and diffracted black hole merger}",
      journal = {arXiv e-prints},
         year = 2025,
        month = dec,
          eid = {arXiv:2512.17631},
        pages = {arXiv:2512.17631},
          doi = {10.48550/arXiv.2512.17631},
archivePrefix = {arXiv},
       eprint = {2512.17631},
 primaryClass = {astro-ph.GA},
       adsurl = {https://ui.adsabs.harvard.edu/abs/2025arXiv251217631G}
}

@ARTICLE{Moncrieff2025,
       author = {{Moncrieff}, Jordan W.~N. and {Panther}, Fiona H.},
        title = "{Connecting the hierarchically merging binary black hole population to their host galaxies}",
      journal = {\mnras},
         year = 2025,
        month = oct,
       volume = {543},
       number = {2},
        pages = {1833-1841},
          doi = {10.1093/mnras/staf1582},
archivePrefix = {arXiv},
       eprint = {2508.18704},
 primaryClass = {astro-ph.CO},
       adsurl = {https://ui.adsabs.harvard.edu/abs/2025MNRAS.543.1833M}
}

@ARTICLE{Guttman2026,
       author = {{Guttman}, Nir and {Payne}, Ethan and {Lasky}, Paul D. and {Thrane}, Eric},
        title = "{Trends in the Population of Binary Black Holes Following the Fourth Gravitational-wave Transient Catalog: A Data-driven Analysis}",
      journal = {\apj},
         year = 2026,
        month = jan,
       volume = {996},
       number = {2},
          eid = {144},
        pages = {144},
          doi = {10.3847/1538-4357/ae17af},
archivePrefix = {arXiv},
       eprint = {2509.09876},
 primaryClass = {astro-ph.HE},
       adsurl = {https://ui.adsabs.harvard.edu/abs/2026ApJ...996..144G}
}

@ARTICLE{Jiang2016,
       author = {{Jiang}, Yan-Fei and {Davis}, Shane W. and {Stone}, James M.},
        title = "{Iron Opacity Bump Changes the Stability and Structure of Accretion Disks in Active Galactic Nuclei}",
      journal = {\apj},
         year = 2016,
        month = aug,
       volume = {827},
       number = {1},
          eid = {10},
        pages = {10},
          doi = {10.3847/0004-637X/827/1/10},
archivePrefix = {arXiv},
       eprint = {1601.06836},
 primaryClass = {astro-ph.HE},
       adsurl = {https://ui.adsabs.harvard.edu/abs/2016ApJ...827...10J}
}

@ARTICLE{Vijaykumar2026,
       author = {{Vijaykumar}, Aditya and {Farah}, Amanda M. and {Fishbach}, Maya},
        title = "{The Maximum Mass Ratio of Hierarchical Binary Black Hole Mergers May Cause the q─{\ensuremath{\chi}}$_{eff}$ Correlation}",
      journal = {\apjl},
         year = 2026,
        month = mar,
       volume = {999},
       number = {2},
          eid = {L30},
        pages = {L30},
          doi = {10.3847/2041-8213/ae4878},
archivePrefix = {arXiv},
       eprint = {2601.03457},
 primaryClass = {astro-ph.HE},
       adsurl = {https://ui.adsabs.harvard.edu/abs/2026ApJ...999L..30V}
}

@ARTICLE{Vitale2025,
       author = {{Vitale}, Salvatore and {Mould}, Matthew and {(Society Of Physicists Interested in Non-Aligned Spins}, Spins)},
        title = "{Long road to alignment: Measuring black hole spin orientation with expanding gravitational-wave datasets}",
      journal = {\prd},
         year = 2025,
        month = oct,
       volume = {112},
       number = {8},
          eid = {083015},
        pages = {083015},
          doi = {10.1103/drsl-n3wz},
archivePrefix = {arXiv},
       eprint = {2505.14875},
 primaryClass = {astro-ph.HE},
       adsurl = {https://ui.adsabs.harvard.edu/abs/2025PhRvD.112h3015V}
}

@ARTICLE{Tanikawa2025,
       author = {{Tanikawa}, Ataru and {Liu}, Shuai and {Wu}, WeiWei and {Fujii}, Michiko S. and {Wang}, Long},
        title = "{GW231123 Formation from Population III Stars: Isolated Binary Evolution}",
      journal = {arXiv e-prints},
         year = 2025,
        month = aug,
          eid = {arXiv:2508.01135},
        pages = {arXiv:2508.01135},
          doi = {10.48550/arXiv.2508.01135},
archivePrefix = {arXiv},
       eprint = {2508.01135},
 primaryClass = {astro-ph.SR},
       adsurl = {https://ui.adsabs.harvard.edu/abs/2025arXiv250801135T}
}

@ARTICLE{Popa2025,
       author = {{Popa}, Silvia A. and {de Mink}, Selma E.},
        title = "{Very Massive, Rapidly Spinning Binary Black Hole Progenitors through Chemically Homogeneous Evolution{\textemdash}The Case of GW231123}",
      journal = {\apjl},
         year = 2025,
        month = dec,
       volume = {995},
       number = {2},
          eid = {L76},
        pages = {L76},
          doi = {10.3847/2041-8213/ae20f1},
archivePrefix = {arXiv},
       eprint = {2509.00154},
 primaryClass = {astro-ph.HE},
       adsurl = {https://ui.adsabs.harvard.edu/abs/2025ApJ...995L..76P}
}

@ARTICLE{Delfavero2025,
       author = {{Delfavero}, V. and {Ray}, S. and {Cook}, H.~E. and {Nathaniel}, K. and {McKernan}, B. and {Ford}, K.~E.~S. and {Postiglione}, J. and {McPike}, E. and {O'Shaughnessy}, R.},
        title = "{Prospects for the formation of GW231123 from the AGN channel}",
      journal = {arXiv e-prints},
         year = 2025,
        month = aug,
          eid = {arXiv:2508.13412},
        pages = {arXiv:2508.13412},
          doi = {10.48550/arXiv.2508.13412},
archivePrefix = {arXiv},
       eprint = {2508.13412},
 primaryClass = {gr-qc},
       adsurl = {https://ui.adsabs.harvard.edu/abs/2025arXiv250813412D}
}

@ARTICLE{DeLuca2025,
       author = {{De Luca}, Valerio and {Franciolini}, Gabriele and {Riotto}, Antonio},
        title = "{GW231123: a Possible Primordial Black Hole Origin}",
      journal = {arXiv e-prints},
         year = 2025,
        month = aug,
          eid = {arXiv:2508.09965},
        pages = {arXiv:2508.09965},
          doi = {10.48550/arXiv.2508.09965},
archivePrefix = {arXiv},
       eprint = {2508.09965},
 primaryClass = {astro-ph.CO},
       adsurl = {https://ui.adsabs.harvard.edu/abs/2025arXiv250809965D}
}

@ARTICLE{Gottlieb2025,
       author = {{Gottlieb}, Ore and {Metzger}, Brian D. and {Issa}, Danat and {Li}, Sean E. and {Renzo}, Mathieu and {Isi}, Maximiliano},
        title = "{Spinning into the Gap: Direct-horizon Collapse as the Origin of GW231123 from End-to-end General-relativistic Magnetohydrodynamic Simulations}",
      journal = {\apjl},
         year = 2025,
        month = nov,
       volume = {993},
       number = {2},
          eid = {L54},
        pages = {L54},
          doi = {10.3847/2041-8213/ae0d81},
archivePrefix = {arXiv},
       eprint = {2508.15887},
 primaryClass = {astro-ph.HE},
       adsurl = {https://ui.adsabs.harvard.edu/abs/2025ApJ...993L..54G}
}

@ARTICLE{Croon2025,
       author = {{Croon}, Djuna and {Gerosa}, Davide and {Sakstein}, Jeremy},
        title = "{Can GW231123 have a stellar origin?}",
      journal = {\mnras},
         year = 2026,
        month = mar,
       volume = {546},
       number = {3},
          eid = {stag073},
        pages = {stag073},
          doi = {10.1093/mnras/stag073},
archivePrefix = {arXiv},
       eprint = {2508.10088},
 primaryClass = {astro-ph.HE},
       adsurl = {https://ui.adsabs.harvard.edu/abs/2026MNRAS.546ag073C}
}

@ARTICLE{LiuBin2025,
       author = {{Liu}, Bin and {Lai}, Dong},
        title = "{Hierarchical Black Hole Mergers in Nuclear Star Clusters: A Combined Dynamical-Secular Channel for GW231123-like Events}",
      journal = {arXiv e-prints},
         year = 2025,
        month = nov,
          eid = {arXiv:2511.13820},
        pages = {arXiv:2511.13820},
          doi = {10.48550/arXiv.2511.13820},
archivePrefix = {arXiv},
       eprint = {2511.13820},
 primaryClass = {astro-ph.HE},
       adsurl = {https://ui.adsabs.harvard.edu/abs/2025arXiv251113820L}
}

@ARTICLE{Liu2025,
       author = {{Liu}, Shuai and {Wang}, Long and {Tanikawa}, Ataru and {Wu}, Weiwei and {Fujii}, Michiko S.},
        title = "{On the Formation of GW231123 in Population III Star Clusters}",
      journal = {\apjl},
         year = 2025,
        month = nov,
       volume = {993},
       number = {1},
          eid = {L30},
        pages = {L30},
          doi = {10.3847/2041-8213/ae1024},
archivePrefix = {arXiv},
       eprint = {2510.05634},
 primaryClass = {astro-ph.GA},
       adsurl = {https://ui.adsabs.harvard.edu/abs/2025ApJ...993L..30L}
}

@ARTICLE{Passenger2025,
       author = {{Passenger}, Lachlan and {Banagiri}, Sharan and {Thrane}, Eric and {Lasky}, Paul D. and {Borchers}, Angela and {Fishbach}, Maya and {Ye}, Claire S.},
        title = "{Is GW231123 a Hierarchical Merger?}",
      journal = {\apj},
         year = 2026,
        month = mar,
       volume = {999},
       number = {2},
          eid = {236},
        pages = {236},
          doi = {10.3847/1538-4357/ae4358},
archivePrefix = {arXiv},
       eprint = {2510.14363},
 primaryClass = {astro-ph.HE},
       adsurl = {https://ui.adsabs.harvard.edu/abs/2026ApJ...999..236P}
}

@ARTICLE{Paiella2025,
       author = {{Paiella}, Lavinia and {Ugolini}, Cristiano and {Spera}, Mario and {Branchesi}, Marica and {Arca Sedda}, Manuel},
        title = "{Assembling GW231123 in Star Clusters through the Combination of Stellar Binary Evolution and Hierarchical Mergers}",
      journal = {\apjl},
         year = 2025,
        month = dec,
       volume = {994},
       number = {2},
          eid = {L54},
        pages = {L54},
          doi = {10.3847/2041-8213/ae1447},
archivePrefix = {arXiv},
       eprint = {2509.10609},
 primaryClass = {astro-ph.GA},
       adsurl = {https://ui.adsabs.harvard.edu/abs/2025ApJ...994L..54P}
}

@ARTICLE{Fabj2025,
       author = {{Fabj}, Gaia and {Tiede}, Christopher and {Rowan}, Connar and {Pessah}, Martin and {Samsing}, Johan},
        title = "{Spin-Orbit Misalignments of Eccentric Black Hole Mergers in AGN Disks}",
      journal = {\mnras},
         year = 2026,
        month = mar,
          doi = {10.1093/mnras/stag427},
archivePrefix = {arXiv},
       eprint = {2510.07952},
 primaryClass = {astro-ph.HE},
       adsurl = {https://ui.adsabs.harvard.edu/abs/2026MNRAS.tmp..408F}
}

@ARTICLE{Adamcewicz2025,
       author = {{Adamcewicz}, Christian and {Guttman}, Nir and {Lasky}, Paul D. and {Thrane}, Eric},
        title = "{Do Both Black Holes Spin in Merging Binaries? Evidence from GWTC-4 and Astrophysical Implications}",
      journal = {\apj},
         year = 2025,
        month = dec,
       volume = {994},
       number = {2},
          eid = {261},
        pages = {261},
          doi = {10.3847/1538-4357/ae1370},
archivePrefix = {arXiv},
       eprint = {2509.04706},
 primaryClass = {astro-ph.HE},
       adsurl = {https://ui.adsabs.harvard.edu/abs/2025ApJ...994..261A}
}

@ARTICLE{Sadowski2016,
       author = {{Sadowski}, Aleksander and {Narayan}, Ramesh},
        title = "{Three-dimensional simulations of supercritical black hole accretion discs - luminosities, photon trapping and variability}",
      journal = {\mnras},
         year = 2016,
        month = mar,
       volume = {456},
       number = {4},
        pages = {3929-3947},
          doi = {10.1093/mnras/stv2941},
archivePrefix = {arXiv},
       eprint = {1509.03168},
 primaryClass = {astro-ph.HE},
       adsurl = {https://ui.adsabs.harvard.edu/abs/2016MNRAS.456.3929S}
}

@ARTICLE{LiYinJie2025_alignment,
       author = {{Li}, Yin-Jie and {Wang}, Yuan-Zhu and {Tang}, Shao-Peng and {Fan}, Yi-Zhong},
        title = "{Aligned Hierarchical Black Hole Mergers in AGN disks revealed by GWTC-4}",
      journal = {arXiv e-prints},
         year = 2025,
        month = sep,
          eid = {arXiv:2509.23897},
        pages = {arXiv:2509.23897},
          doi = {10.48550/arXiv.2509.23897},
archivePrefix = {arXiv},
       eprint = {2509.23897},
 primaryClass = {astro-ph.HE},
       adsurl = {https://ui.adsabs.harvard.edu/abs/2025arXiv250923897L}
}

@ARTICLE{Tagawa2025,
       author = {{Tagawa}, Hiromichi and {Rowan}, Connar and {Tak{\'a}tsy}, J{\'a}nos and {Zwick}, Lorenz and {Hendriks}, Kai and {Han}, Wen-Biao and {Samsing}, Johan},
        title = "{Gravitational Wave Phase Shifts of Black Hole Mergers in AGN Disks}",
      journal = {\apj},
         year = 2026,
        month = feb,
       volume = {998},
       number = {2},
          eid = {244},
        pages = {244},
          doi = {10.3847/1538-4357/ae3a87},
archivePrefix = {arXiv},
       eprint = {2511.15193},
 primaryClass = {astro-ph.HE},
       adsurl = {https://ui.adsabs.harvard.edu/abs/2026ApJ...998..244T}
}

@ARTICLE{Trani2024,
       author = {{Trani}, A.~A. and {Quaini}, S. and {Colpi}, M.},
        title = "{Three-body encounters in black hole discs around a supermassive black hole. The disc velocity dispersion and the Keplerian tidal field determine the eccentricity and spin-orbit alignment of gravitational wave mergers}",
      journal = {\aap},
         year = 2024,
        month = mar,
       volume = {683},
          eid = {A135},
        pages = {A135},
          doi = {10.1051/0004-6361/202347920},
archivePrefix = {arXiv},
       eprint = {2312.13281},
 primaryClass = {astro-ph.HE},
       adsurl = {https://ui.adsabs.harvard.edu/abs/2024A&A...683A.135T}
}

@ARTICLE{Stegmann2025,
       author = {{Stegmann}, Jakob and {Olejak}, Aleksandra and {de Mink}, Selma E.},
        title = "{Resolving Black Hole Family Issues among the Massive Ancestors of Very High-spin Gravitational-wave Events like GW231123}",
      journal = {\apjl},
         year = 2025,
        month = oct,
       volume = {992},
       number = {2},
          eid = {L26},
        pages = {L26},
          doi = {10.3847/2041-8213/ae0e5f},
archivePrefix = {arXiv},
       eprint = {2507.15967},
 primaryClass = {astro-ph.HE},
       adsurl = {https://ui.adsabs.harvard.edu/abs/2025ApJ...992L..26S}
}

@ARTICLE{Olejak2024,
       author = {{Olejak}, Aleksandra and {Klencki}, Jakub and {Xu}, Xiao-Tian and {Wang}, Chen and {Belczynski}, Krzysztof and {Lasota}, Jean-Pierre},
        title = "{Unequal-mass highly spinning binary black hole mergers in the stable mass transfer formation channel}",
      journal = {\aap},
         year = 2024,
        month = sep,
       volume = {689},
          eid = {A305},
        pages = {A305},
          doi = {10.1051/0004-6361/202450480},
archivePrefix = {arXiv},
       eprint = {2404.12426},
 primaryClass = {astro-ph.HE},
       adsurl = {https://ui.adsabs.harvard.edu/abs/2024A&A...689A.305O}
}

@ARTICLE{Banerjee2024,
       author = {{Banerjee}, Sambaran and {Olejak}, Aleksandra},
        title = "{On the effective spin-mass ratio relation of binary black hole mergers that evolved in isolation}",
      journal = {arXiv e-prints},
         year = 2024,
        month = nov,
          eid = {arXiv:2411.15112},
        pages = {arXiv:2411.15112},
          doi = {10.48550/arXiv.2411.15112},
archivePrefix = {arXiv},
       eprint = {2411.15112},
 primaryClass = {astro-ph.HE},
       adsurl = {https://ui.adsabs.harvard.edu/abs/2024arXiv241115112B}
}

@ARTICLE{Zevin2022,
       author = {{Zevin}, Michael and {Bavera}, Simone S.},
        title = "{Suspicious Siblings: The Distribution of Mass and Spin across Component Black Holes in Isolated Binary Evolution}",
      journal = {\apj},
         year = 2022,
        month = jul,
       volume = {933},
       number = {1},
          eid = {86},
        pages = {86},
          doi = {10.3847/1538-4357/ac6f5d},
archivePrefix = {arXiv},
       eprint = {2203.02515},
 primaryClass = {astro-ph.HE},
       adsurl = {https://ui.adsabs.harvard.edu/abs/2022ApJ...933...86Z}
}

@ARTICLE{WangZiYuan2025,
       author = {{Wang}, Zi-Yuan and {Qin}, Ying and {Hu}, Rui-Chong and {Wang}, Yuan-Zhu and {Meynet}, Georges and {Song}, Han-Feng},
        title = "{Reassessing the Spin of Second-born Black Holes in Coalescing Binary Black Holes and Its Connection to the chi\_eff-q Correlation}",
      journal = {arXiv e-prints},
         year = 2025,
        month = sep,
          eid = {arXiv:2509.05976},
        pages = {arXiv:2509.05976},
          doi = {10.48550/arXiv.2509.05976},
archivePrefix = {arXiv},
       eprint = {2509.05976},
 primaryClass = {astro-ph.HE},
       adsurl = {https://ui.adsabs.harvard.edu/abs/2025arXiv250905976W}
}

@ARTICLE{Cook2024,
       author = {{Cook}, Harrison E. and {McKernan}, Barry and {Ford}, K.~E. Saavik and {Delfavero}, Vera and {Nathaniel}, Kaila and {Postiglione}, Jake and {Ray}, Shawn and {McPike}, Emily J. and {O'Shaughnessy}, Richard},
        title = "{McFACTS. II. Mass Ratio─Effective Spin Relationship of Black Hole Mergers in the Active Galactic Nucleus Channel}",
      journal = {\apj},
         year = 2025,
        month = nov,
       volume = {993},
       number = {2},
          eid = {163},
        pages = {163},
          doi = {10.3847/1538-4357/adfd56},
archivePrefix = {arXiv},
       eprint = {2411.10590},
 primaryClass = {astro-ph.HE},
       adsurl = {https://ui.adsabs.harvard.edu/abs/2025ApJ...993..163C}
}

@ARTICLE{McKernan2022,
       author = {{McKernan}, B. and {Ford}, K.~E.~S. and {Callister}, T. and {Farr}, W.~M. and {O'Shaughnessy}, R. and {Smith}, R. and {Thrane}, E. and {Vajpeyi}, A.},
        title = "{LIGO-Virgo correlations between mass ratio and effective inspiral spin: testing the active galactic nuclei channel}",
      journal = {\mnras},
         year = 2022,
        month = aug,
       volume = {514},
       number = {3},
        pages = {3886-3893},
          doi = {10.1093/mnras/stac1570},
archivePrefix = {arXiv},
       eprint = {2107.07551},
 primaryClass = {astro-ph.HE},
       adsurl = {https://ui.adsabs.harvard.edu/abs/2022MNRAS.514.3886M}
}

@ARTICLE{McKernan2024,
       author = {{McKernan}, B. and {Ford}, K.~E.~S.},
        title = "{Constraining the LVK AGN channel with black hole spins}",
      journal = {\mnras},
         year = 2024,
        month = jul,
       volume = {531},
       number = {3},
        pages = {3479-3485},
          doi = {10.1093/mnras/stae1351},
archivePrefix = {arXiv},
       eprint = {2309.15213},
 primaryClass = {astro-ph.HE},
       adsurl = {https://ui.adsabs.harvard.edu/abs/2024MNRAS.531.3479M}
}

@ARTICLE{McKernan2025,
       author = {{McKernan}, Barry and {Ford}, K.~E. Saavik and {Cook}, Harrison E. and {Delfavero}, Vera and {McPike}, Emily and {Nathaniel}, Kaila and {Postiglione}, Jake and {Ray}, Shawn and {O'Shaughnessy}, Richard},
        title = "{McFACTS I: Testing the LVK AGN Channel with Monte Carlo for AGN Channel Testing and Simulation (McFACTS)}",
      journal = {\apj},
         year = 2025,
        month = sep,
       volume = {990},
       number = {2},
          eid = {217},
        pages = {217},
          doi = {10.3847/1538-4357/adf114},
       adsurl = {https://ui.adsabs.harvard.edu/abs/2025ApJ...990..217M}
}

@ARTICLE{LiYinJie2025_Xeffq,
       author = {{Li}, Yin-Jie and {Wang}, Yuan-Zhu and {Tang}, Shao-Peng and {Chen}, Tong and {Fan}, Yi-Zhong},
        title = "{Revealing the {\ensuremath{\chi}}$_{eff}${\textendash}q Correlation among Coalescing Binary Black Holes and Tentative Evidence for AGN-driven Hierarchical Mergers}",
      journal = {\apj},
         year = 2025,
        month = jul,
       volume = {987},
       number = {1},
          eid = {65},
        pages = {65},
          doi = {10.3847/1538-4357/add535},
archivePrefix = {arXiv},
       eprint = {2501.09495},
 primaryClass = {astro-ph.HE},
       adsurl = {https://ui.adsabs.harvard.edu/abs/2025ApJ...987...65L}
}

@ARTICLE{Adamcewicz2023,
       author = {{Adamcewicz}, Christian and {Lasky}, Paul D. and {Thrane}, Eric},
        title = "{Evidence for a Correlation between Binary Black Hole Mass Ratio and Black Hole Spins}",
      journal = {\apj},
         year = 2023,
        month = nov,
       volume = {958},
       number = {1},
          eid = {13},
        pages = {13},
          doi = {10.3847/1538-4357/acf763},
archivePrefix = {arXiv},
       eprint = {2307.15278},
 primaryClass = {astro-ph.HE},
       adsurl = {https://ui.adsabs.harvard.edu/abs/2023ApJ...958...13A}
}

@ARTICLE{Adamcewicz2022,
       author = {{Adamcewicz}, Christian and {Thrane}, Eric},
        title = "{Do unequal-mass binary black hole systems have larger {\ensuremath{\chi}}$_{eff}$? Probing correlations with copulas in gravitational-wave astronomy}",
      journal = {\mnras},
         year = 2022,
        month = dec,
       volume = {517},
       number = {3},
        pages = {3928-3937},
          doi = {10.1093/mnras/stac2961},
archivePrefix = {arXiv},
       eprint = {2208.03405},
 primaryClass = {astro-ph.HE},
       adsurl = {https://ui.adsabs.harvard.edu/abs/2022MNRAS.517.3928A}
}

@ARTICLE{Gangardt2024,
       author = {{Gangardt}, Daria and {Trani}, Alessandro Alberto and {Bonnerot}, Cl{\'e}ment and {Gerosa}, Davide},
        title = "{pAGN: the one-stop solution for AGN disc modelling}",
      journal = {\mnras},
         year = 2024,
        month = jun,
       volume = {530},
       number = {4},
        pages = {3689-3705},
          doi = {10.1093/mnras/stae1117},
archivePrefix = {arXiv},
       eprint = {2403.00060},
 primaryClass = {astro-ph.HE},
       adsurl = {https://ui.adsabs.harvard.edu/abs/2024MNRAS.530.3689G}
}

@ARTICLE{Vaccaro2024,
       author = {{Vaccaro}, Maria Paola and {Mapelli}, Michela and {P{\'e}rigois}, Carole and {Barone}, Dario and {Artale}, Maria Celeste and {Dall'Amico}, Marco and {Iorio}, Giuliano and {Torniamenti}, Stefano},
        title = "{Impact of gas hardening on the population properties of hierarchical black hole mergers in active galactic nucleus disks}",
      journal = {\aap},
         year = 2024,
        month = may,
       volume = {685},
          eid = {A51},
        pages = {A51},
          doi = {10.1051/0004-6361/202348509},
archivePrefix = {arXiv},
       eprint = {2311.18548},
 primaryClass = {astro-ph.HE},
       adsurl = {https://ui.adsabs.harvard.edu/abs/2024A&A...685A..51V}
}

@ARTICLE{LIGO2025_O4aProp,
       author = {{The LIGO Scientific Collaboration} and {the Virgo Collaboration} and {the KAGRA Collaboration}},
        title = "{GWTC-4.0: Population Properties of Merging Compact Binaries}",
      journal = {arXiv e-prints},
         year = 2025,
        month = aug,
          eid = {arXiv:2508.18083},
        pages = {arXiv:2508.18083},
          doi = {10.48550/arXiv.2508.18083},
archivePrefix = {arXiv},
       eprint = {2508.18083},
 primaryClass = {astro-ph.HE},
       adsurl = {https://ui.adsabs.harvard.edu/abs/2025arXiv250818083T}
}

@ARTICLE{LIGO2025_O4aCatalog,
       author = {{The LIGO Scientific Collaboration} and {The Virgo Collaboration} and {the KAGRA Collaboration}},
        title = "{GWTC-4.0: Updating the Gravitational-Wave Transient Catalog with Observations from the First Part of the Fourth LIGO-Virgo-KAGRA Observing Run}",
      journal = {arXiv e-prints},
         year = 2025,
        month = aug,
          eid = {arXiv:2508.18082},
        pages = {arXiv:2508.18082},
          doi = {10.48550/arXiv.2508.18082},
archivePrefix = {arXiv},
       eprint = {2508.18082},
 primaryClass = {gr-qc},
       adsurl = {https://ui.adsabs.harvard.edu/abs/2025arXiv250818082T}
}

@ARTICLE{Bartos2025,
       author = {{Bartos}, Imre and {Haiman}, Zolt{\'a}n},
        title = "{Accretion is All You Need: Black Hole Spin Alignment in Merger GW231123 Indicates Accretion Pathway}",
      journal = {\apjl},
         year = 2026,
        month = jan,
       volume = {996},
       number = {2},
          eid = {L44},
        pages = {L44},
          doi = {10.3847/2041-8213/ae2bff},
archivePrefix = {arXiv},
       eprint = {2508.08558},
 primaryClass = {astro-ph.HE},
       adsurl = {https://ui.adsabs.harvard.edu/abs/2026ApJ...996L..44B}
}

@ARTICLE{LIGO2025_O4_Prop,
       author = {{The LIGO Scientific Collaboration} and {the Virgo Collaboration} and {the KAGRA Collaboration}},
        title = "{GWTC-4.0: Population Properties of Merging Compact Binaries}",
      journal = {arXiv e-prints},
         year = 2025,
        month = aug,
          eid = {arXiv:2508.18083},
        pages = {arXiv:2508.18083},
          doi = {10.48550/arXiv.2508.18083},
archivePrefix = {arXiv},
       eprint = {2508.18083},
 primaryClass = {astro-ph.HE},
       adsurl = {https://ui.adsabs.harvard.edu/abs/2025arXiv250818083T}
}

@ARTICLE{Callister2021,
       author = {{Callister}, Thomas A. and {Haster}, Carl-Johan and {Ng}, Ken K.~Y. and {Vitale}, Salvatore and {Farr}, Will M.},
        title = "{Who Ordered That? Unequal-mass Binary Black Hole Mergers Have Larger Effective Spins}",
      journal = {\apjl},
         year = 2021,
        month = nov,
       volume = {922},
       number = {1},
          eid = {L5},
        pages = {L5},
          doi = {10.3847/2041-8213/ac2ccc},
archivePrefix = {arXiv},
       eprint = {2106.00521},
 primaryClass = {astro-ph.HE},
       adsurl = {https://ui.adsabs.harvard.edu/abs/2021ApJ...922L...5C}
}

@ARTICLE{Pierra2024,
       author = {{Pierra}, G. and {Mastrogiovanni}, S. and {Perri{\`e}s}, S.},
        title = "{The spin magnitude of stellar-mass black holes evolves with the mass}",
      journal = {\aap},
         year = 2024,
        month = dec,
       volume = {692},
          eid = {A80},
        pages = {A80},
          doi = {10.1051/0004-6361/202452545},
       adsurl = {https://ui.adsabs.harvard.edu/abs/2024A&A...692A..80P}
}

@ARTICLE{Antonini2025,
       author = {{Antonini}, Fabio and {Romero-Shaw}, Isobel M. and {Callister}, Thomas},
        title = "{Star Cluster Population of High Mass Black Hole Mergers in Gravitational Wave Data}",
      journal = {\prl},
         year = 2025,
        month = jan,
       volume = {134},
       number = {1},
          eid = {011401},
        pages = {011401},
          doi = {10.1103/PhysRevLett.134.011401},
archivePrefix = {arXiv},
       eprint = {2406.19044},
 primaryClass = {astro-ph.HE},
       adsurl = {https://ui.adsabs.harvard.edu/abs/2025PhRvL.134a1401A}
}

@ARTICLE{LiYinJie2024,
       author = {{Li}, Yin-Jie and {Wang}, Yuan-Zhu and {Tang}, Shao-Peng and {Fan}, Yi-Zhong},
        title = "{Resolving the Stellar-Collapse and Hierarchical-Merger Origins of the Coalescing Black Holes}",
      journal = {\prl},
         year = 2024,
        month = aug,
       volume = {133},
       number = {5},
          eid = {051401},
        pages = {051401},
          doi = {10.1103/PhysRevLett.133.051401},
archivePrefix = {arXiv},
       eprint = {2303.02973},
 primaryClass = {astro-ph.HE},
       adsurl = {https://ui.adsabs.harvard.edu/abs/2024PhRvL.133e1401L}
}

@ARTICLE{Tiwari2024,
       author = {{Tiwari}, Vaibhav},
        title = "{What's in a binary black hole's mass parameter?}",
      journal = {\mnras},
         year = 2024,
        month = jan,
       volume = {527},
       number = {1},
        pages = {298-306},
          doi = {10.1093/mnras/stad3155},
archivePrefix = {arXiv},
       eprint = {2304.03498},
 primaryClass = {astro-ph.HE},
       adsurl = {https://ui.adsabs.harvard.edu/abs/2024MNRAS.527..298T}
}

@ARTICLE{LiYinJie2025,
       author = {{Li}, Yin-Jie and {Tang}, Shao-Peng and {Xue}, Ling-Qin and {Fan}, Yi-Zhong},
        title = "{GW231123: Likely a Product of Successive Mergers from {\ensuremath{\sim}}10 Stellar-mass Black Holes}",
      journal = {\apj},
         year = 2026,
        month = mar,
       volume = {999},
       number = {1},
          eid = {127},
        pages = {127},
          doi = {10.3847/1538-4357/ae4102},
archivePrefix = {arXiv},
       eprint = {2507.17551},
 primaryClass = {astro-ph.HE},
       adsurl = {https://ui.adsabs.harvard.edu/abs/2026ApJ...999..127L}
}

@ARTICLE{WangMengye2025_BSI,
       author = {{Wang}, Mengye and {Ma}, Yiqiu and {Li}, Hui and {Wu}, Qingwen and {Li}, Ya-Ping and {Lei}, Xiangli and {Wu}, Jiancheng},
        title = "{Simulation of Binary-single Interactions in AGN Disk. I. Gas-enhanced Binary Orbital Hardening}",
      journal = {\apj},
         year = 2025,
        month = apr,
       volume = {983},
       number = {2},
          eid = {114},
        pages = {114},
          doi = {10.3847/1538-4357/adbf8e},
archivePrefix = {arXiv},
       eprint = {2501.10703},
 primaryClass = {astro-ph.HE},
       adsurl = {https://ui.adsabs.harvard.edu/abs/2025ApJ...983..114W}
}

@ARTICLE{WangMengye2025_BSII,
       author = {{Wang}, Mengye and {Wu}, Qingwen and {Ma}, Yiqiu},
        title = "{Simulation of Binary─Single Interactions in Active Galactic Nucleus Disks. II. Merger Probability of Binary Black Holes during Chaotic Triple Process}",
      journal = {\apj},
         year = 2025,
        month = nov,
       volume = {993},
       number = {1},
          eid = {139},
        pages = {139},
          doi = {10.3847/1538-4357/ae06a5},
archivePrefix = {arXiv},
       eprint = {2507.07715},
 primaryClass = {astro-ph.HE},
       adsurl = {https://ui.adsabs.harvard.edu/abs/2025ApJ...993..139W}
}

@ARTICLE{Rowan2025_BS,
       author = {{Rowan}, Connar and {Whitehead}, Henry and {Fabj}, Gaia and {Saini}, Pankaj and {Kocsis}, Bence and {Pessah}, Martin and {Samsing}, Johan},
        title = "{Prompt gravitational-wave mergers aided by gas in active galactic nuclei: the hydrodynamics of binary-single black hole scatterings}",
      journal = {\mnras},
         year = 2025,
        month = may,
       volume = {539},
       number = {2},
        pages = {1501-1515},
          doi = {10.1093/mnras/staf547},
archivePrefix = {arXiv},
       eprint = {2501.09017},
 primaryClass = {astro-ph.GA},
       adsurl = {https://ui.adsabs.harvard.edu/abs/2025MNRAS.539.1501R}
}

@ARTICLE{Calcino2024,
       author = {{Calcino}, Josh and {Dempsey}, Adam M. and {Dittmann}, Alexander J. and {Li}, Hui},
        title = "{Runaway Eccentricity Growth: A Pathway for Binary Black Hole Mergers in AGN Disks}",
      journal = {\apj},
         year = 2024,
        month = aug,
       volume = {970},
       number = {2},
          eid = {107},
        pages = {107},
          doi = {10.3847/1538-4357/ad4a53},
archivePrefix = {arXiv},
       eprint = {2311.13727},
 primaryClass = {astro-ph.HE},
       adsurl = {https://ui.adsabs.harvard.edu/abs/2024ApJ...970..107C}
}

@ARTICLE{Dittmann2025,
       author = {{Dittmann}, Alexander J. and {Dempsey}, Adam M. and {Li}, Hui},
        title = "{The Multiple Paths to Merger of Unequal-mass Black Hole Binaries in the Disks of Active Galactic Nuclei}",
      journal = {\apj},
         year = 2025,
        month = sep,
       volume = {990},
       number = {2},
          eid = {137},
        pages = {137},
          doi = {10.3847/1538-4357/adea72},
archivePrefix = {arXiv},
       eprint = {2505.05555},
 primaryClass = {astro-ph.HE},
       adsurl = {https://ui.adsabs.harvard.edu/abs/2025ApJ...990..137D}
}

@ARTICLE{Rowan2025,
       author = {{Rowan}, Connar and {Whitehead}, Henry and {Fabj}, Gaia and {Kirkeberg}, Philip and {Pessah}, Martin E. and {Kocsis}, Bence},
        title = "{Hydrodynamic simulations of black hole evolution in AGN discs I: orbital alignment of highly inclined satellites}",
      journal = {arXiv e-prints},
         year = 2025,
        month = may,
          eid = {arXiv:2505.23739},
        pages = {arXiv:2505.23739},
          doi = {10.48550/arXiv.2505.23739},
archivePrefix = {arXiv},
       eprint = {2505.23739},
 primaryClass = {astro-ph.HE},
       adsurl = {https://ui.adsabs.harvard.edu/abs/2025arXiv250523739R}
}

@ARTICLE{Whitehead2025,
       author = {{Whitehead}, Henry and {Rowan}, Connar and {Kocsis}, Bence},
        title = "{Hydrodynamic simulations of black hole evolution in AGN discs II: inclination damping for partially embedded satellites}",
      journal = {\mnras},
         year = 2025,
        month = nov,
       volume = {543},
       number = {4},
        pages = {3768-3782},
          doi = {10.1093/mnras/staf1686},
archivePrefix = {arXiv},
       eprint = {2505.23899},
 primaryClass = {astro-ph.HE},
       adsurl = {https://ui.adsabs.harvard.edu/abs/2025MNRAS.543.3768W}
}

@ARTICLE{Rodriguez-Ramirez2025,
       author = {{Rodr{\'\i}guez-Ram{\'\i}rez}, J.~C. and {Nemmen}, R. and {Bom}, C.~R.},
        title = "{Optical and UV flares from binary black hole mergers in active galactic nuclei}",
      journal = {\prd},
         year = 2025,
        month = apr,
       volume = {111},
       number = {8},
          eid = {083020},
        pages = {083020},
          doi = {10.1103/PhysRevD.111.083020},
archivePrefix = {arXiv},
       eprint = {2407.09945},
 primaryClass = {astro-ph.HE},
       adsurl = {https://ui.adsabs.harvard.edu/abs/2025PhRvD.111h3020R}
}

@ARTICLE{ChenKen2025,
       author = {{Chen}, Ken and {Dai}, Zi-Gao},
        title = "{Observational Properties of Thermal Emission from Relativistic Jets Embedded in Active Galactic Nucleus Disks}",
      journal = {\apj},
         year = 2025,
        month = jul,
       volume = {987},
       number = {2},
          eid = {214},
        pages = {214},
          doi = {10.3847/1538-4357/addb48},
archivePrefix = {arXiv},
       eprint = {2505.16390},
 primaryClass = {astro-ph.HE},
       adsurl = {https://ui.adsabs.harvard.edu/abs/2025ApJ...987..214C}
}

@ARTICLE{Rodriguez-Ramirez2024,
       author = {{Rodr{\'\i}guez-Ram{\'\i}rez}, J.~C. and {Bom}, C.~R. and {Fraga}, B. and {Nemmen}, R.},
        title = "{Optical emission model for Binary Black Hole merger remnants travelling through discs of Active Galactic Nuclei}",
      journal = {\mnras},
         year = 2024,
        month = jan,
       volume = {527},
       number = {3},
        pages = {6076-6089},
          doi = {10.1093/mnras/stad3575},
archivePrefix = {arXiv},
       eprint = {2304.10567},
 primaryClass = {astro-ph.HE},
       adsurl = {https://ui.adsabs.harvard.edu/abs/2024MNRAS.527.6076R}
}

@ARTICLE{Ma2025,
       author = {{Ma}, Zhi-Peng and {Wang}, Kai and {Wu}, Qingwen and {Wang}, Jian-Min},
        title = "{Electromagnetic flares associated with gravitational waves from binary black hole mergers in AGN accretion disks}",
      journal = {\prd},
         year = 2025,
        month = apr,
       volume = {111},
       number = {8},
          eid = {083033},
        pages = {083033},
          doi = {10.1103/PhysRevD.111.083033},
archivePrefix = {arXiv},
       eprint = {2409.18567},
 primaryClass = {astro-ph.HE},
       adsurl = {https://ui.adsabs.harvard.edu/abs/2025PhRvD.111h3033M}
}

@ARTICLE{Zhang2025_S241125n,
       author = {{Zhang}, Shu-Rui and {Wang}, Yu and {Yuan}, Ye-Fei and {Tagawa}, Hiromichi and {Wei}, Yun-Feng and {Li}, Liang and {Liu}, Zheng-Yan and {Zhao}, Wen and {Cai}, Rong-Gen},
        title = "{LVK S241125n: Massive Binary Black Hole Merger Produces Gamma Ray Burst in Active Galactic Nucleus Disk}",
      journal = {\apj},
         year = 2026,
        month = feb,
       volume = {998},
       number = {1},
          eid = {171},
        pages = {171},
          doi = {10.3847/1538-4357/ae3319},
archivePrefix = {arXiv},
       eprint = {2505.10395},
 primaryClass = {astro-ph.HE},
       adsurl = {https://ui.adsabs.harvard.edu/abs/2026ApJ...998..171Z}
}

@ARTICLE{LIGO2025_GW231123,
       author = {{Abac}, A.~G. and {Abouelfettouh}, I. and {Acernese}, F. and {Ackley}, K. and {Adamcewicz}, C. and {Adhicary}, S. and {Adhikari}, D. and {Adhikari}, N. and {Adhikari}, R.~X. and {Adkins}, V.~K. and {Afroz}, S. and {Agapito}, A. and {Agarwal}, D. and {Agathos}, M. and {Aggarwal}, N. and {Aggarwal}, S. and {Aguiar}, O.~D. and {Ahrend}, I.-L. and {Aiello}, L. and {Ain}, A. and {Ajith}, P. and {Akutsu}, T. and {Albanesi}, S. and {Ali}, W. and {Al-Kershi}, S. and {All{\'e}n{\'e}}, C. and {Allocca}, A. and {Al-Shammari}, S. and {Altin}, P.~A. and {Alvarez-Lopez}, S. and {Amar}, W. and {Amarasinghe}, O. and {Amato}, A. and {Amicucci}, F. and {Amra}, C. and {Ananyeva}, A. and {Anderson}, S.~B. and {Anderson}, W.~G. and {Andia}, M. and {Ando}, M. and {Andr{\'e}s-Carcasona}, M. and {Andri{\'c}}, T. and {Anglin}, J. and {Ansoldi}, S. and {Antelis}, J.~M. and {Antier}, S. and {Aoumi}, M. and {Appavuravther}, E.~Z. and {Appert}, S. and {Apple}, S.~K. and {Arai}, K. and {Alvarez}, C. Araujo and {Araya}, A. and {Araya}, M.~C. and {Arca Sedda}, M. and {Areeda}, J.~S. and {Aritomi}, N. and {Armato}, F. and {Armstrong}, S. and {Arnaud}, N. and {Arogeti}, M. and {Aronson}, S.~M. and {Arun}, K.~G. and {Ashton}, G. and {Aso}, Y. and {Asprea}, L. and {Assiduo}, M. and {Assis de Souza Melo}, S. and {Aston}, S.~M. and {Astone}, P. and {Attadio}, F. and {Aubin}, F. and {AultONeal}, K. and {Avallone}, G. and {Avila}, E.~A. and {Babak}, S. and {Badger}, C. and {Bae}, S. and {Bagnasco}, S. and {Baiotti}, L. and {Bajpai}, R. and {Baka}, T. and {Baker}, A.~M. and {Baker}, K.~A. and {Baker}, T. and {Baldi}, G. and {Baldicchi}, N. and {Ball}, M. and {Ballardin}, G. and {Ballmer}, S.~W. and {Banagiri}, S. and {Banerjee}, B. and {Bankar}, D. and {Baptiste}, T.~M. and {Baral}, P. and {Baratti}, M. and {Barayoga}, J.~C. and {Barish}, B.~C. and {Barker}, D. and {Barman}, N. and {Barneo}, P. and {Barone}, F. and {Barr}, B. and {Barsotti}, L. and {Barsuglia}, M. and {Barta}, D. and {Bartoletti}, A.~M. and {Barton}, M.~A. and {Bartos}, I. and {Basalaev}, A. and {Bassiri}, R. and {Basti}, A. and {Bawaj}, M. and {Baxi}, P. and {Bayley}, J.~C. and {Baylor}, A.~C. and {Baynard}, II, P.~A. and {Bazzan}, M. and {Bedakihale}, V.~M. and {Beirnaert}, F. and {Bejger}, M. and {Belardinelli}, D. and {Bell}, A.~S. and {Bellie}, D.~S. and {Bellizzi}, L. and {Benoit}, W. and {Bentara}, I. and {Bentley}, J.~D. and {Ben Yaala}, M. and {Bera}, S. and {Bergamin}, F. and {Berger}, B.~K. and {Bernuzzi}, S. and {Beroiz}, M. and {Berry}, C.~P.~L. and {Bersanetti}, D. and {Bertheas}, T. and {Bertolini}, A. and {Betzwieser}, J. and {Beveridge}, D. and {Bevilacqua}, G. and {Bevins}, N. and {Bhandare}, R. and {Bhatt}, R. and {Bhattacharjee}, D. and {Bhattacharyya}, S. and {Bhaumik}, S. and {Bhagwat}, S. and {Biancalana}, V. and {Bianchi}, A. and {Bilenko}, I.~A. and {Billingsley}, G. and {Binetti}, A. and {Bini}, S. and {Binu}, C. and {Biot}, S. and {Birnholtz}, O. and {Biscoveanu}, S. and {Bisht}, A. and {Bitossi}, M. and {Bizouard}, M.-A. and {Blaber}, S. and {Blackburn}, J.~K. and {Blagg}, L.~A. and {Blair}, C.~D. and {Blair}, D.~G. and {Bode}, N. and {Boettner}, N. and {Boileau}, G. and {Boldrini}, M. and {Bolingbroke}, G.~N. and {Bolliand}, A. and {Bonavena}, L.~D. and {Bondarescu}, R. and {Bondu}, F. and {Bonilla}, E. and {Bonilla}, M.~S. and {Bonino}, A. and {Bonnand}, R. and {Borchers}, A. and {Borhanian}, S. and {Boschi}, V. and {Bose}, S. and {Bossilkov}, V. and {Bothra}, Y. and {Boudon}, A. and {Bourg}, L. and {Boyle}, M. and {Bozzi}, A. and {Bradaschia}, C. and {Brady}, P.~R. and {Branch}, A. and {Branchesi}, M. and {Braun}, I. and {Briant}, T. and {Brillet}, A. and {Brinkmann}, M. and {Brockill}, P. and {Brockmueller}, E. and {Brooks}, A.~F.},
        title = "{GW231123: A Binary Black Hole Merger with Total Mass 190─265 M$_{{\ensuremath{\odot}}}$}",
      journal = {\apjl},
         year = 2025,
        month = nov,
       volume = {993},
       number = {1},
          eid = {L25},
        pages = {L25},
          doi = {10.3847/2041-8213/ae0c9c},
archivePrefix = {arXiv},
       eprint = {2507.08219},
 primaryClass = {astro-ph.HE},
       adsurl = {https://ui.adsabs.harvard.edu/abs/2025ApJ...993L..25A}
}

@ARTICLE{RomeroShaw2025,
       author = {{Romero-Shaw}, Isobel and {Stegmann}, Jakob and {Tagawa}, Hiromichi and {Gerosa}, Davide and {Samsing}, Johan and {Gupte}, Nihar and {Green}, Stephen R.},
        title = "{GW200208\_222617 as an eccentric black-hole binary merger: Properties and astrophysical implications}",
      journal = {\prd},
         year = 2025,
        month = sep,
       volume = {112},
       number = {6},
          eid = {063052},
        pages = {063052},
          doi = {10.1103/jj7m-x66y},
archivePrefix = {arXiv},
       eprint = {2506.17105},
 primaryClass = {astro-ph.HE},
       adsurl = {https://ui.adsabs.harvard.edu/abs/2025PhRvD.112f3052R}
}

@ARTICLE{Tanikawa2022,
       author = {{Tanikawa}, Ataru and {Yoshida}, Takashi and {Kinugawa}, Tomoya and {Trani}, Alessandro A. and {Hosokawa}, Takashi and {Susa}, Hajime and {Omukai}, Kazuyuki},
        title = "{Merger Rate Density of Binary Black Holes through Isolated Population I, II, III and Extremely Metal-poor Binary Star Evolution}",
      journal = {\apj},
         year = 2022,
        month = feb,
       volume = {926},
       number = {1},
          eid = {83},
        pages = {83},
          doi = {10.3847/1538-4357/ac4247},
archivePrefix = {arXiv},
       eprint = {2110.10846},
 primaryClass = {astro-ph.HE},
       adsurl = {https://ui.adsabs.harvard.edu/abs/2022ApJ...926...83T}
}

@ARTICLE{Coughlin2014,
       author = {{Coughlin}, Eric R. and {Begelman}, Mitchell C.},
        title = "{Hyperaccretion during Tidal Disruption Events: Weakly Bound Debris Envelopes and Jets}",
      journal = {\apj},
         year = 2014,
        month = feb,
       volume = {781},
       number = {2},
          eid = {82},
        pages = {82},
          doi = {10.1088/0004-637X/781/2/82},
archivePrefix = {arXiv},
       eprint = {1312.5314},
 primaryClass = {astro-ph.HE},
       adsurl = {https://ui.adsabs.harvard.edu/abs/2014ApJ...781...82C}
}

@ARTICLE{Begelman2017,
       author = {{Begelman}, Mitchell C. and {Volonteri}, Marta},
        title = "{Hyperaccreting black holes in galactic nuclei}",
      journal = {\mnras},
         year = 2017,
        month = jan,
       volume = {464},
       number = {1},
        pages = {1102-1107},
          doi = {10.1093/mnras/stw2446},
archivePrefix = {arXiv},
       eprint = {1609.07137},
 primaryClass = {astro-ph.HE},
       adsurl = {https://ui.adsabs.harvard.edu/abs/2017MNRAS.464.1102B}
}

@ARTICLE{Sagynbayeva2025,
       author = {{Sagynbayeva}, Sabina and {Li}, Rixin and {Kuznetsova}, Aleksandra and {Zhu}, Zhaohuan and {Jiang}, Yan-Fei and {Armitage}, Philip J.},
        title = "{Circumplanetary Disks are Rare Around Planets at Large Orbital Radii: A Parameter Survey of Flow Morphology Around Giant Planets}",
      journal = {\apj},
         year = 2025,
        month = jul,
       volume = {987},
       number = {2},
          eid = {216},
        pages = {216},
          doi = {10.3847/1538-4357/add934},
archivePrefix = {arXiv},
       eprint = {2410.14896},
 primaryClass = {astro-ph.EP},
       adsurl = {https://ui.adsabs.harvard.edu/abs/2025ApJ...987..216S}
}

@ARTICLE{Fragile2025,
       author = {{Fragile}, P. Chris and {Middleton}, Matthew J. and {Bollimpalli}, Deepika A. and {Smith}, Zach},
        title = "{Long time-scale numerical simulations of large supercritical accretion discs}",
      journal = {\mnras},
         year = 2025,
        month = jul,
       volume = {540},
       number = {3},
        pages = {2820-2829},
          doi = {10.1093/mnras/staf890},
archivePrefix = {arXiv},
       eprint = {2505.08859},
 primaryClass = {astro-ph.HE},
       adsurl = {https://ui.adsabs.harvard.edu/abs/2025MNRAS.540.2820F}
}

@ARTICLE{Alexander2009,
       author = {{Alexander}, Tal and {Hopman}, Clovis},
        title = "{Strong Mass Segregation Around a Massive Black Hole}",
      journal = {\apj},
         year = 2009,
        month = jun,
       volume = {697},
       number = {2},
        pages = {1861-1869},
          doi = {10.1088/0004-637X/697/2/1861},
archivePrefix = {arXiv},
       eprint = {0808.3150},
 primaryClass = {astro-ph},
       adsurl = {https://ui.adsabs.harvard.edu/abs/2009ApJ...697.1861A}
}

@ARTICLE{EpsteinMartin2025,
       author = {{Epstein-Martin}, Marguerite and {Tagawa}, Hiromichi and {Haiman}, Zolt{\'a}n and {Perna}, Rosalba},
        title = "{Time-dependent models of AGN discs with radiation from embedded stellar-mass black holes}",
      journal = {\mnras},
         year = 2025,
        month = mar,
       volume = {537},
       number = {4},
        pages = {3396-3420},
          doi = {10.1093/mnras/staf237},
archivePrefix = {arXiv},
       eprint = {2405.09380},
 primaryClass = {astro-ph.HE},
       adsurl = {https://ui.adsabs.harvard.edu/abs/2025MNRAS.537.3396E}
}

@ARTICLE{DeLaunay2024GCN.38351....1D,
       author = {{DeLaunay}, James and {Tohuvavohu}, Aaron and {Ronchini}, Samuele and {Raman}, Gayathri and {Kennea}, Jamie A. and {Parsotan}, Tyler},
        title = "{LIGO/Virgo/KAGRA S241125n: Preliminary flux estimate and spectral analysis of the Swift/BAT-GUANO candidate counterpart}",
      journal = {GRB Coordinates Network},
         year = 2024,
        month = nov,
       volume = {38351},
        pages = {1},
       adsurl = {https://ui.adsabs.harvard.edu/abs/2024GCN.38351....1D}
}

@ARTICLE{Veronesi2025,
       author = {{Veronesi}, Niccol{\`o} and {van Velzen}, Sjoert and {Rossi}, Elena Maria},
        title = "{AGN flares as counterparts to LIGO/Virgo mergers: no confident causal connection in spatial correlation analysis}",
      journal = {\mnras},
         year = 2025,
        month = jan,
       volume = {536},
       number = {3},
        pages = {3112-3122},
          doi = {10.1093/mnras/stae2787},
archivePrefix = {arXiv},
       eprint = {2405.05318},
 primaryClass = {astro-ph.HE},
       adsurl = {https://ui.adsabs.harvard.edu/abs/2025MNRAS.536.3112V}
}

@ARTICLE{Morton2023,
       author = {{Morton}, Sophia L. and {Rinaldi}, Stefano and {Torres-Orjuela}, Alejandro and {Derdzinski}, Andrea and {Vaccaro}, M. Paola and {Del Pozzo}, Walter},
        title = "{GW190521: A binary black hole merger inside an active galactic nucleus?}",
      journal = {\prd},
         year = 2023,
        month = dec,
       volume = {108},
       number = {12},
          eid = {123039},
        pages = {123039},
          doi = {10.1103/PhysRevD.108.123039},
archivePrefix = {arXiv},
       eprint = {2310.16025},
 primaryClass = {gr-qc},
       adsurl = {https://ui.adsabs.harvard.edu/abs/2023PhRvD.108l3039M}
}

@ARTICLE{Graham2023,
       author = {{Graham}, Matthew J. and {McKernan}, Barry and {Ford}, K.~E. Saavik and {Stern}, Daniel and {Djorgovski}, S.~G. and {Coughlin}, Michael and {Burdge}, Kevin B. and {Bellm}, Eric C. and {Helou}, George and {Mahabal}, Ashish A. and {Masci}, Frank J. and {Purdum}, Josiah and {Rosnet}, Philippe and {Rusholme}, Ben},
        title = "{A Light in the Dark: Searching for Electromagnetic Counterparts to Black Hole-Black Hole Mergers in LIGO/Virgo O3 with the Zwicky Transient Facility}",
      journal = {\apj},
         year = 2023,
        month = jan,
       volume = {942},
       number = {2},
          eid = {99},
        pages = {99},
          doi = {10.3847/1538-4357/aca480},
archivePrefix = {arXiv},
       eprint = {2209.13004},
 primaryClass = {astro-ph.HE},
       adsurl = {https://ui.adsabs.harvard.edu/abs/2023ApJ...942...99G}
}

@ARTICLE{Gilbaum2025,
       author = {{Gilbaum}, Shmuel and {Grishin}, Evgeni and {Stone}, Nicholas C. and {Mandel}, Ilya},
        title = "{How to Escape from a Trap: Outcomes of Repeated Black Hole Mergers in Active Galactic Nuclei}",
      journal = {\apjl},
         year = 2025,
        month = mar,
       volume = {982},
       number = {1},
          eid = {L13},
        pages = {L13},
          doi = {10.3847/2041-8213/adb7dc},
archivePrefix = {arXiv},
       eprint = {2410.19904},
 primaryClass = {astro-ph.HE},
       adsurl = {https://ui.adsabs.harvard.edu/abs/2025ApJ...982L..13G}
}

@ARTICLE{Madau2014,
       author = {{Madau}, Piero and {Dickinson}, Mark},
        title = "{Cosmic Star-Formation History}",
      journal = {\araa},
         year = 2014,
        month = aug,
       volume = {52},
        pages = {415-486},
          doi = {10.1146/annurev-astro-081811-125615},
archivePrefix = {arXiv},
       eprint = {1403.0007},
 primaryClass = {astro-ph.CO},
       adsurl = {https://ui.adsabs.harvard.edu/abs/2014ARA&A..52..415M}
}

@ARTICLE{Zhu2025,
       author = {{Zhu}, Liang-Gui and {Chen}, Xian},
        title = "{Evidence of a Fraction of LIGO/Virgo/KAGRA Events Coming from Active Galactic Nuclei}",
      journal = {\apjl},
         year = 2025,
        month = aug,
       volume = {989},
       number = {1},
          eid = {L15},
        pages = {L15},
          doi = {10.3847/2041-8213/adf31f},
archivePrefix = {arXiv},
       eprint = {2505.02924},
 primaryClass = {astro-ph.HE},
       adsurl = {https://ui.adsabs.harvard.edu/abs/2025ApJ...989L..15Z}
}

@ARTICLE{Fahrion2022,
       author = {{Fahrion}, Katja and {Leaman}, Ryan and {Lyubenova}, Mariya and {van de Ven}, Glenn},
        title = "{Disentangling the formation mechanisms of nuclear star clusters}",
      journal = {\aap},
         year = 2022,
        month = feb,
       volume = {658},
          eid = {A172},
        pages = {A172},
          doi = {10.1051/0004-6361/202039778},
archivePrefix = {arXiv},
       eprint = {2112.05610},
 primaryClass = {astro-ph.GA},
       adsurl = {https://ui.adsabs.harvard.edu/abs/2022A&A...658A.172F}
}

@ARTICLE{Secunda2019,
       author = {{Secunda}, Amy and {Bellovary}, Jillian and {Mac Low}, Mordecai-Mark and {Ford}, K.~E. Saavik and {McKernan}, Barry and {Leigh}, Nathan W.~C. and {Lyra}, Wladimir and {S{\'a}ndor}, Zsolt},
        title = "{Orbital Migration of Interacting Stellar Mass Black Holes in Disks around Supermassive Black Holes}",
      journal = {\apj},
         year = 2019,
        month = jun,
       volume = {878},
       number = {2},
          eid = {85},
        pages = {85},
          doi = {10.3847/1538-4357/ab20ca},
archivePrefix = {arXiv},
       eprint = {1807.02859},
 primaryClass = {astro-ph.HE},
       adsurl = {https://ui.adsabs.harvard.edu/abs/2019ApJ...878...85S}
}

@ARTICLE{Secunda2020,
       author = {{Secunda}, Amy and {Bellovary}, Jillian and {Mac Low}, Mordecai-Mark and {Ford}, K.~E. Saavik and {McKernan}, Barry and {Leigh}, Nathan W.~C. and {Lyra}, Wladimir and {S{\'a}ndor}, Zsolt and {Adorno}, Jose I.},
        title = "{Orbital Migration of Interacting Stellar Mass Black Holes in Disks around Supermassive Black Holes. II. Spins and Incoming Objects}",
      journal = {\apj},
         year = 2020,
        month = nov,
       volume = {903},
       number = {2},
          eid = {133},
        pages = {133},
          doi = {10.3847/1538-4357/abbc1d},
archivePrefix = {arXiv},
       eprint = {2004.11936},
 primaryClass = {astro-ph.HE},
       adsurl = {https://ui.adsabs.harvard.edu/abs/2020ApJ...903..133S}
}

@ARTICLE{Miralda-Escude2005,
       author = {{Miralda-Escud{\'e}}, Jordi and {Kollmeier}, Juna A.},
        title = "{Star Captures by Quasar Accretion Disks: A Possible Explanation of the M-{\ensuremath{\sigma}} Relation}",
      journal = {\apj},
         year = 2005,
        month = jan,
       volume = {619},
       number = {1},
        pages = {30-40},
          doi = {10.1086/426467},
archivePrefix = {arXiv},
       eprint = {astro-ph/0310717},
 primaryClass = {astro-ph},
       adsurl = {https://ui.adsabs.harvard.edu/abs/2005ApJ...619...30M}
}

@ARTICLE{Bartos2017,
       author = {{Bartos}, Imre and {Kocsis}, Bence and {Haiman}, Zolt{\'a}n and {M{\'a}rka}, Szabolcs},
        title = "{Rapid and Bright Stellar-mass Binary Black Hole Mergers in Active Galactic Nuclei}",
      journal = {\apj},
         year = 2017,
        month = feb,
       volume = {835},
       number = {2},
          eid = {165},
        pages = {165},
          doi = {10.3847/1538-4357/835/2/165},
archivePrefix = {arXiv},
       eprint = {1602.03831},
 primaryClass = {astro-ph.HE},
       adsurl = {https://ui.adsabs.harvard.edu/abs/2017ApJ...835..165B}
}

@ARTICLE{2021ApJ...920L..42G,
       author = {{Gayathri}, V. and {Yang}, Y. and {Tagawa}, H. and {Haiman}, Z. and {Bartos}, I.},
        title = "{Black Hole Mergers of AGN Origin in LIGO-Virgo's O1-O3a Observing Periods}",
      journal = {\apjl},
         year = 2021,
        month = oct,
       volume = {920},
       number = {2},
          eid = {L42},
        pages = {L42},
          doi = {10.3847/2041-8213/ac2cc1},
archivePrefix = {arXiv},
       eprint = {2104.10253},
 primaryClass = {gr-qc},
       adsurl = {https://ui.adsabs.harvard.edu/abs/2021ApJ...920L..42G}
}

@ARTICLE{ChenY2023,
       author = {{Chen}, Yi-Xian and {Jiang}, Yan-Fei and {Goodman}, Jeremy and {Ostriker}, Eve C.},
        title = "{3D Radiation Hydrodynamic Simulations of Gravitational Instability in AGN Accretion Disks: Effects of Radiation Pressure}",
      journal = {\apj},
         year = 2023,
        month = may,
       volume = {948},
       number = {2},
          eid = {120},
        pages = {120},
          doi = {10.3847/1538-4357/acc023},
archivePrefix = {arXiv},
       eprint = {2302.10868},
 primaryClass = {astro-ph.HE},
       adsurl = {https://ui.adsabs.harvard.edu/abs/2023ApJ...948..120C}
}

@ARTICLE{Ford2022,
       author = {{Ford}, K.~E. Saavik and {McKernan}, Barry},
        title = "{Binary black hole merger rates in AGN discs versus nuclear star clusters: loud beats quiet}",
      journal = {\mnras},
         year = 2022,
        month = dec,
       volume = {517},
       number = {4},
        pages = {5827-5834},
          doi = {10.1093/mnras/stac2861},
archivePrefix = {arXiv},
       eprint = {2109.03212},
 primaryClass = {astro-ph.HE},
       adsurl = {https://ui.adsabs.harvard.edu/abs/2022MNRAS.517.5827F}
}

@article{Derdzinski23,
       author = {{Derdzinski}, Andrea and {Mayer}, Lucio},
        title = "{In situ extreme mass ratio inspirals via subparsec formation and migration of stars in thin, gravitationally unstable AGN discs}",
      journal = {\mnras},
         year = 2023,
        month = may,
       volume = {521},
       number = {3},
        pages = {4522-4543},
          doi = {10.1093/mnras/stad749},
archivePrefix = {arXiv},
       eprint = {2205.10382},
 primaryClass = {astro-ph.GA},
       adsurl = {https://ui.adsabs.harvard.edu/abs/2023MNRAS.521.4522D}
}

@article{Derdzinski19,
       author = {{Derdzinski}, A.~M. and {D'Orazio}, D. and {Duffell}, P. and {Haiman}, Z. and {MacFadyen}, A.},
        title = "{Probing gas disc physics with LISA: simulations of an intermediate mass ratio inspiral in an accretion disc}",
      journal = {\mnras},
         year = 2019,
        month = jun,
       volume = {486},
       number = {2},
        pages = {2754-2765},
          doi = {10.1093/mnras/stz1026},
archivePrefix = {arXiv},
       eprint = {1810.03623},
 primaryClass = {astro-ph.HE},
       adsurl = {https://ui.adsabs.harvard.edu/abs/2019MNRAS.486.2754D}
}

@ARTICLE{Han_Yang2024,
       author = {{Yang}, Shu-Cheng and {Han}, Wen-Biao and {Tagawa}, Hiromichi and {Li}, Song and {Zhang}, Chen},
        title = "{Indication for a Compact Object Next to a LIGO{\textendash}Virgo Binary Black Hole Merger}",
      journal = {\apjl},
         year = 2025,
        month = aug,
       volume = {988},
       number = {2},
          eid = {L41},
        pages = {L41},
          doi = {10.3847/2041-8213/adeaad},
archivePrefix = {arXiv},
       eprint = {2401.01743},
 primaryClass = {astro-ph.HE},
       adsurl = {https://ui.adsabs.harvard.edu/abs/2025ApJ...988L..41Y}
}

@ARTICLE{Wong2019,
       author = {{Wong}, Kaze W.~K. and {Baibhav}, Vishal and {Berti}, Emanuele},
        title = "{Binary radial velocity measurements with space-based gravitational-wave detectors}",
      journal = {\mnras},
         year = 2019,
        month = oct,
       volume = {488},
       number = {4},
        pages = {5665-5670},
          doi = {10.1093/mnras/stz2077},
archivePrefix = {arXiv},
       eprint = {1902.01402},
 primaryClass = {astro-ph.HE},
       adsurl = {https://ui.adsabs.harvard.edu/abs/2019MNRAS.488.5665W}
}

@article{Meiron17, 
       author = {{Meiron}, Yohai and {Kocsis}, Bence and {Loeb}, Abraham},
        title = "{Detecting Triple Systems with Gravitational Wave Observations}",
      journal = {\apj},
         year = 2017,
        month = jan,
       volume = {834},
       number = {2},
          eid = {200},
        pages = {200},
          doi = {10.3847/1538-4357/834/2/200},
archivePrefix = {arXiv},
       eprint = {1604.02148},
 primaryClass = {astro-ph.HE},
       adsurl = {https://ui.adsabs.harvard.edu/abs/2017ApJ...834..200M}
}

@ARTICLE{Samsing2022,
       author = {{Samsing}, J. and {Bartos}, I. and {D'Orazio}, D.~J. and {Haiman}, Z. and {Kocsis}, B. and {Leigh}, N.~W.~C. and {Liu}, B. and {Pessah}, M.~E. and {Tagawa}, H.},
        title = "{AGN as potential factories for eccentric black hole mergers}",
      journal = {\nat},
         year = 2022,
        month = mar,
       volume = {603},
       number = {7900},
        pages = {237-240},
          doi = {10.1038/s41586-021-04333-1},
archivePrefix = {arXiv},
       eprint = {2010.09765},
 primaryClass = {astro-ph.HE},
       adsurl = {https://ui.adsabs.harvard.edu/abs/2022Natur.603..237S}
}

@ARTICLE{Whitehead_2023b_novae,
       author = {{Whitehead}, Henry and {Rowan}, Connar and {Boekholt}, Tjarda and {Kocsis}, Bence},
        title = "{Disc novae: thermodynamics of gas-assisted binary black hole formation in AGN discs}",
      journal = {\mnras},
         year = 2024,
        month = sep,
       volume = {533},
       number = {2},
        pages = {1766-1781},
          doi = {10.1093/mnras/stae1866},
archivePrefix = {arXiv},
       eprint = {2312.14431},
 primaryClass = {astro-ph.HE},
       adsurl = {https://ui.adsabs.harvard.edu/abs/2024MNRAS.533.1766W}
}

@ARTICLE{Qian2023,
       author = {{Qian}, Kecheng and {Li}, Jiaru and {Lai}, Dong},
        title = "{Dynamical Friction Models for Black Hole Binary Formation in Active Galactic Nucleus Disks}",
      journal = {\apj},
         year = 2024,
        month = feb,
       volume = {962},
       number = {2},
          eid = {143},
        pages = {143},
          doi = {10.3847/1538-4357/ad1b53},
archivePrefix = {arXiv},
       eprint = {2310.12208},
 primaryClass = {astro-ph.HE},
       adsurl = {https://ui.adsabs.harvard.edu/abs/2024ApJ...962..143Q}
}

@ARTICLE{Rowan2023,
       author = {{Rowan}, Connar and {Whitehead}, Henry and {Boekholt}, Tjarda and {Kocsis}, Bence and {Haiman}, Zolt{\'a}n},
        title = "{Black hole binaries in AGN accretion discs - II. Gas effects on black hole satellite scatterings}",
      journal = {\mnras},
         year = 2024,
        month = feb,
       volume = {527},
       number = {4},
        pages = {10448-10468},
          doi = {10.1093/mnras/stad3641},
archivePrefix = {arXiv},
       eprint = {2309.14433},
 primaryClass = {astro-ph.HE},
       adsurl = {https://ui.adsabs.harvard.edu/abs/2024MNRAS.52710448R}
}

@ARTICLE{Whitehead2023,
       author = {{Whitehead}, Henry and {Rowan}, Connar and {Boekholt}, Tjarda and {Kocsis}, Bence},
        title = "{Gas assisted binary black hole formation in AGN discs}",
      journal = {\mnras},
         year = 2024,
        month = jul,
       volume = {531},
       number = {4},
        pages = {4656-4680},
          doi = {10.1093/mnras/stae1430},
archivePrefix = {arXiv},
       eprint = {2309.11561},
 primaryClass = {astro-ph.GA},
       adsurl = {https://ui.adsabs.harvard.edu/abs/2024MNRAS.531.4656W}
}

@ARTICLE{Rozner2023,
       author = {{Rozner}, Mor and {Generozov}, Aleksey and {Perets}, Hagai B.},
        title = "{Binary formation through gas-assisted capture and the implications for stellar, planetary and compact-object evolution}",
      journal = {\mnras},
         year = 2023,
        month = feb,
          doi = {10.1093/mnras/stad603},
archivePrefix = {arXiv},
       eprint = {2212.00807},
 primaryClass = {astro-ph.GA},
       adsurl = {https://ui.adsabs.harvard.edu/abs/2023MNRAS.tmp..587R}
}

@ARTICLE{DeLaurentiis2023,
       author = {{DeLaurentiis}, Stanislav and {Epstein-Martin}, Marguerite and {Haiman}, Zolt{\'a}n},
        title = "{Gas dynamical friction as a binary formation mechanism in AGN discs}",
      journal = {\mnras},
         year = 2023,
        month = jul,
       volume = {523},
       number = {1},
        pages = {1126-1139},
          doi = {10.1093/mnras/stad1412},
archivePrefix = {arXiv},
       eprint = {2212.02650},
 primaryClass = {astro-ph.HE},
       adsurl = {https://ui.adsabs.harvard.edu/abs/2023MNRAS.523.1126D}
}

@ARTICLE{LiJiaru2023,
       author = {{Li}, Jiaru and {Dempsey}, Adam M. and {Li}, Hui and {Lai}, Dong and {Li}, Shengtai},
        title = "{Hydrodynamical Simulations of Black Hole Binary Formation in AGN Disks}",
      journal = {\apjl},
         year = 2023,
        month = feb,
       volume = {944},
       number = {2},
          eid = {L42},
        pages = {L42},
          doi = {10.3847/2041-8213/acb934},
archivePrefix = {arXiv},
       eprint = {2211.10357},
 primaryClass = {astro-ph.HE},
       adsurl = {https://ui.adsabs.harvard.edu/abs/2023ApJ...944L..42L}
}

@ARTICLE{Rowan2022,
       author = {{Rowan}, Connar and {Boekholt}, Tjarda and {Kocsis}, Bence and {Haiman}, Zolt{\'a}n},
        title = "{Black hole binary formation in AGN discs: from isolation to merger}",
      journal = {\mnras},
         year = 2023,
        month = sep,
       volume = {524},
       number = {2},
        pages = {2770-2796},
          doi = {10.1093/mnras/stad1926},
archivePrefix = {arXiv},
       eprint = {2212.06133},
 primaryClass = {astro-ph.GA},
       adsurl = {https://ui.adsabs.harvard.edu/abs/2023MNRAS.524.2770R}
}

@ARTICLE{Boekholt2023,
       author = {{Boekholt}, Tjarda C.~N. and {Rowan}, Connar and {Kocsis}, Bence},
        title = "{On the Jacobi capture origin of binaries with applications to the Earth-Moon system and black holes in galactic nuclei}",
      journal = {\mnras},
         year = 2023,
        month = feb,
       volume = {518},
       number = {4},
        pages = {5653-5669},
          doi = {10.1093/mnras/stac3495},
archivePrefix = {arXiv},
       eprint = {2203.09646},
 primaryClass = {astro-ph.EP},
       adsurl = {https://ui.adsabs.harvard.edu/abs/2023MNRAS.518.5653B}
}

@ARTICLE{LiJiaru2022,
       author = {{Li}, Jiaru and {Lai}, Dong and {Rodet}, Laetitia},
        title = "{Long-term Evolution of Tightly Packed Stellar Black Holes in AGN Disks: Formation of Merging Black Hole Binaries via Close Encounters}",
      journal = {\apj},
         year = 2022,
        month = aug,
       volume = {934},
       number = {2},
          eid = {154},
        pages = {154},
          doi = {10.3847/1538-4357/ac7c0d},
archivePrefix = {arXiv},
       eprint = {2203.05584},
 primaryClass = {astro-ph.HE},
       adsurl = {https://ui.adsabs.harvard.edu/abs/2022ApJ...934..154L}
}

@ARTICLE{Grishin2023,
       author = {{Grishin}, Evgeni and {Gilbaum}, Shmuel and {Stone}, Nicholas C.},
        title = "{The effect of thermal torques on AGN disc migration traps and gravitational wave populations}",
      journal = {\mnras},
         year = 2024,
        month = may,
       volume = {530},
       number = {2},
        pages = {2114-2132},
          doi = {10.1093/mnras/stae828},
archivePrefix = {arXiv},
       eprint = {2307.07546},
 primaryClass = {astro-ph.HE},
       adsurl = {https://ui.adsabs.harvard.edu/abs/2024MNRAS.530.2114G}
}

@article{Bellovary16, 
       author = {{Bellovary}, Jillian M. and {Mac Low}, Mordecai-Mark and {McKernan}, Barry and {Ford}, K.~E. Saavik},
        title = "{Migration Traps in Disks around Supermassive Black Holes}",
      journal = {\apjl},
         year = 2016,
        month = mar,
       volume = {819},
       number = {2},
          eid = {L17},
        pages = {L17},
          doi = {10.3847/2041-8205/819/2/L17},
archivePrefix = {arXiv},
       eprint = {1511.00005},
 primaryClass = {astro-ph.GA},
       adsurl = {https://ui.adsabs.harvard.edu/abs/2016ApJ...819L..17B}
}

@ARTICLE{Milosavljevic2004,
       author = {{Milosavljevi{\'c}}, Milo{\v{s}} and {Loeb}, Abraham},
        title = "{The Link between Warm Molecular Disks in Maser Nuclei and Star Formation near the Black Hole at the Galactic Center}",
      journal = {\apjl},
         year = 2004,
        month = mar,
       volume = {604},
       number = {1},
        pages = {L45-L48},
          doi = {10.1086/383467},
archivePrefix = {arXiv},
       eprint = {astro-ph/0401221},
 primaryClass = {astro-ph},
       adsurl = {https://ui.adsabs.harvard.edu/abs/2004ApJ...604L..45M}
}

@ARTICLE{Goodman2004,
       author = {{Goodman}, J. and {Tan}, Jonathan C.},
        title = "{Supermassive Stars in Quasar Disks}",
      journal = {\apj},
         year = 2004,
        month = jun,
       volume = {608},
       number = {1},
        pages = {108-118},
          doi = {10.1086/386360},
archivePrefix = {arXiv},
       eprint = {astro-ph/0307361},
 primaryClass = {astro-ph},
       adsurl = {https://ui.adsabs.harvard.edu/abs/2004ApJ...608..108G}
}

@ARTICLE{Wang2023_capture,
       author = {{Wang}, Yihan and {Zhu}, Zhaohuan and {Lin}, Douglas N.~C.},
        title = "{Stellar/BH population in AGN discs: direct binary formation from capture objects in nuclei clusters}",
      journal = {\mnras},
         year = 2024,
        month = mar,
       volume = {528},
       number = {3},
        pages = {4958-4975},
          doi = {10.1093/mnras/stae321},
archivePrefix = {arXiv},
       eprint = {2308.09129},
 primaryClass = {astro-ph.GA},
       adsurl = {https://ui.adsabs.harvard.edu/abs/2024MNRAS.528.4958W}
}

@ARTICLE{Generozov2023,
       author = {{Generozov}, A. and {Perets}, H.~B.},
        title = "{Capture of stars into gaseous discs around massive black holes: alignment, circularization, and growth}",
      journal = {\mnras},
         year = 2023,
        month = jun,
       volume = {522},
       number = {2},
        pages = {1763-1778},
          doi = {10.1093/mnras/stad1016},
archivePrefix = {arXiv},
       eprint = {2212.11301},
 primaryClass = {astro-ph.GA},
       adsurl = {https://ui.adsabs.harvard.edu/abs/2023MNRAS.522.1763G}
}

@ARTICLE{Masset2017,
       author = {{Masset}, Fr{\'e}d{\'e}ric S.},
        title = "{Coorbital thermal torques on low-mass protoplanets}",
      journal = {\mnras},
         year = 2017,
        month = dec,
       volume = {472},
       number = {4},
        pages = {4204-4219},
          doi = {10.1093/mnras/stx2271},
archivePrefix = {arXiv},
       eprint = {1708.09807},
 primaryClass = {astro-ph.EP},
       adsurl = {https://ui.adsabs.harvard.edu/abs/2017MNRAS.472.4204M}
}

@ARTICLE{Gilbaum2022,
       author = {{Gilbaum}, Shmuel and {Stone}, Nicholas C.},
        title = "{Feedback-dominated Accretion Flows}",
      journal = {\apj},
         year = 2022,
        month = apr,
       volume = {928},
       number = {2},
          eid = {191},
        pages = {191},
          doi = {10.3847/1538-4357/ac4ded},
archivePrefix = {arXiv},
       eprint = {2107.07519},
 primaryClass = {astro-ph.HE},
       adsurl = {https://ui.adsabs.harvard.edu/abs/2022ApJ...928..191G}
}

@ARTICLE{Peng2021,
       author = {{Peng}, Peng and {Chen}, Xian},
        title = "{The last migration trap of compact objects in AGN accretion disc}",
      journal = {\mnras},
         year = 2021,
        month = jul,
       volume = {505},
       number = {1},
        pages = {1324-1333},
          doi = {10.1093/mnras/stab1419},
archivePrefix = {arXiv},
       eprint = {2104.07685},
 primaryClass = {astro-ph.HE},
       adsurl = {https://ui.adsabs.harvard.edu/abs/2021MNRAS.505.1324P}
}

@ARTICLE{Pan2021,
       author = {{Pan}, Zhen and {Yang}, Huan},
        title = "{Formation rate of extreme mass ratio inspirals in active galactic nuclei}",
      journal = {\prd},
         year = 2021,
        month = may,
       volume = {103},
       number = {10},
          eid = {103018},
        pages = {103018},
          doi = {10.1103/PhysRevD.103.103018},
archivePrefix = {arXiv},
       eprint = {2101.09146},
 primaryClass = {astro-ph.HE},
       adsurl = {https://ui.adsabs.harvard.edu/abs/2021PhRvD.103j3018P}
}

@ARTICLE{Tagawa2023_EM,
       author = {{Tagawa}, Hiromichi and {Kimura}, Shigeo S. and {Haiman}, Zolt{\'a}n and {Perna}, Rosalba and {Bartos}, Imre},
        title = "{Observable Signature of Merging Stellar-mass Black Holes in Active Galactic Nuclei}",
      journal = {\apj},
         year = 2023,
        month = jun,
       volume = {950},
       number = {1},
          eid = {13},
        pages = {13},
          doi = {10.3847/1538-4357/acc4bb},
archivePrefix = {arXiv},
       eprint = {2301.07111},
 primaryClass = {astro-ph.HE},
       adsurl = {https://ui.adsabs.harvard.edu/abs/2023ApJ...950...13T}
}

\end{document}